\documentclass{cernyrep}
\usepackage{graphicx}
\usepackage{colordvi}
\usepackage{color}
\usepackage{latexsym}
\usepackage{epsfig}
\usepackage{amsmath}
\newcommand{\nc}{\newcommand}
\nc{\beq}{\begin{equation}} \nc{\eeq}{\end{equation}}
\nc{\beqa}{\begin{eqnarray}} \nc{\eeqa}{\end{eqnarray}}
\nc{\tb}{\tan\beta}
\nc{\bsg}{b\to s\gamma}
\nc{\btaunu}{b\to\tau\nu}
\nc{\bsmm}{B_s\to\mu\mu}
\nc{\gl}{\tilde g}
\nc{\sq}{\tilde q}
\nc{\mzero}{m_0}
\nc{\mhalf}{m_{1/2}}

\begin{document} \thispagestyle{empty}
\begin{center}
{\Large \bf   Beyond the Standard Model' 17}
\renewcommand{\thefootnote}{\fnsymbol{footnote}}\footnote[2]{Lectures
given at the European School of High-Energy Physics, September 2017, Evola, Portugal}
\renewcommand{\thefootnote}{\arabic{footnote}}
\vspace{10mm}

{\large \bf Dmitry Kazakov } \\[5mm]

{\it Bogoliubov Laboratory of Theoretical Physics, \\Joint Institute for Nuclear Research,
Dubna, Russia }

\vspace{12mm}

\begin{center}
\includegraphics[width=0.40\textwidth]{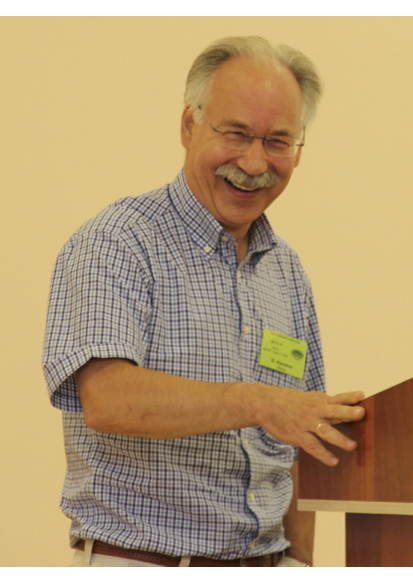}
\end{center}
\vspace{8mm}

\begin{abstract} We discuss the status of the SM - The principles
- The Lagrangian
- The problems
- Open questions
- The ways beyond. Then we consider
possible physics beyond the SM
- New symmetries (Gauge, SUSY, etc)
- New particles (gauge, axion, superpartners)
- New dimensions (extra, large, compact, etc)
- New Paradigm (strings, branes, gravity). In conclusion, we formulate the first priority tasks for the future HEP program.
\end{abstract}
\end{center}

\thispagestyle{empty}


\tableofcontents
\vspace{8mm}

\section{Introduction: The Standard Model}

Physics of elementary particles today is perfectly described by the Standard  Model of fundamental interactions which accumulates all achievements of the recent years. It is usually said that with the discovery of the Higgs boson the Standard Model is completed. Nevertheless, it still contains many puzzles and possibly requires some modification in future. The search for new physics beyond the Standard Model  is inevitably  based on comparison of experimental data with predictions of the Standard Model since the particles observed in the final states are the well-known stable ones and new physics as a rule manifests itself in the form of excess above the SM background. 

It is instructive to remind the main principles in the foundation of the Standard Model and possible ways to go beyond it. They are:
\begin{itemize}
\item  Three groups of gauged symmetries $SU(3)\times SU(2)\times U(1)$
\item  Three families of quarks and leptons in representations  $(3\times 2, 3\times 1, 1\times 2, 1\times 1)$
\item  Brout-Englert-Higgs mechanism of spontaneous EW symmetry breaking accompanied by the Higgs boson
\item  Mixing of flavours with the help of the Cabibbo-Kobayashi-Maskawa (CKM) and the Pontecorvo-Maki-Nakagava-Sakato (PMNS)  matrices
\item  CP violation via the phase factors in the flavour mixing matrices
\item Confinement of quarks and gluons inside hadrons
\item Baryon and lepton number conservation
\item CPT invariance which leads to the existence of antimatter
\end{itemize}

The principles of the Standard Model allow its small modifications with respect  to the minimal scheme. Thus, for instance, it is possible to add new families of matter particles, additional Higgs bosons, the presence or absence of right-handed neutrino, Dirac or  Majorana nature of neutrino
is fully acceptable.

The formalism of the Standard Model is based on local quantum field theory. The SM is described by Lagrangian which is built in accordance with the Lorentz invariance and invariance under three gauged groups of symmetry and also obeys the principle of renormalizability, which means that it contains only the operators of dimension 2, 3 and 4~\cite{PDG}. 
\begin{equation}
{\cal L} ={\cal L}_{gauge} + {\cal L}_{Yukawa} + {\cal L}_{Higgs},
 \end{equation}
\begin{eqnarray*}
{\cal L}_{gauge} & = & -\frac{1}{4} G_{\mu\nu}^aG_{\mu\nu}^a - \frac{1}{4} A_{\mu\nu}^iA_{\mu\nu}^i
-\frac{1}{4} B_{\mu\nu}B_{\mu\nu}\\
 & &  + i\overline{L}_{\alpha}\gamma^{\mu}D_{\mu}L_{\alpha}
+ i\overline{Q}_{\alpha}\gamma^{\mu}D_{\mu}Q_{\alpha} + i\overline{l}_{\alpha}
\gamma^{\mu}D_{\mu}l_{\alpha}\\
 & &  + i\overline{U}_{\alpha}\gamma^{\mu}D_{\mu}U_{
\alpha} + i\overline{D}_{\alpha}\gamma^{\mu}D_{\mu}D_{\alpha} + (D_{\mu}H)^{\dagger}
(D_{\mu}H)\\
&& + i \bar N_\alpha\gamma^\mu\partial_\mu N_\alpha \ \ \  \leftarrow \mbox{possible  rigth-handed neutrino}
\end{eqnarray*}
\begin{eqnarray*}
{\cal L}_{Yukawa} &=& y_{\alpha\beta}^l\overline{L}_{\alpha}l_{\beta}H + y_{\alpha
\beta}^d\overline{Q}_{\alpha}D_{\beta}H + y_{\alpha\beta}^u\overline{Q}_{\alpha}
U_{\beta}\tilde{H} + h.c.,\\
&&+ y^N_{\alpha\beta}\bar L_\alpha N_\beta \tilde H \ \ \  \leftarrow \mbox{possible  rigth-handed neutrino}
\end{eqnarray*}
where $\tilde{H}=i\tau_2H^{\dagger}$.
$${\cal L}_{Higgs} = - V = m^2H^{\dagger}H - \frac{\lambda}{2}(H^{\dagger}H)^2.$$
Here $y$ are the Yukawa and $\lambda$ is the Higgs coupling constants, respectively,
both dimensionless and $m$ is the only dimensional mass parameter.

The symmetries of the SM allow one to fix all the interactions of quarks and leptons which are performed by the exchange of the force carriers, namely, by gluons, W and Z bosons, photons and the Higgs boson in the case of strong, weak, electromagnetic and Yukawa interactions, respectively. The only freedom is the choice of parameters: 3 gauge couplings $g_i$, 3 (or 4) Yukawa matrices $y^k_{\alpha\beta}$, the Higgs coupling $\lambda$, and the mass parameter $m$. All of them are not predicted by the SM but are measured experimentally. The existence of the right-handed neutrino leads to two additional terms in the Lagrangian, the kinetic one and the interaction with the Higgs boson. If the neutrino is a Majorana particle, then one should also add the Majorana mass term.

The Standard model has some drawbacks which, however, are manifested at very high energies where it can possibly be replaced by a new theory. Below, we list some of them.

1) The running couplings of the SM tend to infinity at finite energies (the Landau pole~\cite{Landau}). This is true for the $U(1)$ and the Higgs couplings (see Fig.1, left). 
\begin{figure}[h!]
\begin{center}
\leavevmode
\includegraphics[width=0.39\textwidth]{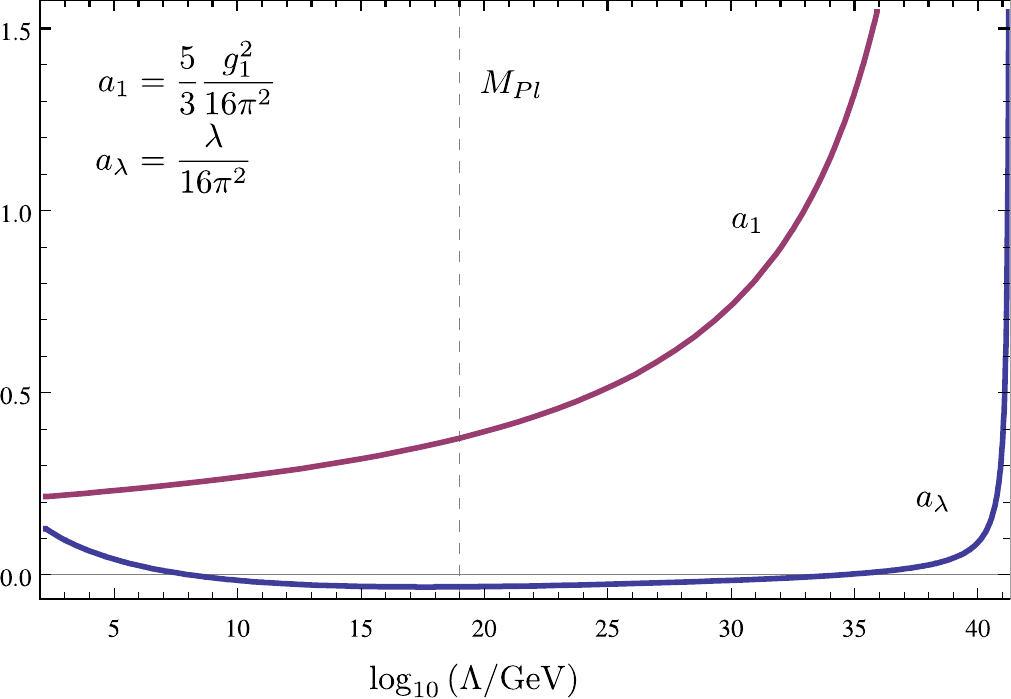}\hspace{1cm}
\includegraphics[width=0.43\textwidth]{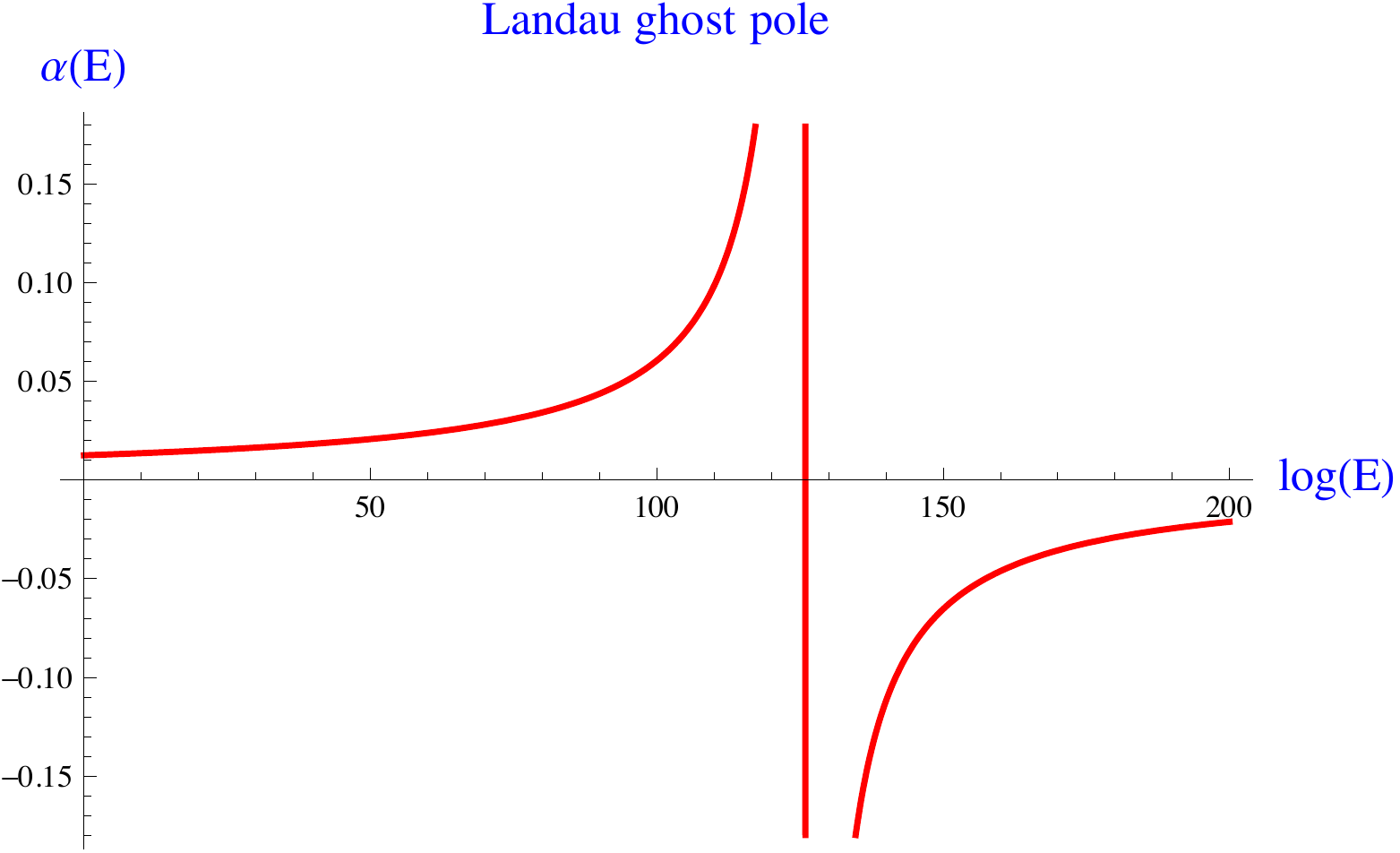}
\end{center}
\caption{The dependence of the abelian gauge and the Higgs couplings on momenta transfer (left). The behaviour of the coupling in the vicinity of the Landau pole (right).}
\label{Sigma_a}
\end{figure}
Thus, the running of the $U(1)$ coupling in the leading order is described by the formula
\begin{equation}
\alpha_1(Q^2)=\frac{\alpha_{10}}{1-\frac{41}{10}\frac{\alpha_{10}}{4\pi}\log(Q^2/M_Z^2)}
\end{equation}
and goes to infinity at $Q^*=M_Zexp(\frac{20\pi}{41\alpha_{10}})\sim 10^{41}$ GeV  (see Fig.1 right).
 The Landau pole has a wrong sign residue that indicates the presence of unphysical ghost fields - intrinsic problem and inconsistency of a theory, which leads to the violation of causality.
And though it takes place at energies much higher than the Planck mass where, as we assume,  quantum gravity might change everything, formally a theory with the Landau pole is not self consistent.
 
2) Radiative corrections lead to the violation of  stability of the electroweak vacuum. The whole construction of the SM may be in trouble being metastable or even unstable. This is also related to the behaviour of the Higgs coupling which crosses zero and then becomes negative at the energies close to $10^{11}$ GeV (see Fig 2.~\cite{Degrassi}) 
\begin{figure}[htb]
\begin{center}
\leavevmode
\includegraphics[width=0.35\textwidth]{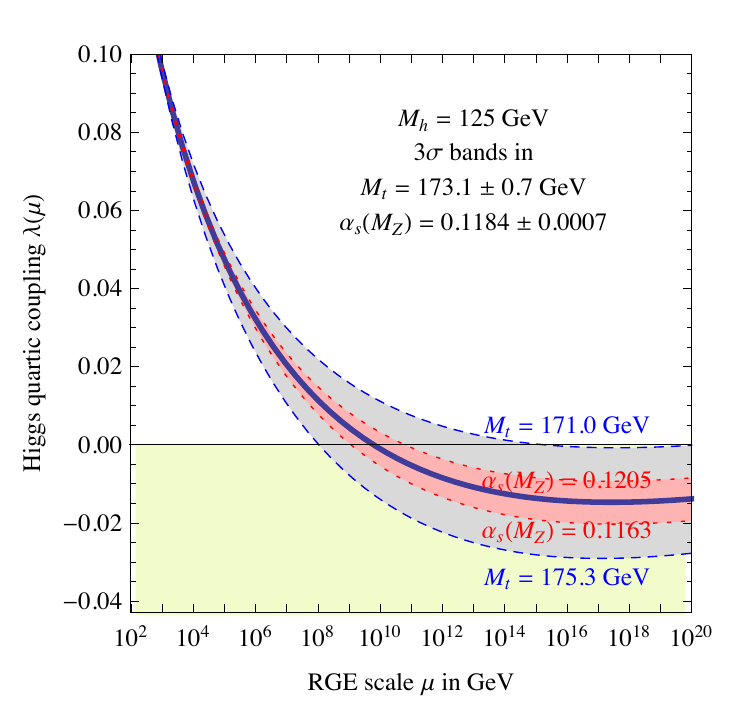}\hspace{1cm}
\includegraphics[width=0.34\textwidth]{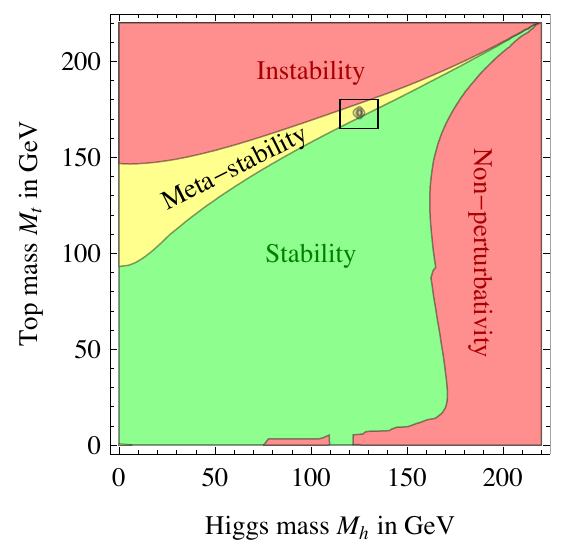}
\end{center}
\caption{Dependence of the Higgs coupling on energy scale for various values of the top quark mass  in the region where it crosses zero and becomes negative (left) and the regions of stability of the Higgs potential as functions of the top quark and the Higgs boson masses (right).}
\label{Sigma_b}
\end{figure}
However, the situation strongly depends on the accuracy of the measurement of the top quark and the Higgs boson masses and on the order of perturbation theory. The tendency when accounting for higher orders is that with increasing accuracy the instability point moves toward higher energies and possibly might reach the Planck scale (see Fig.~\ref{Sigma_c}~\cite{Espinosa}). 
\begin{figure}[ht!]
\begin{center}
\leavevmode
\includegraphics[width=0.45\textwidth]{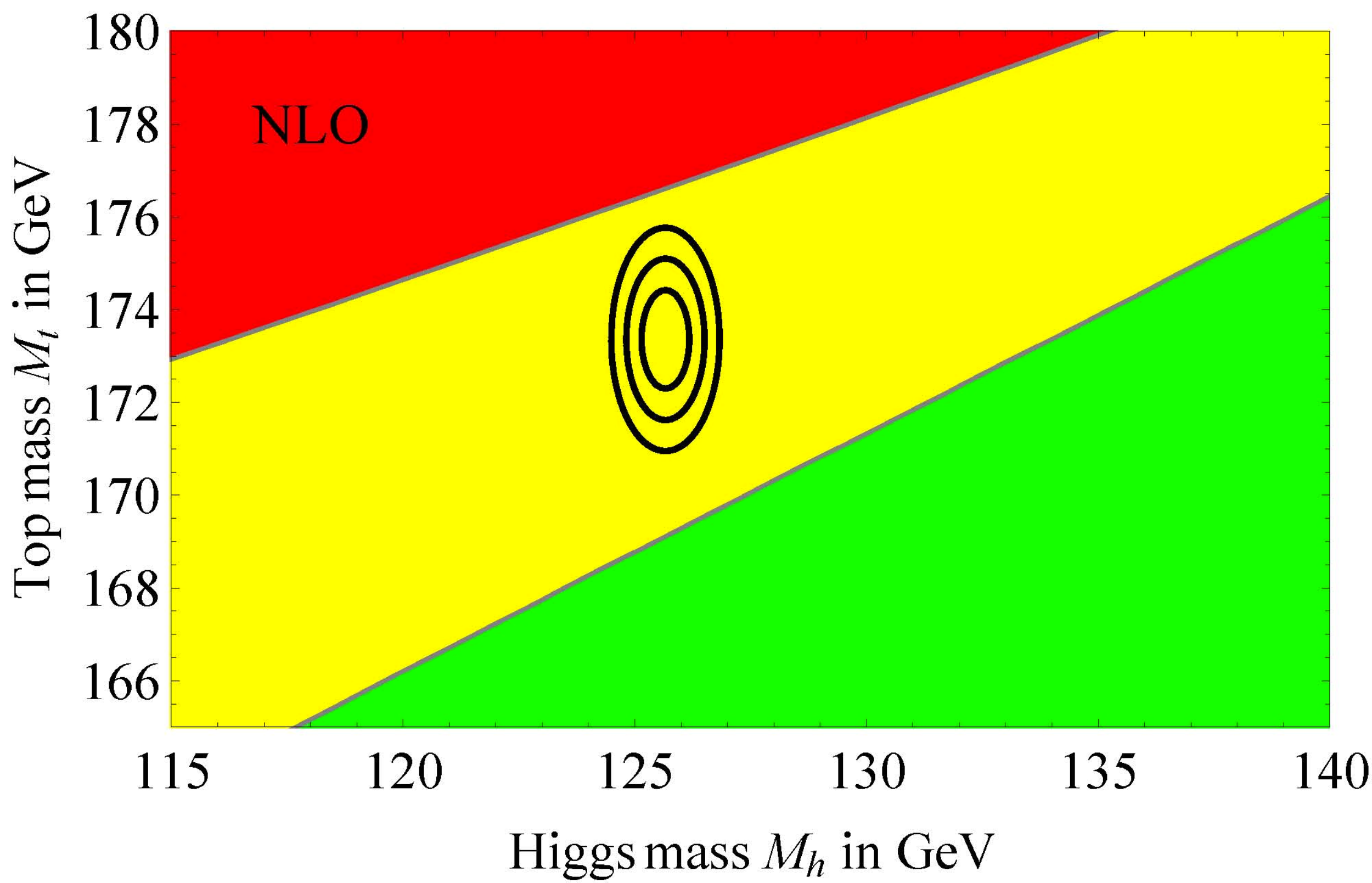}\hspace{1cm}
\includegraphics[width=0.45\textwidth]{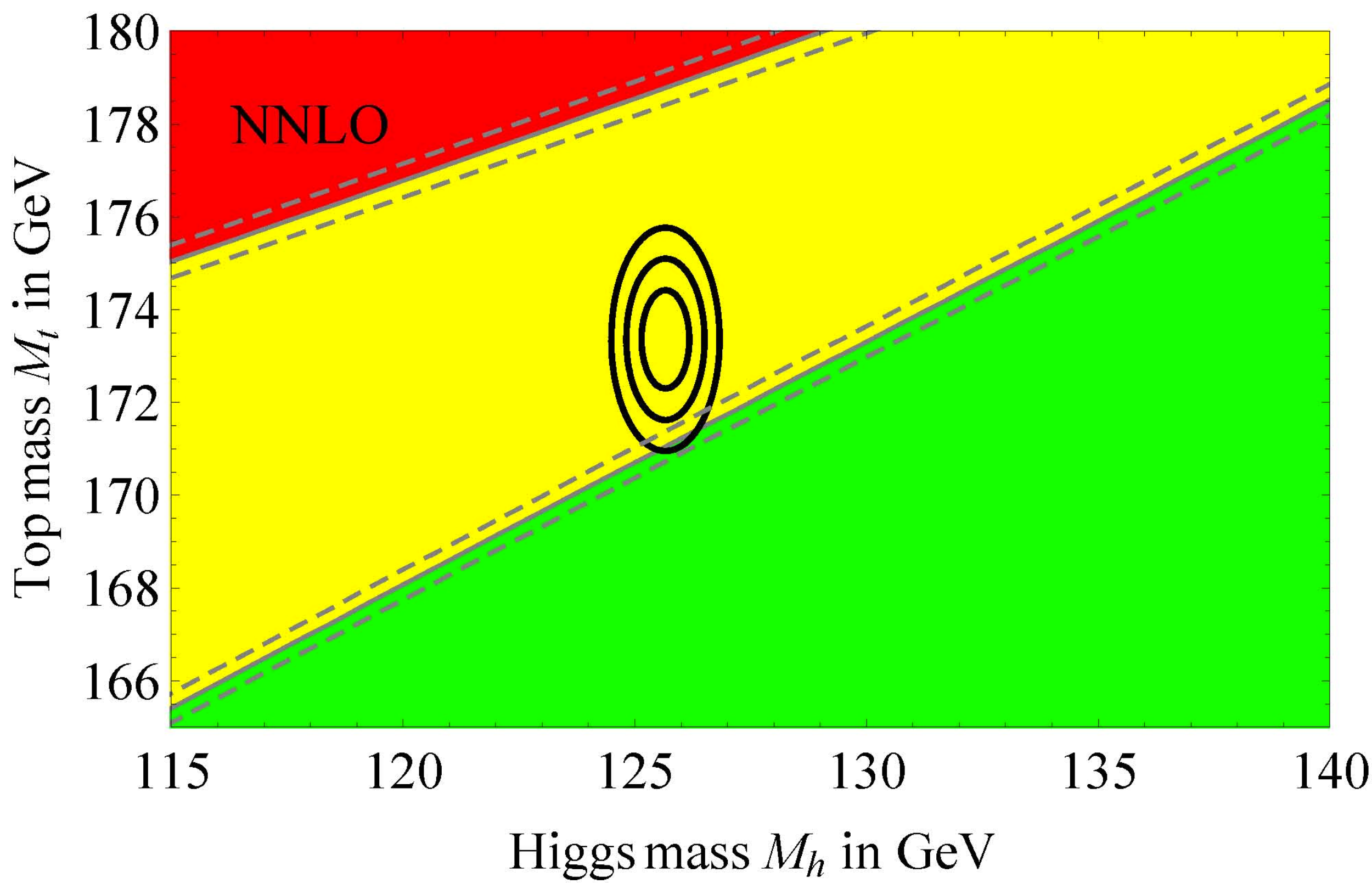}
\end{center}
\caption{The same as Fig.\ref{Sigma_b} (right) but with bigger resolution. The left panel corresponds to the NLO corrections while the right panel to the NNLO ones. One can see that the allowed spot moves towards the stability border line}
\label{Sigma_c}
\end{figure}
The situation may change if there are new heavy particles beyond the SM.

\begin{figure}[ht!]
\begin{center}
\leavevmode
\includegraphics[width=0.45\textwidth]{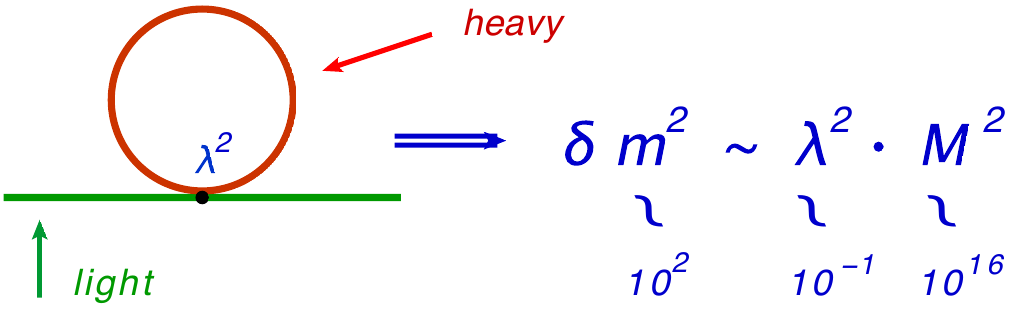}\hspace{1cm}
\end{center}
\caption{The one loop diagram which gives the contribution to the renormalization of the Higgs boson mass due to the interaction with hypothetical heavy particles}
\label{corr}
\end{figure}

3) New physics at the high energy scale might destroy the electroweak scale of the Standard Model due to radiative corrections. This is because contrary to quarks, leptons and intermediate weak bosons the mass of the Higgs boson is not protected by any symmetry. For this reason the radiative correction to the mass of the Higgs boson due to the interaction with hypothetical heavy particles, which are proportional to their mass squared, destroy the electroweak scale. The example of such interaction in the Grand Unified theories  is shown in Fig.\ref{corr}. The existing mass hierarchy $M_W/M_{GUT}\sim 10^{-14}$ might be broken. This is called the hierarchy problem.

Notice that this is not a problem of the SM itself (the quadratic divergences are absorbed into the redefinition of the bare mass which is unobservable), but leads to a quadratic dependence of  low energy physics on unknown high energy one that is not acceptable. The way out of this situation might be a new physics at intermediate energies.

The Standard Model puts some questions, the answers to which might lie beyond it. They are:
\begin{itemize}
\item why is the symmetry group $SU(3)\times SU(2)\times U(1)$?
\item why are there 3 generations of matter particles?
\item why does the SM obey the quark-lepton symmetry?
\item why does the weak interaction have a $V-A$ structure?
\item why is the SM left-right asymmetric?
\item why are the baryon and lepton numbers conserved?
\item etc.
\end{itemize}
It is not clear also how some mechanisms inside the SM work. In particular, it is not clear
\begin{itemize}
\item how confinement actually works
\item how the quark-hadron phase transition happens
\item how neutrinos get a mass
\item how CP violation occurs in the Universe
\item how to protect the SM from would be heavy scale  physics
\end{itemize}
There are other questions to the Standard Model:
\begin{itemize}
\item Is it self consistent quantum field theory?
\item Does it describe \underline{all} experimental data?
\item  Are there any indications for physics beyond the SM?
\item Is there another scale except for the EW and the Planck ones?
\item  Is it compatible with Cosmology? (Where is Dark Matter?)
\end{itemize}

\section{Possible Physics Beyond the Standard Model}

\begin{figure}[ht!]
\begin{center}\vspace{0.3cm}
\leavevmode
\includegraphics[width=0.95\textwidth]{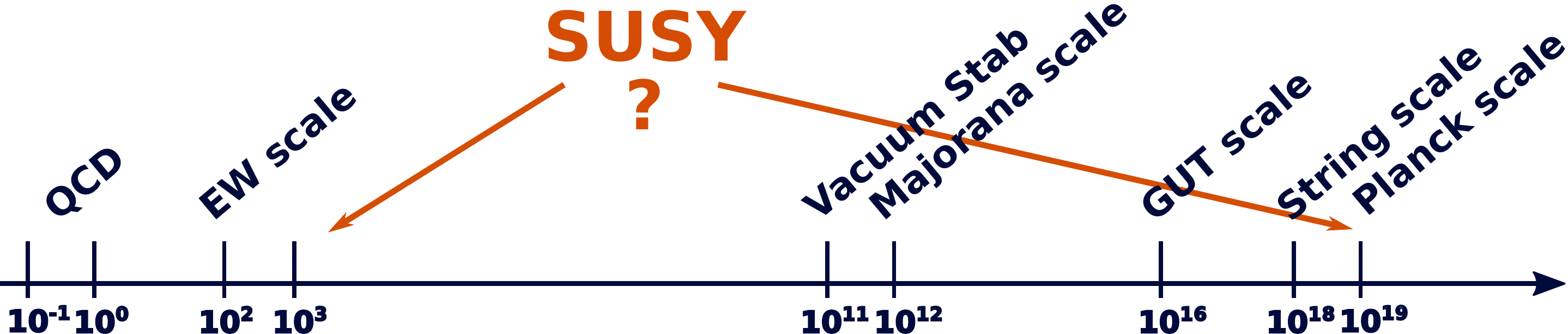}
\end{center}
\caption{The high energy physics panorama from the point of view of the energy scale}
\label{EScale}
\end{figure}

Let us look at the high energy physics panorama from the point of view of the energy scale (see Fig.\ref{EScale}).
Besides the electroweak scale $\sim 10^2$  GeV and the Planck scale $\sim 10^{19}$  GeV there is a scale of quantum chromodynamics $\Lambda_{КХД}\sim 200$ MeV, the whole spectra of quark,  lepton, intermediate vector boson and the Higgs boson masses, all related to the electroweak scale. 
Presumably, there is also a string scale $\sim 10^{18}$ GeV, the Grand unification scale $\sim 10^{16}$ GeV, the Majorana mass scale $\sim 10^{12}$ GeV, the vacuum stability scale $\sim 10^{11}$ GeV and finally somewhere in the interval from $10^{3}$ to $10^{19}$ GeV there is a supersymmetry scale.

So far there are no indications that all these scales and new physics related to them exist and high energy physics today stays in a kind of fog masking the horizon of knowledge. But sooner or later the fog will clear away and we will see the ways of future science. At the moment we live in the era of data when theory suggests various ways of development  and only experiment can show the right road.

The way out beyond the Standard Model is performed along the following directions:

1. Extension of  the \underline{symmetry} group of the SM : supersymmetry, Grand Unified Theories, new U(1) factors, etc. This way one  may solve the problem of the Landau pole, the problem of stability, 
the hierarchy problem, and also the Dark Matter problem.

2. Addition of new   \underline{particles}:  extra generations of matter, extra gauge bosons, extra Higgs bosons, extra neutrinos, etc. This way one may solve the problem of stability and the Dark Matter problem.

3. Introduction of extra  \underline{dimensions} of space: compact or flat extra dimensions. 
 This opportunity opens a whole new world of possibilities, one may solve the problem of 
   stability and  the hierarchy problem, get a new insight into gravity.
   
4.  Transition to a new  \underline{paradigm} beyond the local QFT: string theory,  brane world, etc.
 The main hope here  is the unification of gravity with other interactions and the construction of quantum gravity.  
 
 Note the paradox in modern high energy physics. If  usually a new theory emerges as a reply to experimental data which are not explained in an old theory, in our case we try to construct a new theory and persistently look for experimental data which go beyond the Standard Model but cannot find them so far. The existing small deviations from the SM at the level of a few sigma such as in the forward-backward asymmetries  in electron-positron scattering or in the anomalous magnetic moment of muon are possibly due to uncertainty of the experiment or data processing. The neutrino oscillations indicating that neutrinos have a mass will probably require a slight modification of the SM: however, there might also be described inside it. Dark Matter, almost the only indication of incompleteness of the SM, yet might be related to heavy Majorana neutrinos and require nothing else.
 
 Nevertheless, there is a vast field of theoretical models of physics beyond the Standard Model. The question is which of these models is correct and adequate to Nature. Note that the prevailing paradigm in most of the attempts to go beyond the SM is the idea of unification. It dates back to the unification of electricity and magnetism in Maxwell theory, unification of electromagnetic and weak forces in electroweak theory, merging of three forces in Grand unified theory, attempts to unify with gravity and creation of the theory of everything on the basis of a string theory.  This scenario, though it did not find any experimental verification, still seems possible and has no reasonable alternative.
 
 \section{New Symmetries}
 Extension of the symmetry group of the SM can be performed along two directions: extension of the Lorentz group and extension of the internal symmetry group. In the first case, we are talking about supersymmetric extension.
 
 \subsection{Supersymmetry}
 
 {\it Supersymmetry} is a {\it boson-fermion} symmetry that is
aimed to unify all forces in Nature including gravity within a
singe framework~\cite{super,Rev,WessB,sspace,Books}.
Supersymmetry emerged from attempts to generalize the Poincar\'e
algebra to mix representations with different spin~\cite{super}.
It happened to be a problematic task due to ``no-go'' theorems
preventing such generalizations~\cite{theorem}. The way out was
found by introducing the so-called graded Lie algebras, i.~e. adding
anti-commutators to usual commutators of the Lorentz algebra.
Such a generalization, described below, appeared to be the only
possible one within the relativistic field theory.

If $Q$ is a generator of the SUSY algebra, then acting on a
boson state it produces a fermion one and vice versa
$$
\bar Q \, | \text{boson} \rangle = | \text{fermion} \rangle , \ \
Q \, | \text{fermion} \rangle = | \text{boson} \rangle.
$$
Combined with the usual Poincar\'e and internal symmetry
algebra the Super-Poincar\'e Lie algebra contains additional
SUSY generators
$Q_{\alpha}^i$ and $\bar Q_{\dot\alpha}^i$~\cite{WessB}
\begin{equation}
\begin{split}
&[P_\mu, P_\nu] = 0, \\
&[P_\mu, M_{\rho\sigma}] = i \,(g_{\mu\rho} P_\sigma - g_{\mu\sigma} P_\rho), \\
&[M_{\mu\nu}, M_{\rho\sigma}] =
    i \,(g_{\nu\rho} M_{\mu\sigma} - g_{\nu\sigma} M_{\mu\rho}
    - g_{\mu\rho} M_{\nu \sigma} + g_{\mu\sigma} M_{\nu\rho}), \\
     &[B_r, B_s] = i \, C_{rs}^t B_{t}, \ \ 
[B_r , P_{\mu}] = [B_r, M_{\mu\sigma}] = 0, \\
&[Q_{\alpha}^i , P_{\mu}] = [\bar Q_{\dot\alpha}^i, P_{\mu}] = 0, \\
&[Q_{\alpha}^i , M_{\mu\nu}] = \frac12 \, (\sigma_{\mu\nu})_{\alpha}^{\beta}Q_{\beta}^i, \ \ \
 [\bar Q_{\dot\alpha}^i, M_{\mu\nu}] = -\frac12 \, \bar Q_{\dot\beta}^i
   (\bar\sigma_{\mu\nu})_{\dot\alpha}^{\dot \beta}, \\
   &\{ Q_{\alpha}^i, \bar Q_{\dot\beta}^j \} = 2 \, \delta^{ij} (\sigma^\mu)_{\alpha \dot\beta} P_\mu, \\
    &[Q_{\alpha}^i, B_r] = (b_r)_j^i Q_{\alpha}^j, \ \  \bar Q_{\dot\alpha}^i, B_r] = - \bar Q_{\dot\alpha}^j (b_r)_j^i, \\   
    &\{ Q_{\alpha}^i, Q_{\beta}^j \} = 2 \, \epsilon_{\alpha \beta}Z^{ij}, \ \ \
    Z_{ij} = a_{ij}^r b_r, \ \ \ Z^{ij} = Z_{ij}^+, \\
 \end{split}
\label{group}
\end{equation}
 \begin{equation}
\begin{split} 
&\{ \bar Q_{\dot\alpha}^i, \bar Q_{\dot\beta}^j \} =
    - 2 \, \epsilon_{\dot\alpha \dot\beta} Z^{ij}, \ \ \ [Z_{ij}, anything] = 0, \\
&\alpha, \dot\alpha  =  1,2 \ \ \ \ i,j = 1,2, \ldots , N.
\end{split}
\label{group}
\end{equation}
Here $P_{\mu}$ and $M_{\mu \nu}$ are the four-momentum and
angular momentum operators, respectively, $B_r$ are the
internal symmetry generators, $Q^i$ and $\bar Q^i$ are the
spinorial SUSY generators and $Z_{ij}$ are the so-called
central charges; $\alpha, \dot\alpha, \beta, \dot\beta$ are the
spinorial indices. In the simplest case, one has one spinor
generator $Q_\alpha$ (and the conjugated one
$\bar Q_{\dot{\alpha}}$) that corresponds to the ordinary or
$N=1$ supersymmetry. When $N>1$ one has the extended
supersymmetry. 

Motivation for supersymmetry in particle physics is based  on the following remarkable features of SUSY theories:

\underline{Unification with gravity}
 The representations of the Super-Poincar\'e algebra
contain particles with different spin contrary to the Poincar\'e algebra where spin is a conserved quantity. This opens the way to unification of  all other forces with gravity since the carriers of the gauge interactions have spin 1 and  of gravity - spin2, and in the case of supersymmetry, they might be in the same multiplet. 
Starting with the graviton
state of  spin 2 and acting by the SUSY generators, we get the
following chain of states:
$$
\text{spin} \ 2 \ \rightarrow \
\text{spin} \ \frac 32 \ \rightarrow \
\text{spin} \ 1 \ \rightarrow \
\text{spin} \ \frac 12 \ \rightarrow \
\text{spin} \ 0.
$$
Thus, the partial unification of matter (the fermions) with
forces (the bosons) naturally arises from an attempt to unify
gravity with other interactions.

Taking infinitesimal transformations
$\delta_\epsilon = \epsilon^\alpha Q_\alpha, \
\bar{\delta}_{\bar \epsilon} =
\bar{Q}_{\dot \alpha}{\bar \epsilon}^{\dot \alpha},$
and using Eqn.~(\ref{group}), one gets
\begin{equation}
\{\delta_\epsilon,\bar{\delta}_{\bar \epsilon} \} =
2 \, (\epsilon \sigma^\mu \bar \epsilon )P_\mu ,
\label{com}
\end{equation}
where $\epsilon, \bar\epsilon$ are the transformation parameters.
Choosing $\epsilon$ to be local, i.~e. the function of the
space-time point $\epsilon = \epsilon(x)$, one finds from
Eqn.~(\ref{com}) that the anticommutator of two SUSY
transformations is a local coordinate translation, and the
theory, which is invariant under the local coordinate
transformation is the General Relativity. Thus, making SUSY
local, one naturally obtains the General Relativity, or the
theory of gravity, or supergravity~\cite{Rev}.

\begin{figure}[htb]
\begin{center}
\leavevmode
\includegraphics[width=0.64\textwidth]{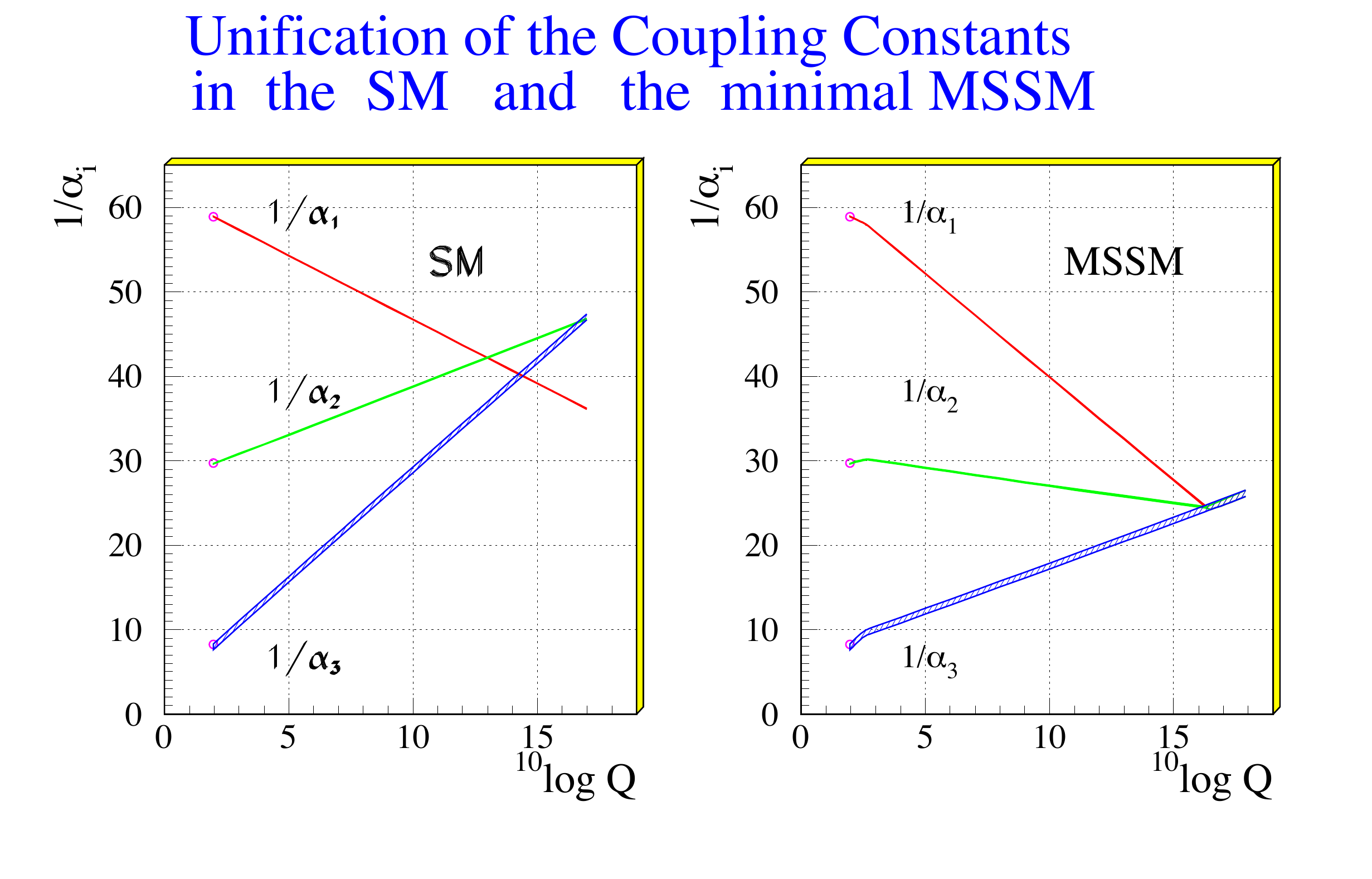}
\vspace*{-10mm}
\end{center}
\caption{The evolution of the inverse of the three coupling
constants in the Standard Model (left) and in the supersymmetric
extension of the SM (MSSM) (right).}
\label{unif}
\end{figure}

\underline{Unification of gauge couplings}
To see how the couplings change with energy, one has to consider the renormalization group equations. They are well known in the leading orders of perturbation theory in any given model.
Besides, one has to know the initial conditions at low energy which are measured experimentally.
After the precise measurement of the $SU(3)\times SU(2) \times
U(1)$ coupling constants at LEP, it became possible to check the
unification numerically. Using these numbers as input and running the RG equations one can check the unification hypothesis.  Taking first just the SM, one can see that the couplings do not unify with an offset of 8 sigma. On the contrary, if one switches  to supersymmetric generalization of the SM at some energy threshold, unification is perfectly possible with the SUSY scale around 1 TeV that gives additional indication at the low energy supersymmetry. 
The result is demonstrated in Fig.~\ref{unif}~\cite{ABF} showing the
evolution of the inverse of the couplings as a function of the
logarithm of energy. In this presentation, the evolution becomes
a straight line in the first order. The second order corrections
are small and do not cause any visible deviation from the
straight line. 
%

\begin{figure}[htb]
\begin{center}
\leavevmode
\includegraphics[width=0.38\textwidth]{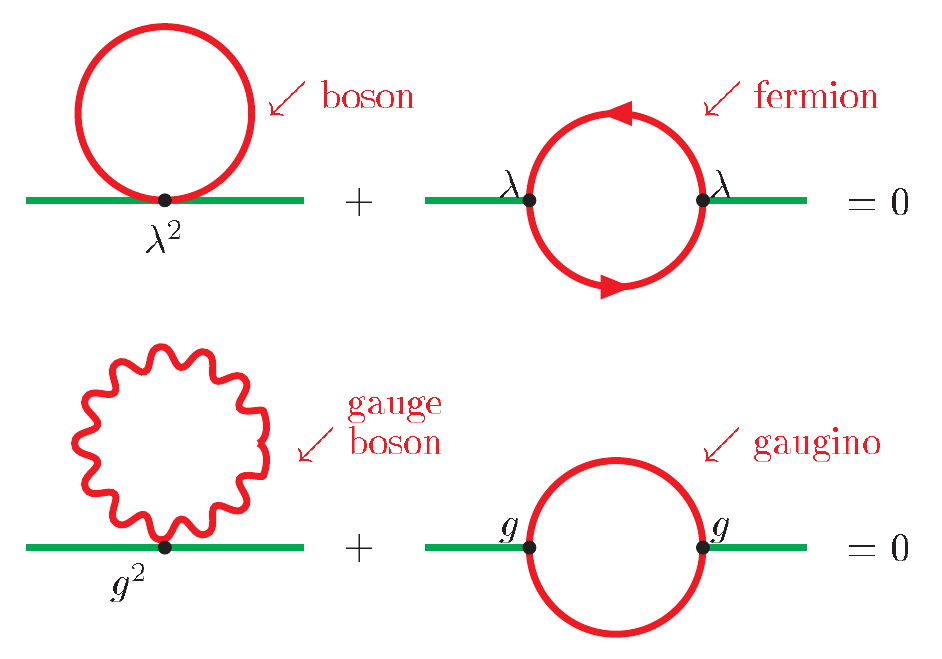}
\end{center}
\vspace{-6mm}
\caption{Cancellation of the quadratic terms (divergencies).}
\label{fig:cancel}
\end{figure}

\underline{Protection of the hierarchy}
Supersymmetry provides natural preservation of the hierarchy and
protection of the low energy scale against radiative corrections. 
Moreover, SUSY
automatically cancels the quadratic corrections in all orders
of perturbation theory. This is due to the contributions
of superpartners of ordinary particles. The contribution from
boson loops cancels those from the fermion ones because of an
additional factor ($-1$) coming from the Fermi statistics,
as shown in Fig.~\ref{fig:cancel}.

One can see here two types of contribution. The first line is
the contribution of the heavy Higgs boson and its superpartner
(higgsino). The strength of the interaction is given by the
Yukawa coupling constant $\lambda$. The second line represents
the gauge interaction proportional to the gauge coupling
constant $g$ with the contribution from the heavy gauge boson
and its heavy superpartner (gaugino).

\begin{figure}[htb]
\begin{center}
\leavevmode
\includegraphics[width=0.6\textwidth]{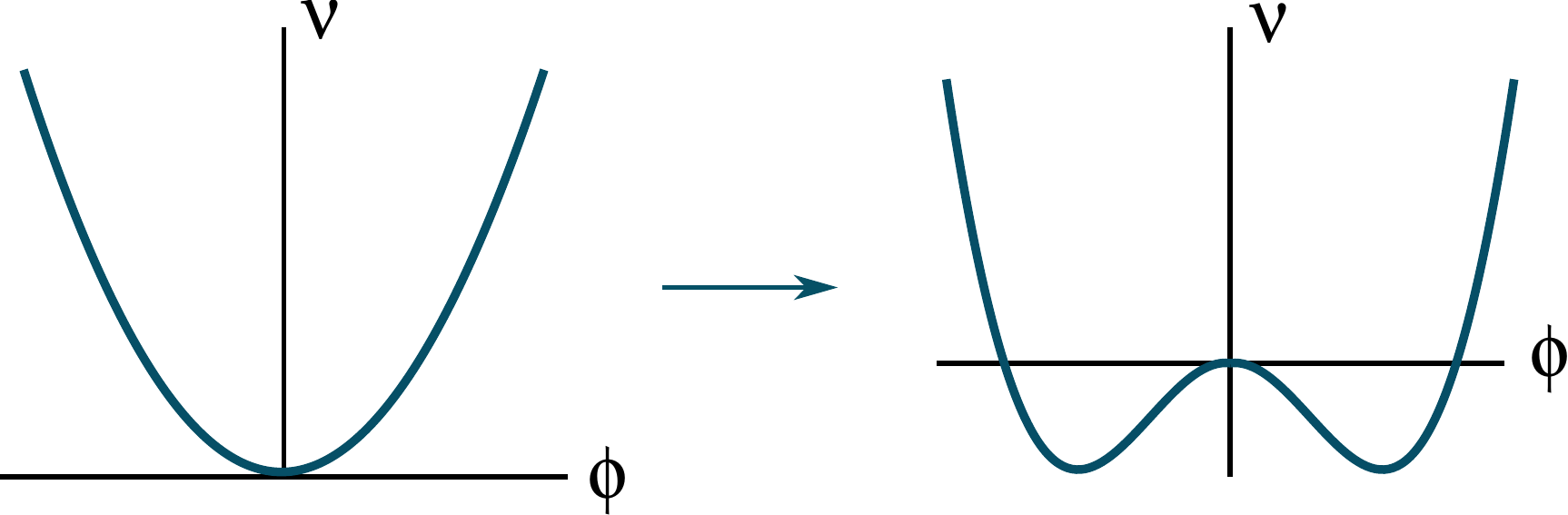}
\end{center}
\vspace{-6mm}
\caption{EW symmetry breaking}
\label{potential}
\end{figure}

\underline{Explanation of the EW symmetry breaking}
To break the Electroweak symmetry, we use the Brout-Englert-Higgs mechanism
of spontaneous symmetry breaking. However, the form of the scalar field potential
is taken ad hoc. On the contrary SUSY models provide
such an explanation. One originally starts with  unbroken potential
shown in Fig.\ref{potential} (left) 
and then arrives at the famous Mexican hat potential Fig.\ref{potential} (right)
as a result of radiative corrections~\cite{RC}. Thus, supersymmetry provides the mechanism of radiative EW symmetry breaking in a natural way. 

\underline{Provides the DM particle}
Supersymmetry provides an excellent candidate
for the cold dark matter, namely, the neutralino, the
lightest superparticle which is the lightest combination of 
superparnters of the photon, Z-boson and two neutral Higgses.
$$
|\tilde \chi^0_1\rangle =N_1|B_0\rangle
+N_2|W^3_0\rangle +N_3|H_1\rangle +N_4|H_2\rangle.
$$
It is neutral, heavy, stable
and takes part in weak interactions, precisely what is needed
for a WIMP. Besides, one can easily get the right amount of DM
with the electroweak annihilation cross-section.


A natural question arises: what is the content of SUSY theory, what kind of states is possible?
To answer this
question, consider massless states. Let us start with the
ground state labeled by the energy and the helicity, 
the projection of the spin on the direction of momenta, and
let it be annihilated by $Q_i$~\cite{WessB}
\begin{equation*}
{\rm Vacuum} =|E,\lambda \rangle, \ \ \ \ \ \ \
Q_i|E,\lambda \rangle = 0.
\end{equation*}
Then one- and many-particle states can be constructed with the
help of creation operators as
\begin{equation*}
\begin{array}{llc}
\mbox{\underline{State}~~~~~~~~~~~~~~} &
\mbox{\underline{Expression}~~~~~~~~~~~~~~~~~~~~~~~~~~} &
\# \ \mbox{\underline{of states}} \\  \\
\mbox{vacuum} & |E,\lambda \rangle & 1 \\
\mbox{1-particle} & \bar Q_i |E,\lambda \rangle =
|E,\lambda + \frac12 \rangle_i & N \\
\mbox{2-particle} & \bar Q_i \bar Q_j |E,\lambda  =
|E,\lambda + 1 \rangle_{ij} & \frac{N(N-1)}{2} \\
\ldots & \ldots & \ldots \\[1mm]
N\mbox{-particle} & \bar Q_1 \ldots \bar Q_N |E,\lambda \rangle =
|E,\lambda + \frac{N}{2} \rangle & 1
\end{array}
\end{equation*}
The total $\#$ of states is:
$\displaystyle
\sum_{k=0}^{N}\left(
\begin{array}{c}
N \\ k
\end{array}
\right) =2^N=2^{N-1}$
bosons + $2^{N-1}$ fermions.
The energy $E$ is not changed, since according to~(\ref{group})
the operators $\bar Q_i$ commute with the Hamiltonian.

Thus, one has a sequence of bosonic and fermionic states
and the total number of the bosons equals that of the
fermions. This is a generic property of any supersymmetric
theory. However, in CPT invariant theories the number of states
is doubled since CPT transformation changes the sign of 
helicity. Hence, in the CPT invariant theories, one has to add
the states with the opposite helicity to the above mentioned
ones.

Let us consider some examples. We take $N=1$ and $\lambda=0$.
Then one has the following set of states:
\begin{equation*}
\begin{array}{lllllc}
N=1 & \lambda=0 & & & & \\
\mbox{helicity} & 0 \ \frac12 & & \mbox{helicity} & 0 \ -\frac 12 \\
& & \stackrel{CPT}{\Longrightarrow} & & \\
\# \ \mbox{of states} & 1 \ 1 & & \# \ \mbox{of states} & 1 \ \ \ \ 1
\end{array}
\end{equation*}
Hence, the complete $N=1$ multiplet is
\begin{equation*}
\begin{array}{llccc} N=1 \ \ &
\mbox{helicity}&-1/2&0&1/2  \\ &
\# \ \mbox{of states}&\ 1& 2&1
\end{array}
\end{equation*}
which contains one complex scalar and one spinor with two
helicity states.

This is an example of the so-called self-conjugated multiplet.
There are also self-con\-ju\-gated multiplets with $N>1$
corresponding to the extended supersymmetry. Two particular
examples are the $N=4$ super Yang-Mills multiplet and the $N=8$
supergravity multiplet
\begin{equation*}
N=4 \ \ \
\mbox{SUSY YM} \  \ \  \lambda=-1
\end{equation*}
\begin{equation*}
\begin{array}{cccccccccc}
\mbox{helicity}&& &-1&-1/2&0&1/2&1 &&  \\
\# \ \mbox{of states}&&&\ \ 1&\ 4&6 & 4 & 1 &&\\
\end{array}
\end{equation*}
\begin{equation*}
N=8  \ \ \  \mbox{SUGRA} \ \ \   \lambda=-2
\end{equation*}
\begin{equation*}
\begin{array}{ccccccccc}
-2&-3/2&-1&-1/2&0&1/2&1&3/2&2 \\
 1&8&28 & 56 & 70 &56&28&  8& 1
\end{array}
\end{equation*}
One can see that the multiplets of extended supersymmetry
are very rich and contain a vast number of particles.

In what follows, we shall consider simple supersymmetry, or
the $N=1$ supersymmetry, contrary to extended
supersymmetries with $N > 1$. 
In this case, one has the
following types of the supermultiplets with lower spins:
 
  - chiral supermultiplet   $(\phi,\psi)$ containing the scalar state $\phi$ and the chiral fermion  $\psi$;
  
- vector supermultiplet $(\lambda, A_\mu)$  containing the Majorana spinor $\lambda$   and the vector field $A_\mu$;

- gravity supermultiplet $(\tilde g, g)$ containing graviton $g$ of spin 2 and gravitino $\tilde g$ of spin 3/2. 

Each of multiplets contains two physical states, one boson and one
fermion. From these multiplets one constructs all supersymmetric models with N=1 supersymmetry.

\begin{figure}[htb]
\begin{center}
\leavevmode
\includegraphics[width=0.70\textwidth]{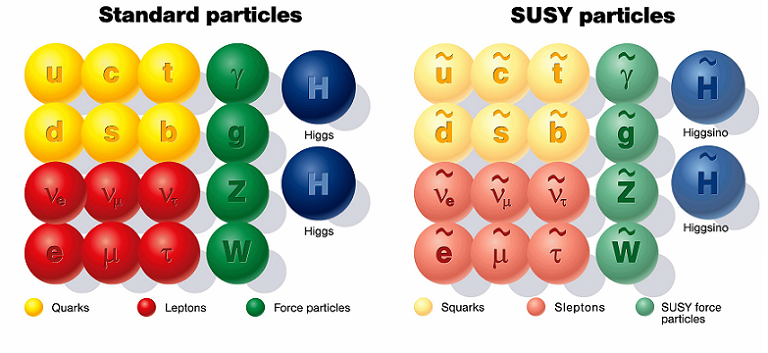}
\end{center}\vspace{-0.1cm}
\caption{The minimal supersymmetric generalization of the standard Model}
\label{content}
\end{figure}

To construct a supersymmetric generalization of the SM~\cite{MSSM}, one
has to put all the particles into these multiplets. For
instance, the quarks should go into the chiral multiplet and
the photon into the vector multiplet. The members of the same multiplet have the same quantum numbers and differ only by spin. Since in the SM there are no particles of different spin having the same quantum numbers, one has to add the corresponding partner for all particles of the SM, thus doubling the number of particles (see fig. \ref{content}~\cite{content})
The particle content of the MSSM then appears as shown in 
Table~\ref{tab:mssm}. Hereafter, the tilde denotes the
superpartner of the ordinary particle. In the last line an extra singlet field is added which corresponds to the so-called Next-to-Minimal model  (NMSSM)~\cite{NMSSM}.
\begin{table}[htb]
\begin{center}
\renewcommand{\tabcolsep}{0.03cm}
\begin{tabular}{|lllccc|}\hline
Superfield & \ \ \ \ \ \ \ Bosons & \ \ \ \ \ \ \ Fermions &
$SU(3)$& $SU(2$ & $U_Y(1)$ \\
\hline \hline
Gauge  &&&&& \\
${\bf G^a}$   & gluon \ \ \ \ \ \ \ \ \ \ \ \ \ \ \  $g^a$ &
gluino$ \ \ \ \ \ \ \ \ \ \ \ \ \tilde{g}^a$ & 8 & 0 & 0 \\
${\bf V^k}$ & Weak \ \ \ $W^k$ \ $(W^\pm, Z)$ & wino, zino \
$\tilde{w}^k$ \ $(\tilde{w}^\pm, \tilde{z})$ & 1 & 3& 0 \\
${\bf V'}$   & Hypercharge  \ \ \ $B\ (\gamma)$ & bino
\ \ \ \ \ \ \ \ \ \ \ $\tilde{b}(\tilde{\gamma})$ & 1 & 1& 0 \\
\hline \hline
Matter &&&&& \\
$\begin{array}{c}
{\bf L_i} \\ {\bf E_i}
\end{array}$ & sleptons
\ $\left\{
\begin{array}{l}
\tilde{L}_i=(\tilde{\nu},\tilde e)_L \\
\tilde{E}_i =\tilde e_R
\end{array} \right. $
& leptons \ $\left\{
\begin{array}{l}
L_i=(\nu,e)_L \\
E_i=e_R^c
\end{array} \right.$
& $\begin{array}{l}
1 \\ 1
\end{array} $
& $\begin{array}{l}
2 \\ 1
\end{array} $ & $
\begin{array}{r} -1 \\ 2
\end{array} $ \\
$\begin{array}{c} {\bf Q_i} \\ {\bf U_i} \\ {\bf D_i}
\end{array}$
& squarks \ $\left\{
\begin{array}{l}
\tilde{Q}_i=(\tilde{u},\tilde d)_L \\ \tilde{U}_i =\tilde u_R \\
\tilde{D}_i =\tilde d_R
\end{array}  \right. $
& quarks \ $\left\{
\begin{array}{l}
Q_i=(u,d)_L \\ U_i=u_R^c \\ D_i=d_R^c
\end{array} \right. $
& $\begin{array}{l}
3 \\ 3^* \\ 3^*
\end{array} $
& $\begin{array}{l}
2 \\ 1 \\ 1
\end{array} $
& $\begin{array}{r}
1/3 \\ -4/3 \\ 2/3
\end{array} $ \\
\hline  \hline
Higgs &&&&& \\
$\begin{array}{c} {\bf H_1} \\ {\bf H_2}
\end{array}$
& Higgses \ $\left\{
\begin{array}{l}
H_1 \\ H_2
\end{array}  \right. $
& higgsinos \ $\left\{
\begin{array}{l}
\tilde{H}_1 \\ \tilde{H}_2
\end{array} \right. $
&
$\begin{array}{l}
1 \\ 1 \end{array} $
& $\begin{array}{l}
2 \\ 2
\end{array} $
& $\begin{array}{r} -1 \\ 1
\end{array} $  \\
\hline \hline
\ \ \ S & Singlet \ \ \ \ \ \  s & singlino \ \  \ \ \ \ s & 1 & 1 & 0\\
\hline \hline
\end{tabular}
\end{center}
\label{tab:mssm}
\caption{Particle content of the MSSM and the NMSSM (the last line)}
\end{table}

The presence of the extra Higgs doublet in the SUSY model is
a novel feature of the theory. In the MSSM one has two doublets
with the quantum numbers (1,2,-1) and (1,2,1). Thus, in
the MSSM, as actually in any two Higgs doublet model, one has
five physical Higgs bosons: two $CP$-even neutral Higgs, one
$CP$-odd neutral Higgs and two charged ones.

\begin{figure}[htb]\vspace{-0.5cm}
\begin{center}
\leavevmode
\includegraphics[width=0.45\textwidth]{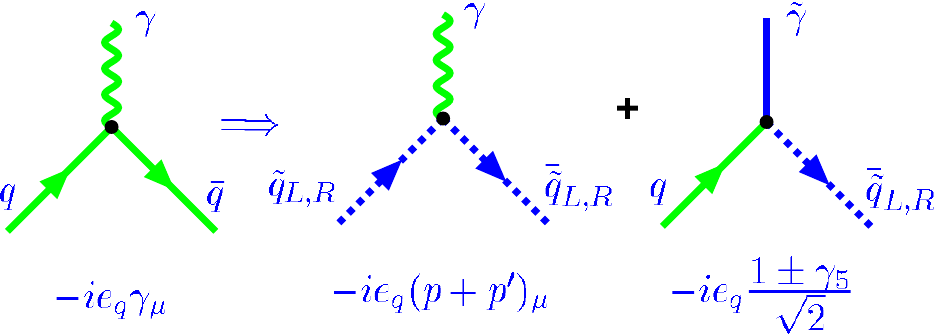}
\hspace*{10mm}
\includegraphics[width=0.45\textwidth]{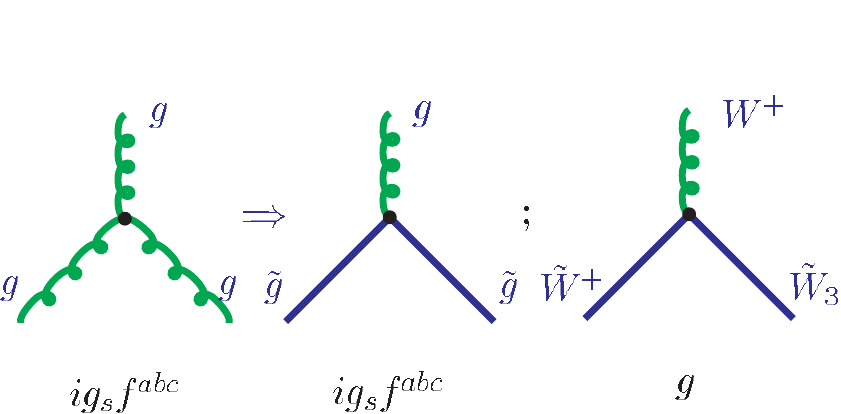}\\
\vspace*{5mm}
\includegraphics[width=0.45\textwidth]{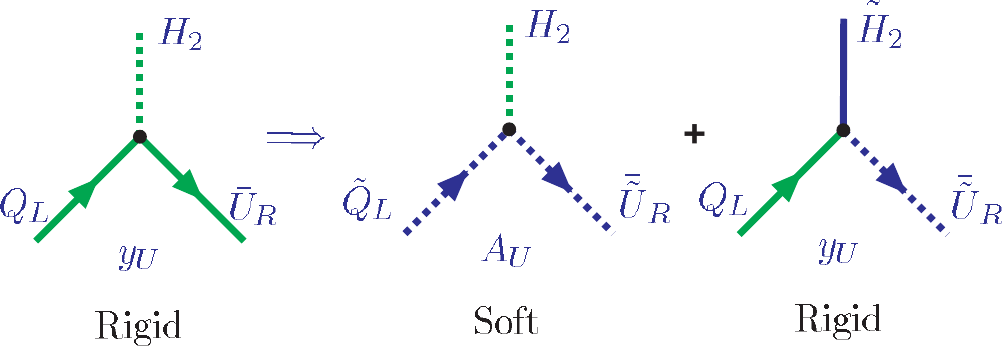}
\end{center}
\caption{The gauge-matter interaction, the gauge
self-interaction and the Yukawa interaction.}
\label{yukint}
\end{figure}

The
interactions of the superpartners are essentially the same as
in the SM, but two of three particles involved into the
interaction at any vertex are replaced by the superpartners.
Typical vertices are shown in Fig.~\ref{yukint}. The tilde
above the letter denotes the corresponding superpartner. Note
that the coupling  in all the vertices involving
the superpartners is the same as in the SM as dictated by supersymmtry. 

\begin{figure}[htb]\vspace*{-2mm}
\begin{center}
\leavevmode
\includegraphics[width=0.40\textwidth]{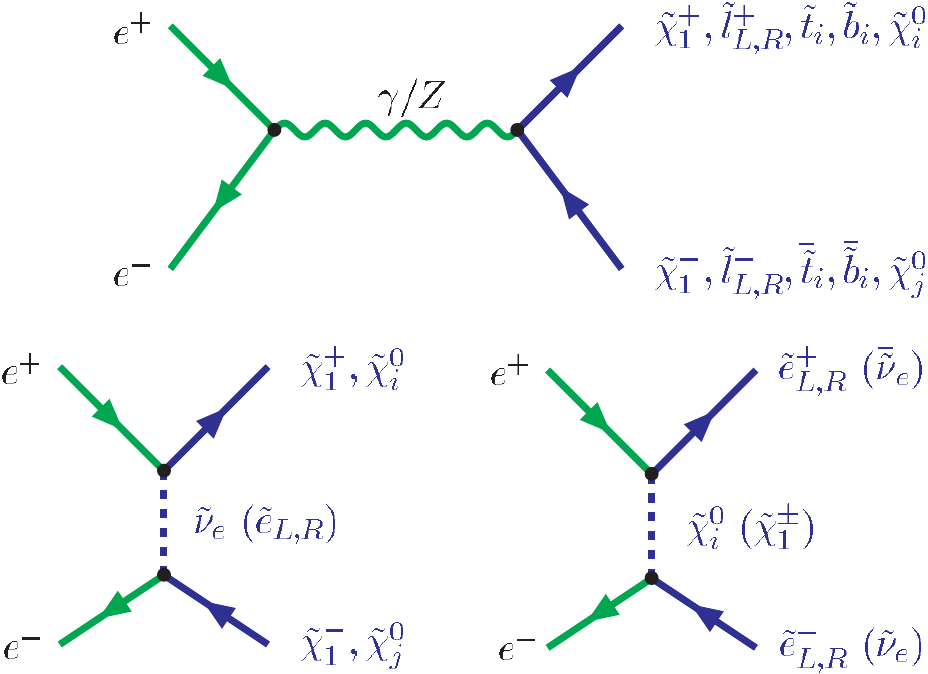}
\end{center}
\vspace*{-3mm}
\caption{Creation of the superpartners at electron-positron colliders.}
\label{creation}
\end{figure}

The above-mentioned rule together with the Feynman rules for the
SM enables one to draw diagrams describing creation of the
superpartners. One of the most promising processes is the $e^+e^-$
annihilation (see Fig.~\ref{creation}). The usual kinematic restriction is given by the c.m. energy
$m^{max}_{sparticle} \leq \sqrt{s}/2$.  

\begin{figure}[htb]
\begin{center}
\leavevmode \hspace*{-2mm}
\includegraphics[width=0.22\textwidth,height=0.1\textwidth]{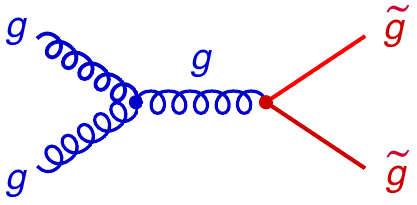}
\includegraphics[width=0.22\textwidth,height=0.1\textwidth]{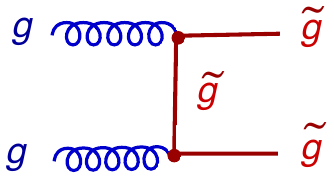}
\includegraphics[width=0.22\textwidth,height=0.1\textwidth]{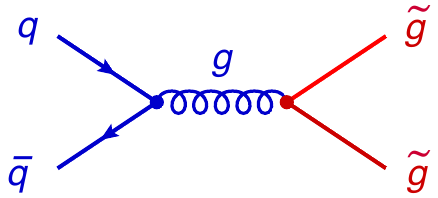}
\includegraphics[width=0.22\textwidth,height=0.1\textwidth]{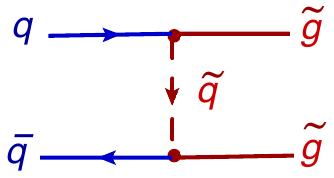}\\[6mm]
\includegraphics[width=0.22\textwidth,height=0.1\textwidth]{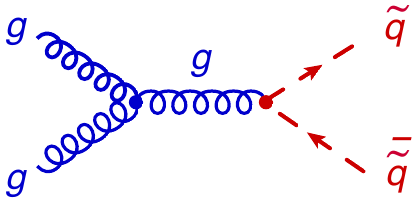}
\includegraphics[width=0.17\textwidth,height=0.1\textwidth]{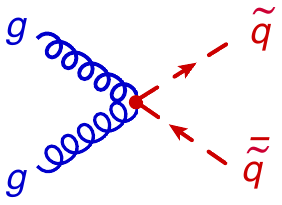}
\includegraphics[width=0.22\textwidth,height=0.1\textwidth]{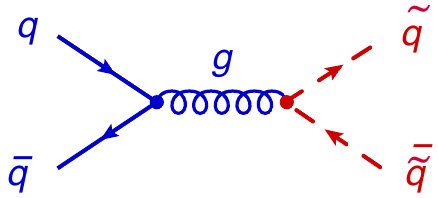}
\includegraphics[width=0.22\textwidth,height=0.1\textwidth]{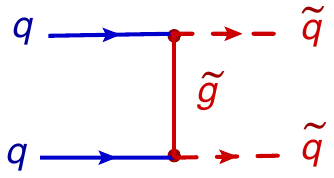}\\[6mm]
\includegraphics[width=0.22\textwidth,height=0.1\textwidth]{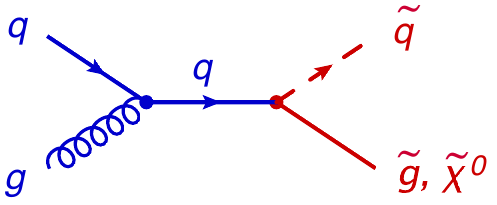}
\includegraphics[width=0.22\textwidth,height=0.1\textwidth]{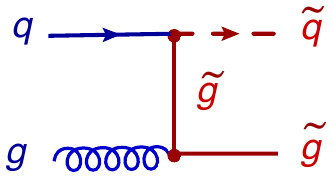}
\includegraphics[width=0.22\textwidth,height=0.1\textwidth]{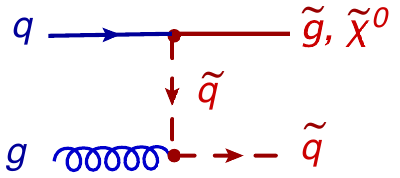}\\[6mm]
\includegraphics[width=0.22\textwidth,height=0.1\textwidth]{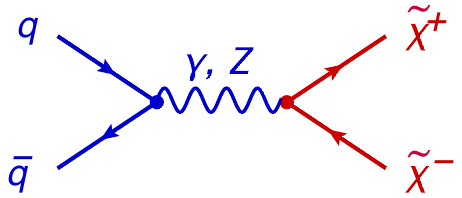}
\includegraphics[width=0.22\textwidth,height=0.1\textwidth]{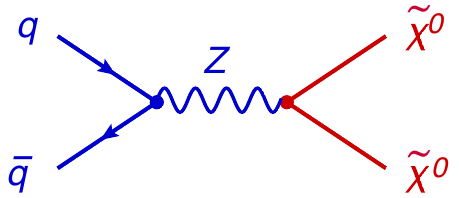}
\includegraphics[width=0.22\textwidth,height=0.1\textwidth]{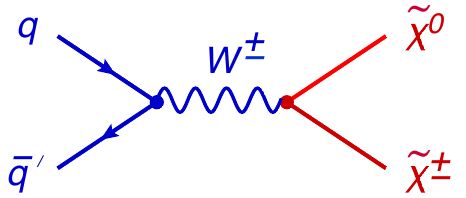}
\includegraphics[width=0.22\textwidth,height=0.1\textwidth]{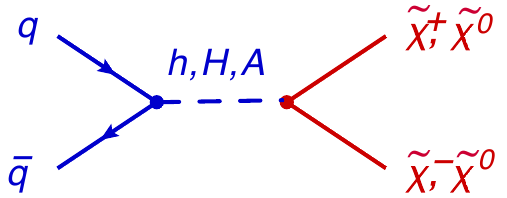}
\end{center}
\caption{Examples of diagrams for the SUSY particle production
via the strong interactions (top rows for $\gl\gl$,
$\sq\overline{\sq}$ and $\gl\sq$, respectively) and the electroweak
interactions (the lowest row).}
\label{f1}
\end{figure}

\noindent At the hadron colliders the signatures  are similar to
those at the $e^+e^-$ machines; however, here one has wider
possibilities. Besides the usual annihilation channel, one has
numerous processes of gluon fusion, quark-antiquark and
quark-gluon scattering (see Fig.~\ref{f1})~\cite{SUSYLHC_GK}. The creation of  superpartners can be accompanied by the creation
of ordinary particles as well. They crucially depend on the
SUSY breaking pattern and on the mass spectrum of the
superpartners.

\begin{figure}[htb]
\begin{center}
$
\begin{array}{llll}
\mbox{squarks} & \tilde q_{L,R}\to q+\tilde \chi^0_i  &&\\
& \tilde q_{L}\to q'+\tilde \chi^\pm_i &&\\
& \tilde q_{L,R}\to q+\tilde g&&\\[4mm]
\mbox{sleptons} & \tilde l\to l+\tilde \chi^0_i &&\\
& \tilde l_{L}\to \nu_l+\tilde \chi^\pm_i &&\\[4mm]
\mbox{chargino}
& \tilde \chi^\pm_i\to e+\nu_e+\tilde \chi^0_i &&\\
& \chi^\pm_i\to q+\bar q'+\tilde \chi^0_i &&\\[4mm]
\mbox{gluino} & \tilde g\to q=\bar q +\tilde \gamma &&\\
& \tilde g\to g+\tilde \gamma &&\\[4mm]
\mbox{neutralino} & \tilde \chi^0_i\to \tilde \chi^0_1+l^+ +l^-
& \hspace{15mm} \mbox{final states} & l^+l^- + \slash \hspace{-0.24cm}E_T\\
& \tilde \chi^0_i\to \tilde \chi^0_1+q +\bar q' & & 2 \mbox{jets} + \slash \hspace{-0.24cm}E_T\\
& \tilde \chi^0_i\to \tilde \chi^\pm_1+l^\pm +\nu_l &  &\gamma + \slash \hspace{-0.24cm}E_T\\
& \tilde \chi^0_i\to \tilde \chi^0_1+\nu_l+\bar \nu_l & & \slash \hspace{-0.24cm}E_T
\end{array}
$
\hspace*{-50mm}
\leavevmode
\raisebox{-8mm}{\includegraphics[width=0.45\textwidth]{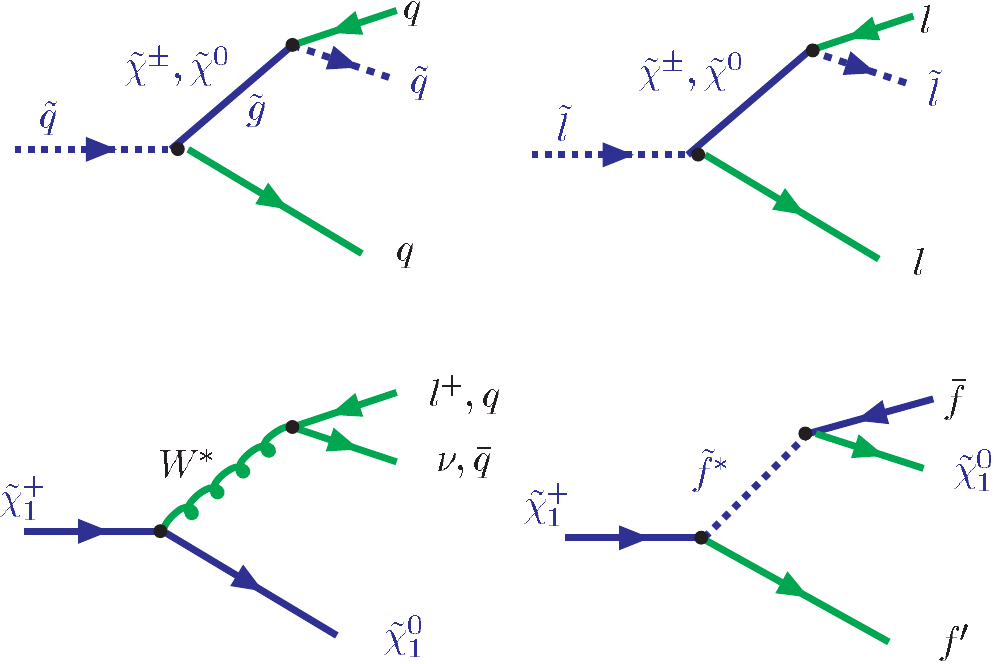}}
\end{center}
\vspace{0cm}
\caption{Decay of superpartners}\label{decay}
\end{figure}

The decay properties of the superpartners also depend on their
masses. For the quark and lepton superpartners the main
processes are shown in Fig.~\ref{decay}. One can notice that the line of superpatners shown in blue is never broken. At the final state one always has a lighter superpartner. This is a consequence of additional  new symmetry.

\clearpage
The interactions of superpartners in the MSSM obey 
new $U(1)$ symmetry called $R$-symmetry~\cite{r-symmetry}
which is reduced to the discrete group $Z_2$ and is called
$R$-parity. The $R$-parity quantum number  is
\begin{equation}
R=(-1)^{3(B-L)+2S}
\label{par}
\end{equation}
for the particles with the spin $S$. Thus, all the ordinary
particles have the $R$-parity quantum number equal to $R=+1$,
while all the superpartners have the $R$-parity quantum number
equal to $R=-1$. 
Conservation of the $R$-parity has two important consequences:
 the superpartners are created in pairs and 
the lightest superparticle (LSP) is stable.
Usually, it is the photino $\tilde \gamma $, the superpartner
of the photon with some admixture of the neutral higgsino. This
is the candidate for the DM particle which should be neutral
and has survived since the Big Bang.

\underline{Breaking of SUSY in the MSSM}
Usually, it is assumed that supersymmetry is broken
spontaneously via the v.e.v.s of some fields.  However, in
the case of supersymmetry, one can not use scalar fields
like the Higgs field, but rather the auxiliary fields present
in any SUSY  multiplet. There are two basic mechanisms of
spontaneous SUSY breaking: the Fayet-Iliopoulos (or $D$-type)
mechanism~\cite{Fayet} based on the  $D$ auxiliary field from
the vector multiplet and the O'Raifeartaigh (or $F$-type)
mechanism~\cite{O'R} based on the $F$ auxiliary field from the
chiral multiplet. Unfortunately, one can not explicitly use
these mechanisms within the MSSM since none of the fields of
the MSSM can develop the nonzero v.e.v. without spoiling the
gauge invariance. Therefore, the spontaneous SUSY breaking
should take place via some other fields.

The most common scenario for producing low-energy supersymmetry
breaking is called the {\em hidden sector
scenario}~\cite{hidden}. According to this scenario, there exist
two sectors: the usual matter belongs to the "visible" one, while
the second, "hidden" sector, contains the fields which lead to
breaking of supersymmetry. These two sectors interact with
each other by an exchange of some fields called {\em messengers},
which mediate SUSY breaking from the hidden to the visible sector.
There might be various types of the messenger fields: gravity,
gauge, etc. The hidden sector is the weakest part of the MSSM.
It contains a lot of ambiguities and leads to uncertainties of
the MSSM predictions.

All mechanisms of the soft SUSY breaking are different in
details but are common in the results. To make certain predictions, one usually introduces the
so-called soft supersymmetry breaking terms that violate supersymmetry by the operators of dimension lower than four. For the MSSM without the $R$-parity violation one has
in general
\begin{align}
&- {\cal L}_{Breaking} = \label{soft} \\
&=  \sum_{i}^{} m^2_{0i} \left|\varphi_i\right|^2 +
\biggl( \frac 12 \sum_{\alpha}^{}
M_\alpha \tilde\lambda_\alpha \tilde\lambda_\alpha +
B H_1 H_2 + A^U_{ab} \tilde Q_a \tilde U^c_b H_2 +
A^D_{ab} \tilde Q_a \tilde D^c_b H_1 +
A^L_{ab} \tilde L_a \tilde E^c_b H_1 \biggr), \notag
\end{align}
where we have suppressed the $SU(2)$ indices. Here $\varphi_i$
are all the scalar fields, $\tilde \lambda_\alpha $ are the
gaugino fields, $\tilde Q, \tilde U, \tilde D$ and
$\tilde L, \tilde E$ are the squark and slepton fields,
respectively, and $H_{1,2}$ are the SU(2) doublet Higgs fields.

Equation.~(\ref{soft}) contains a vast number of free parameters
which spoils the predictiive power of the model. To reduce
their number, we adopt the so-called {\em universality
hypothesis}, i.~e., we assume the universality or equality of
various soft parameters at the high energy scale, namely, we
put all the spin-0 particle masses to be equal to the universal
value $m_0$, all the spin-1/2 particle (gaugino) masses to be
equal to $m_{1/2}$ and all the cubic and quadratic terms
proportional to $A$ and $B$, to repeat the structure of the
Yukawa superpotential. This is an additional
requirement motivated by the supergravity mechanism of SUSY
breaking. The universality is not a necessary requirement and
one may consider nonuniversal soft terms as well. 
In this case, Eqn.~(\ref{soft}) takes the form
\begin{align}
&-{\cal L}_{Breaking}= \\
&=m_0^2\sum_{i}^{}|\varphi_i|^2+\left( \frac{ m_{1/2}}{2}
\sum_{\alpha}^{} \tilde \lambda_\alpha\tilde \lambda_\alpha+  B\mu H_1H_2
+A\bigl[y^U_{ab}\tilde Q_a\tilde U^c_bH_2+y^D_{ab}\tilde Q_a
\tilde D^c_bH_1+ y^L_{ab}\tilde L_a\tilde E^c_bH_1\bigr]
\right).\notag
\label{soft2}
\end{align}

\underline{Manifestation of SUSY}  Search for supersymmetry was and still is one of the main 
tasks in high energy physics. In particle physics this is direct production at colliders at high energies, indirect manifestation at low energies in high precision observables like rare decays or $g-2$ of the muon  and search for long-lived SUSY particles.   In astrophysics this is a measurement of the dark matter abundance in the Universe, search for the DM annihilation signal in cosmic rays and direct interaction of DM with the nucleon target in underground experimrnts. So far there is no positive signal anywhere.

\begin{figure}[htb]
\begin{center}
\leavevmode
\includegraphics[width=0.8\textwidth]{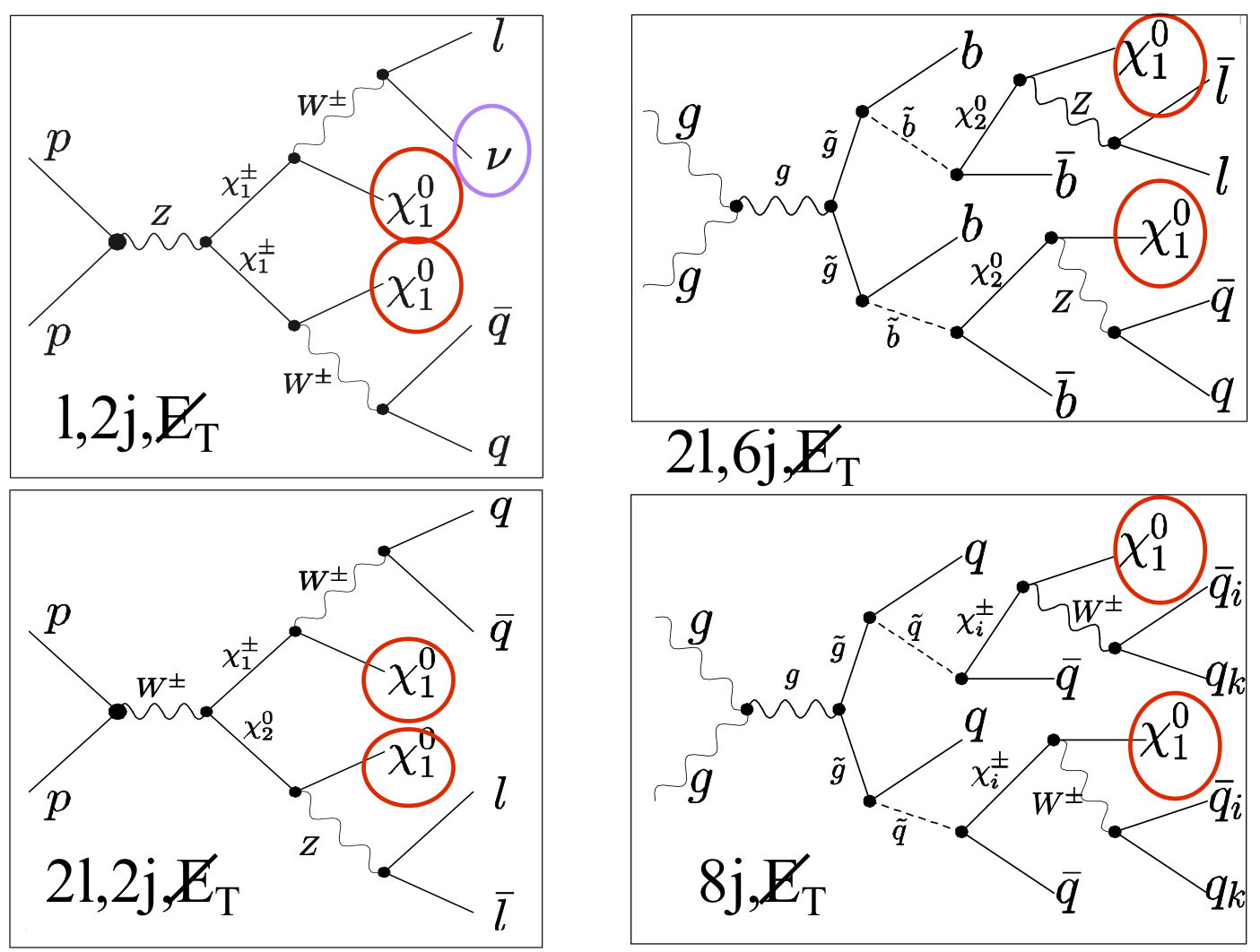}
\end{center}
\caption{Creation of superpartners in weak (left) and strong (right) interactions. The expected final states are also shown}
\label{sprod}
\end{figure}

Under the assumption that super\-symmetry exists at the TeV scale the superpartners of ordinary particles have to be produced at the LHC. Typical processes of creation of superpartners in strong and weak interaction are shown in Fig.\ref{sprod}~\cite{susyprob}.
A typical signature of supersymmetry is the presence of missing energy and missing transverse momentum carried away by the lightest supersymmetric particle  $\chi^0_1$ which is neutral and stable.

So far the creation of superpartners at the LHC is not found, there are only limits on the masses of hypothetical new particles. To present and analyze the data, two different approaches are used:
the high energy input and the low energy input. In the first case, one introduces universal high energy parameters like $m_0,m_{1/2}, A_0, \tan\beta$  of the MSSM~\cite{MSSM} and performs the analysis  in this universal parameter space. The advantage of this approach is that one has a small number of universal parameters for all particles. The disadvantage is that this set it model dependent (MSSM, NMSSM, etc). In the second case one uses the low energy parameters like masses of superpartners, $\tilde m_g, \tilde m_q, \tilde m_\chi$ or $ m_A,\tan\beta$. The advantage is that it is model independent, the disadvantage is that one has many parameters and they are process dependent. Both the approaches are used in practice. 

\begin{figure}[htb]
\begin{center}
\centering
\includegraphics[width=0.45\textwidth]{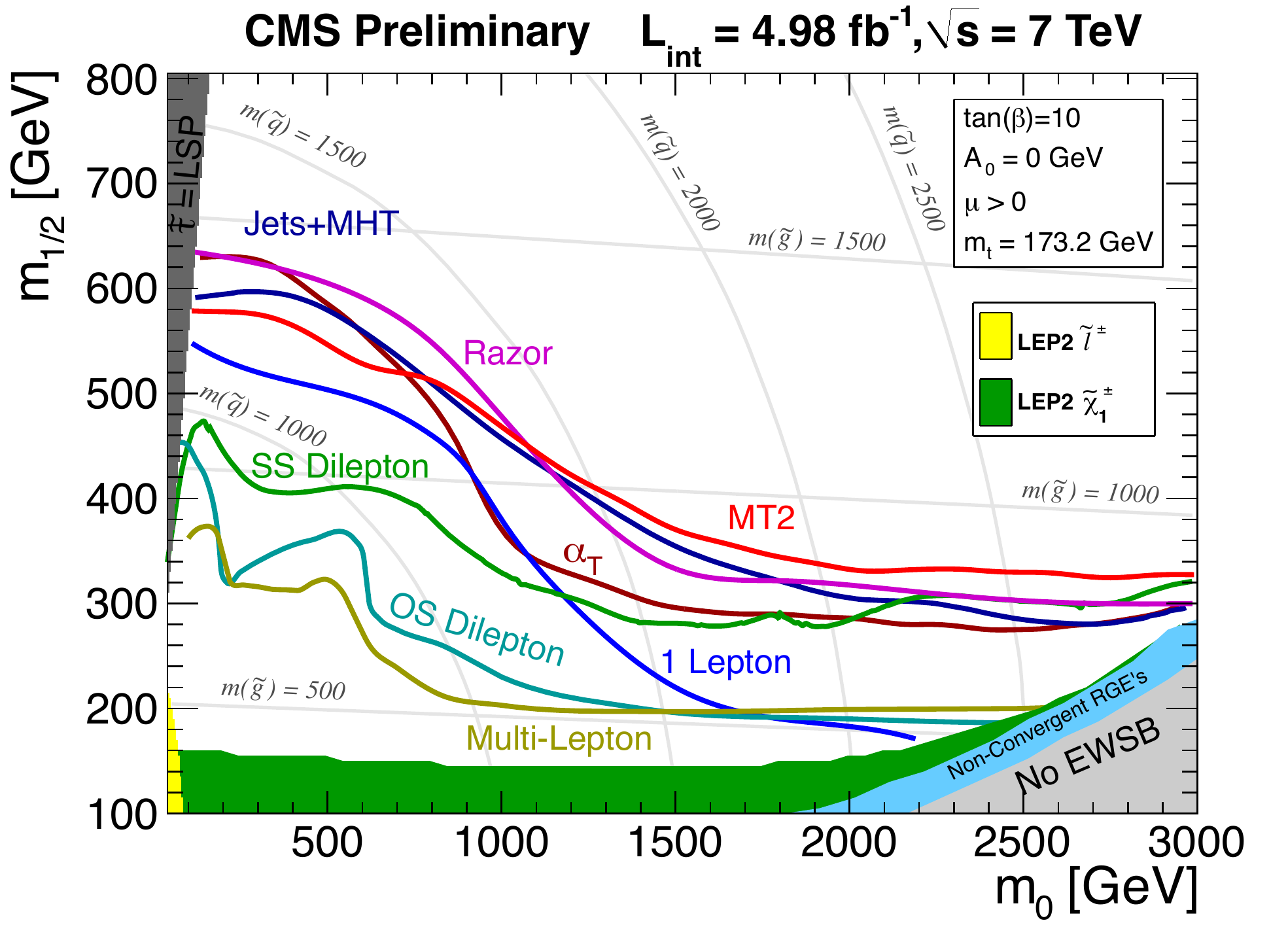}
\includegraphics[width=0.48\textwidth]{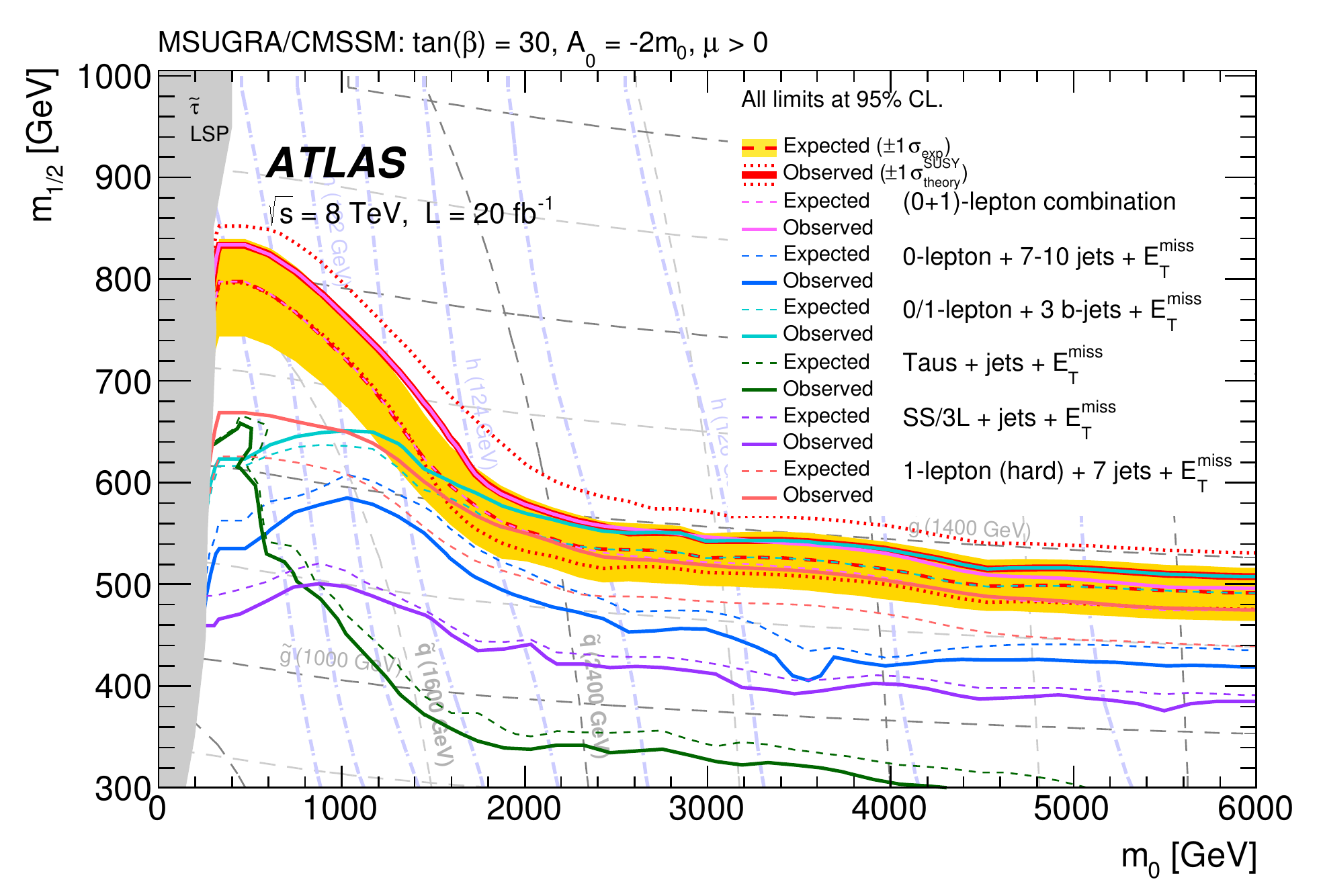}\vspace{0.5cm}
\includegraphics[width=0.31\textwidth]{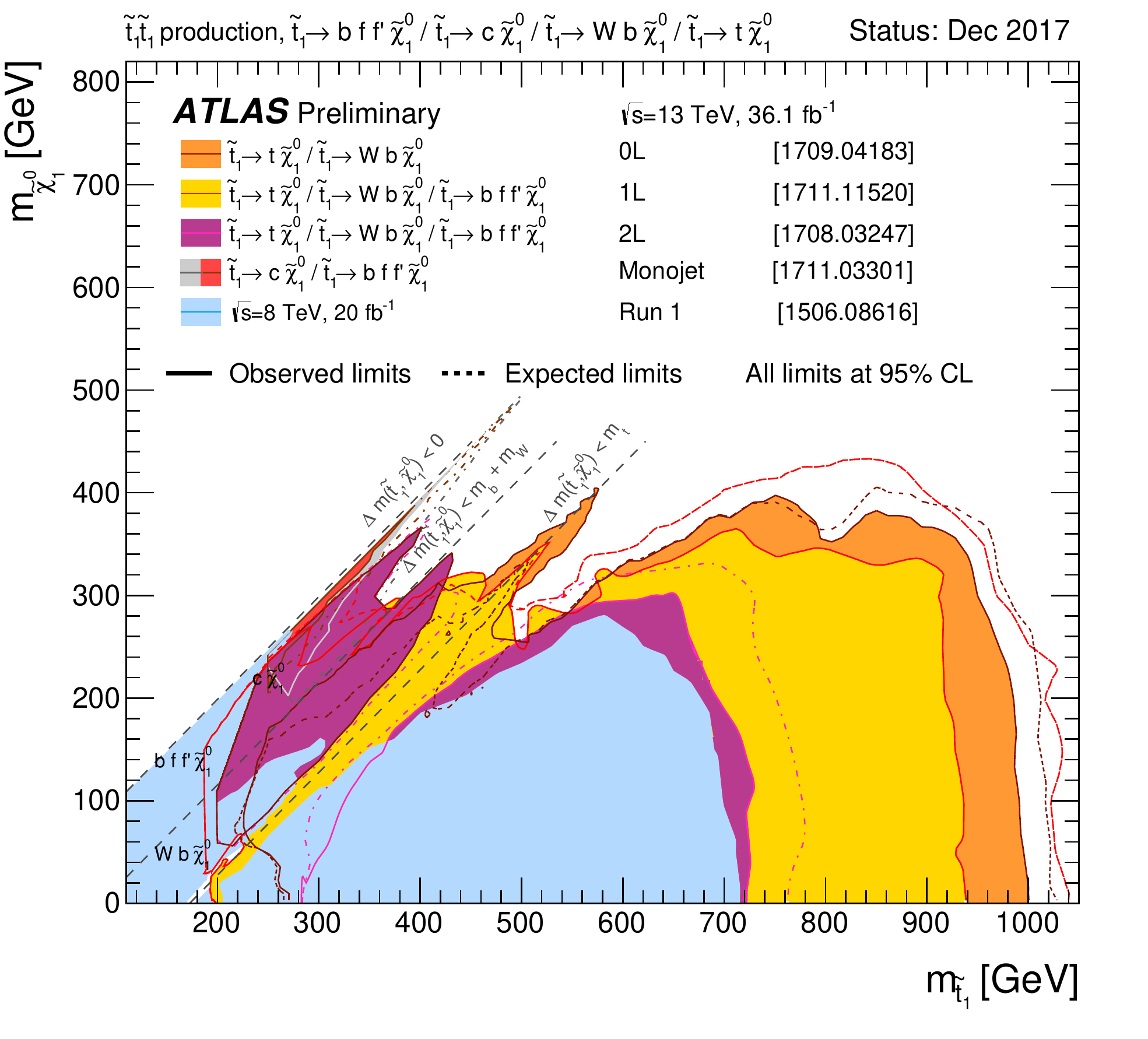}
\includegraphics[width=0.3\textwidth]{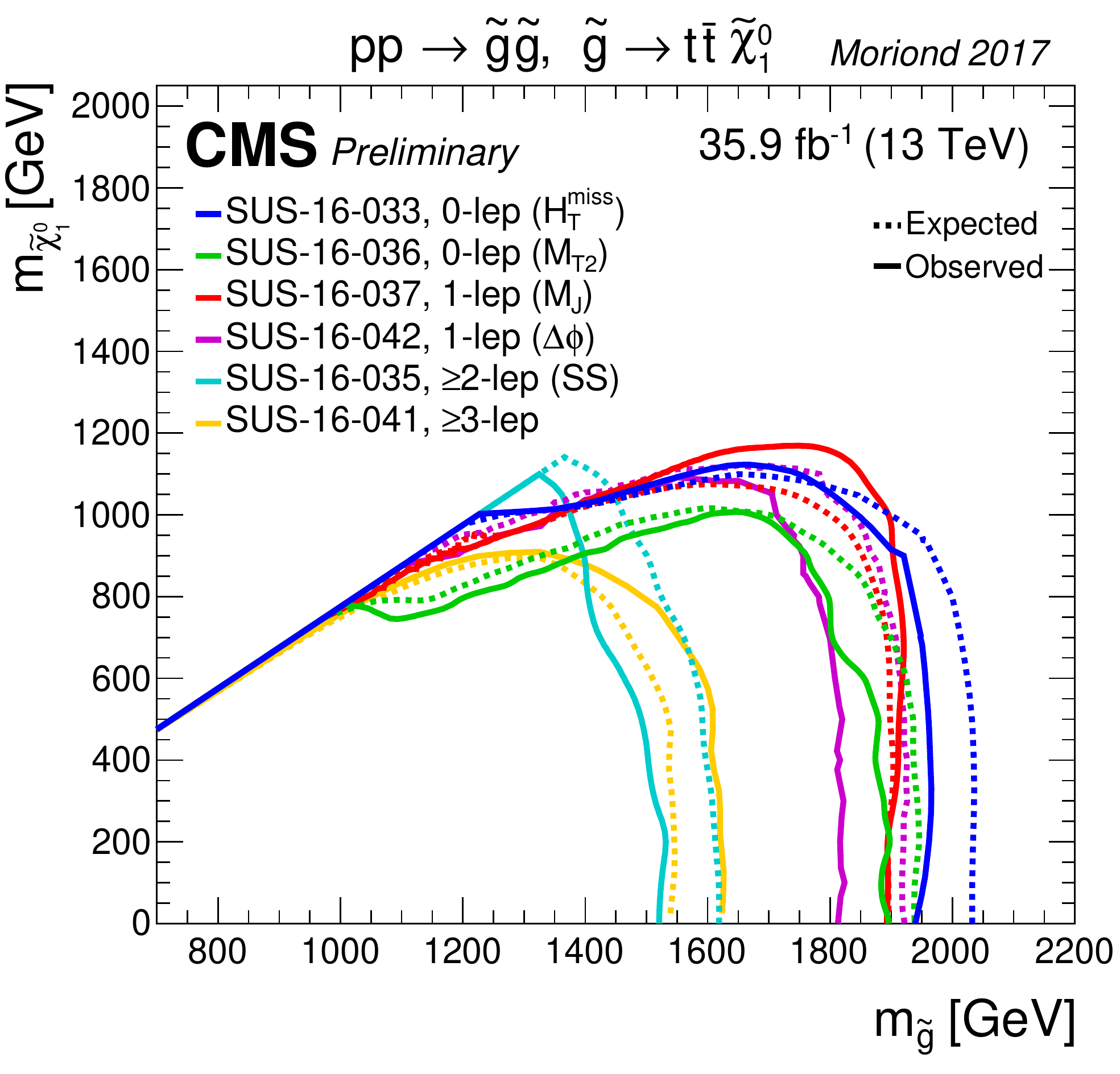}
\includegraphics[width=0.33\textwidth]{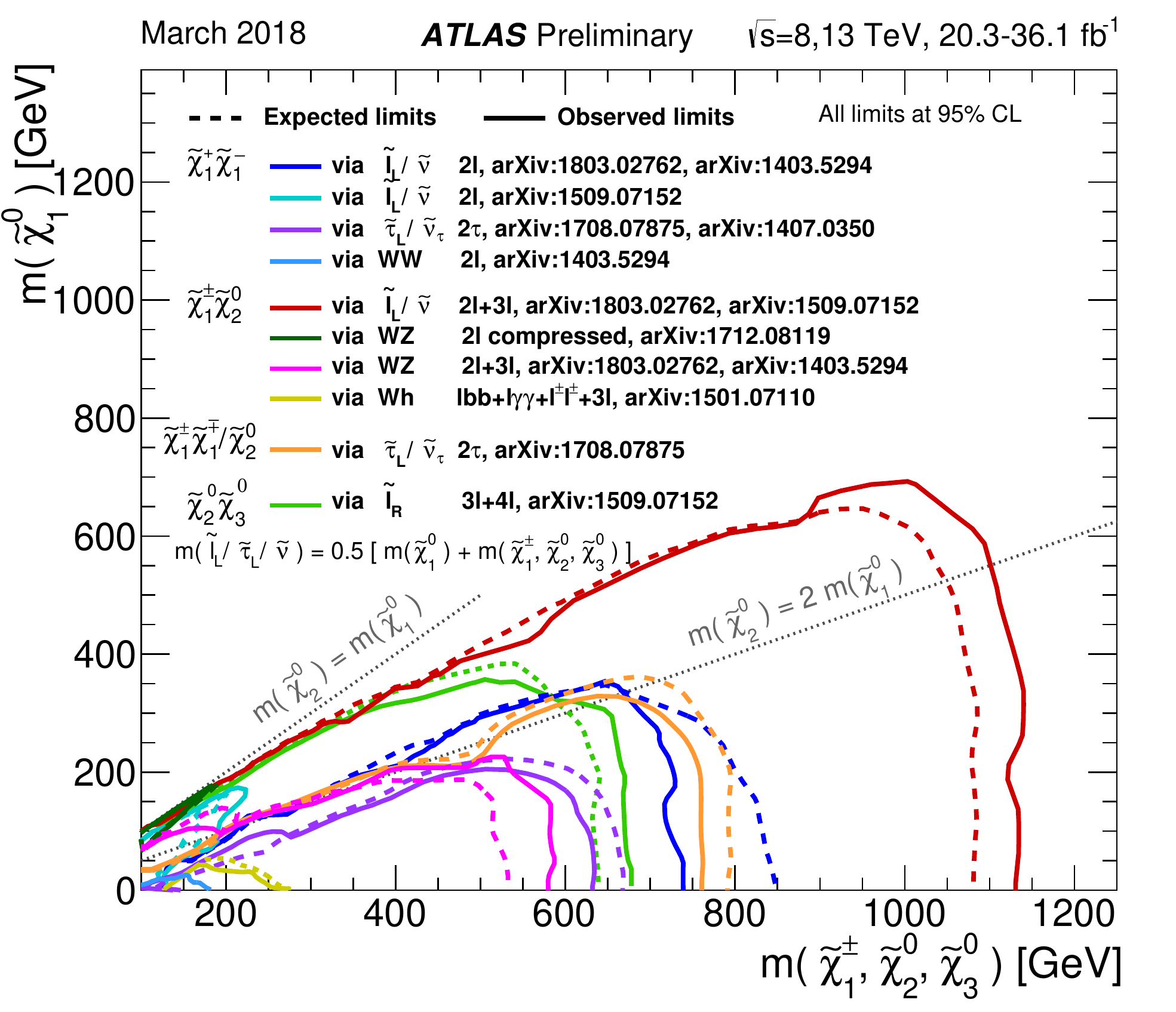}
\end{center}
\caption{Search for supersymmetry: the universal parameter plot (upper row) and the superpartner mass plot (lower row)}
\label{susysearch}
\end{figure}

As one can see from Fig.\ref{susysearch}~\cite{SUSY_serches}, the progress achieved at the LHC run is rather remarkable. The boundary of possible values of masses of the scalar quarks and gluino have reached approximately 1500 and 1000 GeV, respectively.  For the stop quarks it is almost two times lower. This is because the created squark always decays into the corresponding quark and in the case of the top quark, due to its heaviness, the phase space decreases and so does the resulting branching ratio. For the lightest neutralino the mass boundary varies between 100 and 400 GeV depending on the values of other masses. The constraints on the masses of charged weakly interacting particles  are almost two times higher than those for the neutral ones but depend on the decay mode. Let us stress once more that the obtained mass limits depend on the assumed decay modes which in their turn depend on the mass spectrum  of superpartners, which is unknown.  The presented constraints refer to the natural scenario. 

\begin{figure}[htb]
\begin{center}
\includegraphics[width=0.9\textwidth]{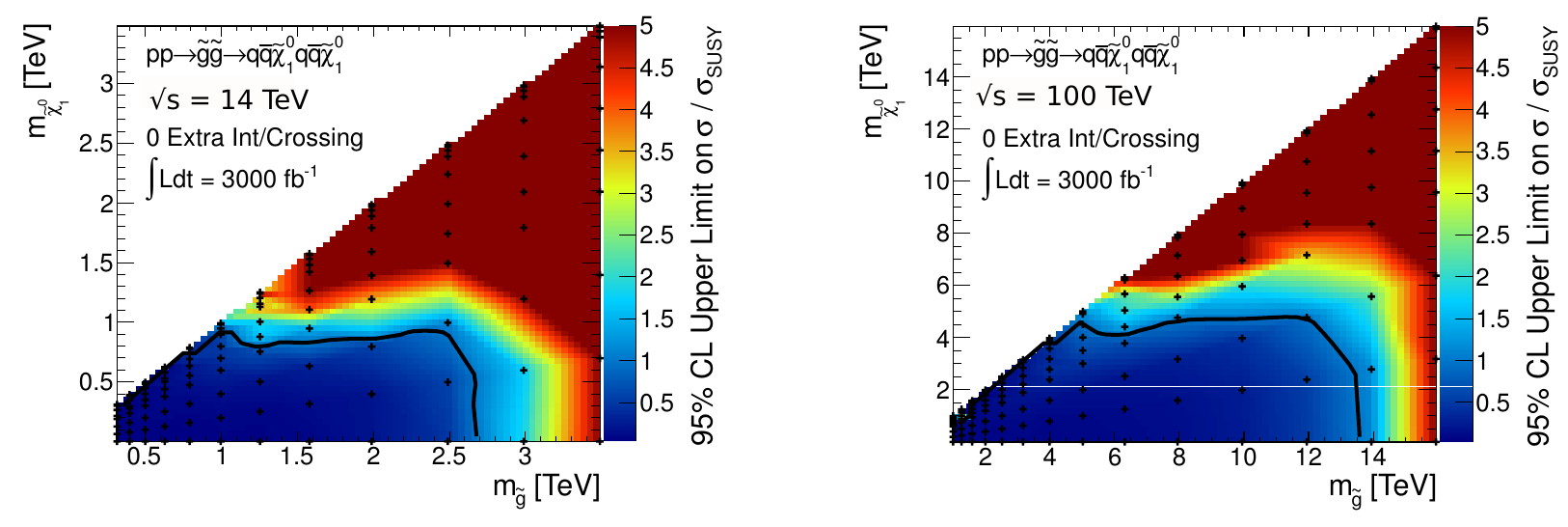}
\end{center}
\caption{Search for supersymmetry at the LHC and FCC~\cite{FCC}}
\label{future}
\end{figure}

The enormous progress reached by the LHC is slightly disappointing.  The natural question arises: Are we looking in the right direction? Or maybe we have not yet reached the needed mass interval? The answers to these questions can be obtained at the next runs of the accelerator. For the doubled energy the cross-sections of the particle production  with the masses around 1 TeV rise almost by an order of magnitude, and one might expect much higher statistics. Taking the gauge coupling unification seriously, SUSY may have some chance to be seen at the LHC, and a good chance at the
FCC. The mass range reach of the high luminosity LHC and the FCC collider are shown in Fig.\ref{future}.

\subsection{Grand Unification}

Grand Unification is an extension of the Gauge symmetry of the SM. Grand Unified Theories (GUT)
unify strong, weak and electromagnetic interactions in the framework of a single theory based on a simple symmetry group~\cite{GUT}. In this case the internal symmetry group of the SM, namely, $SU(3)\times SU(2)\times U(1)$ becomes a part of a wider group $G_{GUT}$.
All known interactions are considered as different branches of a unique
interaction associated with a simple gauge group. The unification
 (or splitting) occurs at high energy
$$\begin{array}{ccccl}
 & \mbox{Low energy} & & \Rightarrow & \mbox{High energy} \\
SU_c(3) \otimes & SU_L(2) \otimes & U_Y(1) & \Rightarrow &
 G_{GUT} \ \ (or \ G^n + \mbox{discrete  symmetry}) \\
\mbox{gluons} & W , Z & \mbox{photon} & \Rightarrow & \mbox{
gauge bosons} \\
\mbox{quarks} & \mbox{leptons} & & \Rightarrow & \mbox{fermions} \\
g_3 & g_2 & g_1 & \Rightarrow & g_{GUT} \end{array} $$

At first sight this is impossible due to a big difference in
the values of the couplings of strong, weak and electromagnetic
interactions. The crucial point here is the running coupling
constants. According to the renormalization group equations,
all the couplings depend on the energy scale. 
\begin{figure}[htb]
\begin{center}
\includegraphics[width=0.4\textwidth]{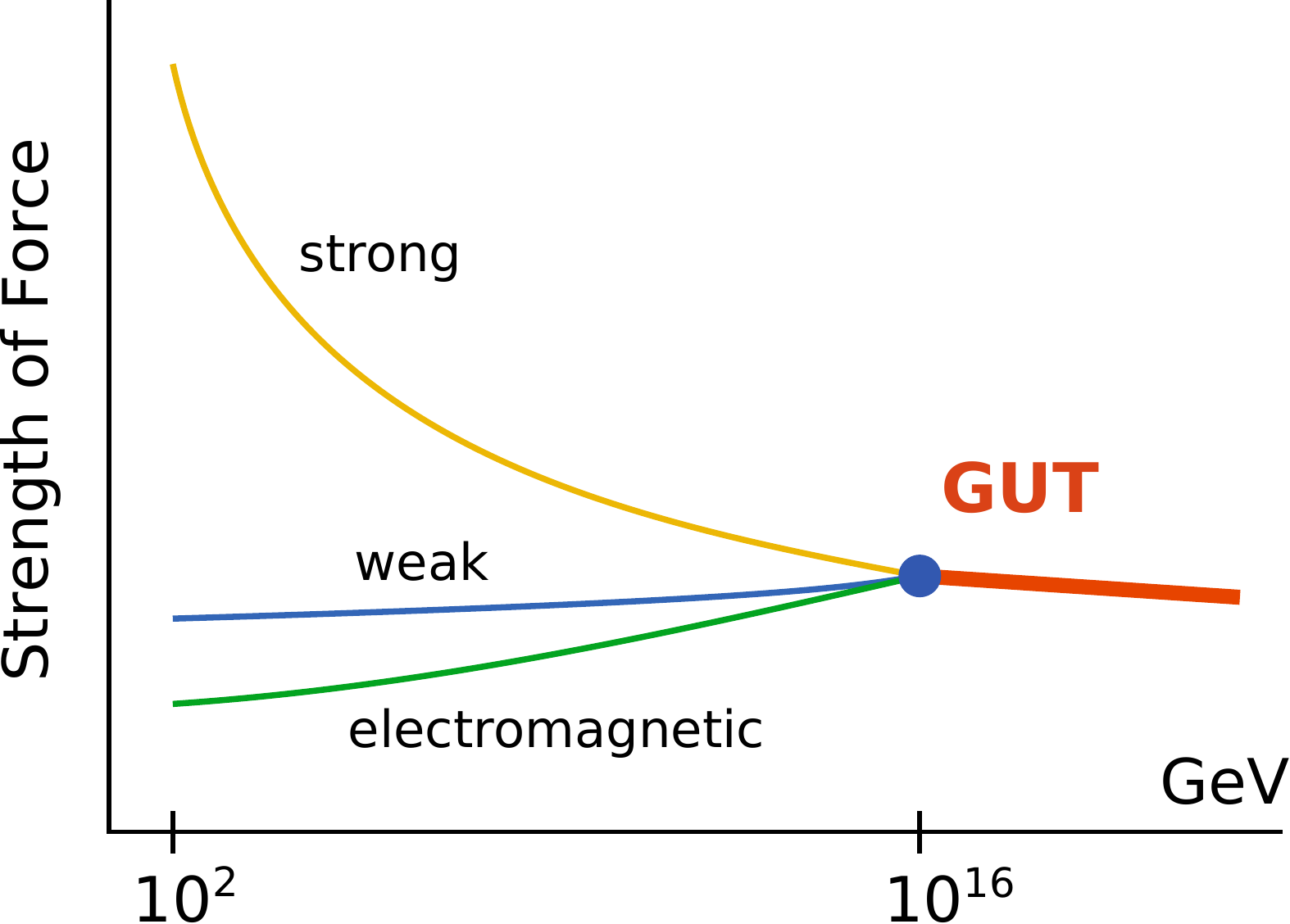}
\caption{The running coupling constants in the GUT scenario}
\label{151}
\end{center}
\end{figure}
 In the SM the strong and weak couplings associated with
non-abelian gauge groups decrease with energy, while the
electromagnetic one associated with the abelian group on the
contrary increases. Thus, it becomes possible that on some
energy scale they become equal (see Fig.\ref{151}).

According to the GUT idea, this equality is not occasional
but is a manifestation of unique origin of these three
interactions. As a result of spontaneous symmetry breaking,
the unifying group is broken and unique interaction
is splitted into three branches which we call strong, weak
 and electromagnetic interactions.

 The symmetry group of a Grand Unified Theory should be sufficiently wide to include the group of the SM and should have appropriate complex representations to fit quarks and leptons inside them.
 This means that the rank of this group (the maximal number of linearly independent generators that commute with each other) should be equal or larger to that of the SM group, i.e. 4. Remind the classical groups of rank $l$: $SU_{l+1},SO_{2l+1},SO_{2l},Sp_{2l}$. Thus, the minimal group  of rank 4 is $SU(5)$. 
 \underline{SU(5) GUT - Minimal GUT}

$SU(5)$ is a minimal group (rank 4) into which $SU(3) \otimes
 SU(2) \otimes U(1)$ can be embedded and which has complex
representations needed for chiral fermions. This group
satisfies all the requirements mentioned above. 
Particle content of the $SU(5)$ GUT is the following:

{\em Gauge sector}. \ \ $W_{\mu} = W_{\mu}^A T^A, \ \ \ A = 1,2, \ldots
 24 , \ \ T^A $ are the generators of $SU(5)$. It is a $24-plet$
which can be represented as a traceless $5 \times 5$ matrix
$$ W_{\mu} = \left ( \begin{array}{cccccc}
 &  &  & \vdots & X_{\mu}^1 & Y_{\mu}^1 \\
G_{\mu}^a \frac{\lambda^a}{2} & - & \frac{1}{\sqrt{15}}B_{\mu}
 {\bf 1}_3 & \vdots & X_{\mu}^2 & Y_{\mu}^2  \\
 &  &  & \vdots & X_{\mu}^3 & Y_{\mu}^3 \\
\cdots & \cdots & \cdots & \vdots & \cdots & \cdots \\
X_{\mu}^{*1} & X_{\mu}^{*2} & X_{\mu}^{*3} & \vdots & \frac{1}{2}
A_{\mu}^3 + \sqrt{\frac{3}{20}}B_{\mu} & W_{\mu}^+ \\
Y_{\mu}^{*1} & Y_{\mu}^{*2} & Y_{\mu}^{*3} & \vdots & W_{\mu}^- &
 - \frac{1}{2}A_{\mu}^3 + \sqrt{\frac{3}{20}}B_{\mu}  \end{array}
 \right ) $$
Among 24 gauge bosons there are 8 gluons $G_{\mu}^a$, 3 weak bosons
$W_{\mu}^{\pm}$ and $A_{\mu}^3$ and 1 $U(1)$ boson $B_{\mu}$. There
 are also 12 {\em new} fields $X_{\mu}$ and $Y_{\mu}$. They are usually called
 lepto-quarks because they mediate lepto-quark transition
leading to baryon No violation. The gauge multiplet has the
following $SU(3) \otimes \ SU(2)$ decomposition
$$\begin{array}{cccc} \underline {24} = & (\underline 8 , \underline 1 )
 & + ( \underline 1 , \underline 3 ) & + ( \underline 3 , \underline 2 )
 + ( \underline 3 , \underline 2 ) \\
 & gluons & W \ and \ Z & leptoquarks \end{array}$$
All fermions are taken to be left-handed.
 Right-handed particles are replaced by the corresponding left-handed
conjugated ones. The minimal fundamental representation of $SU(5)$ is
 $\underline 5$. However, it is more convenient to use the conjugated
one which has appropriate $SU(3) \otimes SU(2) \otimes U(1)$
quantum numbers
$$\underline 5^* = (\underline 3 , \underline 1 , -2/3 ) + (
\underline 1 , \underline 2 ,1) $$
It is naturally identified with d-quark and electron-neutrino
 doublet
$$\underline 5^* = (d_1^c , d_2^c , d_3^c , e^- , \nu_e )_{Left} $$

To find place for the other members of the same family, we have
 to go beyond the fundamental representation. Surprisingly, the
 next (after $\underline 5$) representation, $\underline {10} =
 ( 5 \times 5)_{asym} $ has precisely correct quantum numbers
$$\underline {10} = (\underline 3 , \underline 2 , 1/3) +
 (\underline 3^* , \underline 1 , -4/3) + (\underline 1 ,
 \underline 1 , -2) $$
It is a $5 \times 5$ antisymmetric matrix and its fermion
 assignment is
$$ \underline {10} = \left ( \begin{array}{ccccc}
0 & u_3^c & -u_2^c & u_1 & d_1 \\
  & 0 & u_1^c & u_2 & d_2 \\
  &  & 0 & u_3 & d_3 \\   &  &  & 0 & e^+ \\   &  &  &  &  0
\end{array} \right )_{Left} , \ \ \  \begin{array}{c}
u_L^c \ \rightarrow \ u_R \\  e_L^+ \ \rightarrow \ e_R
\end{array}. $$

Thus, all known fermions exactly fit to $(\underline 5^* + \underline
 10 ) $ representations of $SU(5)$. Now new fermions appear. Note
that there is no room for the right-handed neutrino $\nu_R$. Hence
either neutrino is massless in the $SU(5)$ model or it could be a singlet
that does not take part in gauge interaction. In spite of the left-
right asymmetry of the model there are no anomalies in the gauge
currents. They are automatically cancelled between contributions of $
\underline 5^* $ and $\underline {10} $.

\underline{SO(10) - optimal GUT}.  The next popular GUT is based on the $SO(10)$ group of rank 5.
The advantage of this model is that
 all the fermions of the same generation belong to a single
irreducible representation $\underline {16}$
$$\underline {16} = (u_1 \ u_2 \ u_3 \ d_1 \ d_2 \ d_3 \ \nu_e \ e^- \
 u_1^c \ u_2^c \ u_3^c \ d_1^c \ d_2^c \ d_3^c \ \nu_e^c \ e^+ )_{Left}$$
Note that contrary to the $SU(5)$ model the right-handed neutrino
(left-handed antineutrino) is present now. This means that the neutrino
in the $SO(10)$ model is massive. The gauge field in the $SO(10)$ model has dimension 45.
The $SO(10)$ multiplets find their
natural decomposition in terms of that of $SU(5)$
\begin{eqnarray*}
\underline {16} & = & \underline 5^* + \underline {10} + \underline
 1  \ \ \ \ \ \ fermions , \\
\underline {45} & = & \underline {24} + \underline {10} + \underline
 {10^*} + \underline 1 \ \ \ \ gauge \ bosons . \end{eqnarray*}
 
\underline{E(6)  GUT}  The next example is the  model based on the exceptional group $E(6)$ of rank 6. It is
left-right symmetric
$$E(6) \ \ \supset \ \ SU_C(3) \otimes SU_L(3) \otimes SU_R(3).$$
Fermions belong to a single fundamental representation $\underline
 {27}$ which has the following decomposition under $SO(10)$
$$\underline {27} = \underline {16} + \underline {10} +
\underline 1 ,$$
while the gauge bosons form an adjoint representation $\underline
 {78}$.
This model contains a lot of new particles. Its attractiveness is
mainly due to the appearance of $E(6)$ GUT in superstring inspired
models.

The GUT symmetry is broken spontaneously via the same Brout-Englert-Higgs mechanism.
In the case of $SU(5)$ it occurs in two stages: one introduces two Higgs multiplets:
$\underline {24} $ which breaks $SU(5)$ down to $SU(3) \otimes SU(2) \otimes U(1) $ and $\underline 5$ which breaks $SU(3) \otimes SU(2) \otimes U(1)$ down to $SU(3) \otimes U(1)$. The v.e.v are chosen to be
$$ <\Phi_{24}> = \left ( \begin{array}{ccccc}
V & & & &  \\  & V & & &  \\  &  & V &  &   \\  &  &  & -3/2 \ V &  \\
  &  &  &  &  -3/2 \ V  \end{array} \right ), \ \  <H_5> = \left ( \begin{array}{c} 0 \\ 0 \\ 0 \\ 0 \\
v/ \sqrt{2} \end{array} \right ) , $$
where $V \sim  M_{GUT} \sim 10^{15} $ Gev  and  $v \sim 250 $ Gev .

The symmetry breaking in the $SO(10)$ model can be achieved in two different ways and needs at least three different scales
$M_1 \gg M_2 \gg \cdots M_W$
$$\begin{array}{ccl}
 & \nearrow & SU(5) \ \stackrel{M_2}{\rightarrow} \ SU(3) \
\otimes SU(2) \otimes U(1) \ \stackrel{M_W}{\rightarrow} \ SU(3) \
\otimes U(1) \\
SO(10) & M_1 &  \\
 & \searrow & SO(6) \otimes SO(4) \ \sim \ SU(4) \otimes SU_L(2)
\otimes SU_R(2)  \end{array} $$

The Grand Unified Theories solve many problems of the SM, for instance, the absence of the Landau pole, reduction of the number of parameters, 
all particles might sit in a single representation (16 of SO(10)), 
unification of quarks and leptons, open the way to baryon and lepton number violation, etc. However, they produce new problems. This is first of all the hierarchy problem. Indeed, the unification of the couplings takes place at the GUT scale $\sim 10^{15}-10^{16}$ GeV where spontaneous symmetry breaking takes place. The new heavy particles acquire masses of the order of this scale. Interacting with the Higgs boson of the SM they create the radiative corrections to its mass of the order of their own, thus destroying the hierarchy (see Fig.\ref{corr}). The solution of this problem might be obtained in SUSY GUTs where these unwanted corrections are canceled
with  the contributions of superpartners in all orders of PT. This way supersymmetry stabilizes GUTs 
eliminating the influence of unknown heavy physics on low energy observables preserving hierarchy.

\begin{figure}[htb]
\centering
\includegraphics[width=0.4\textwidth]{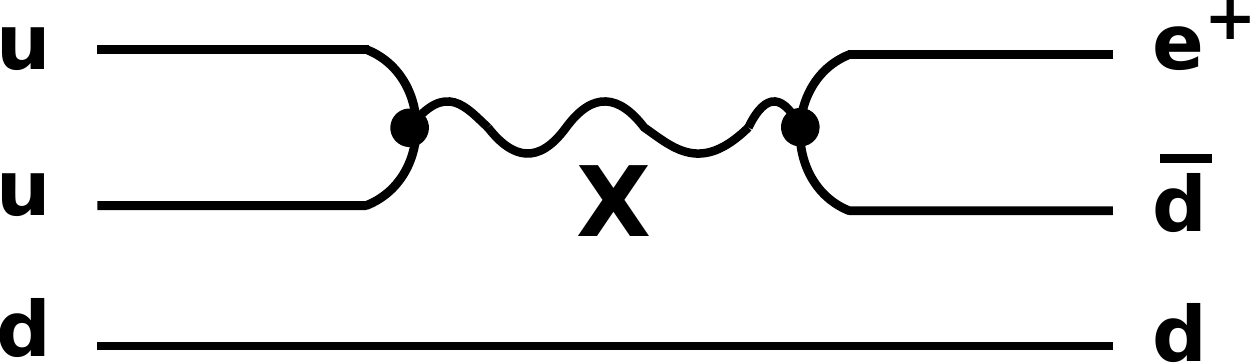}\hspace{15pt}
\includegraphics[width=0.36\textwidth]{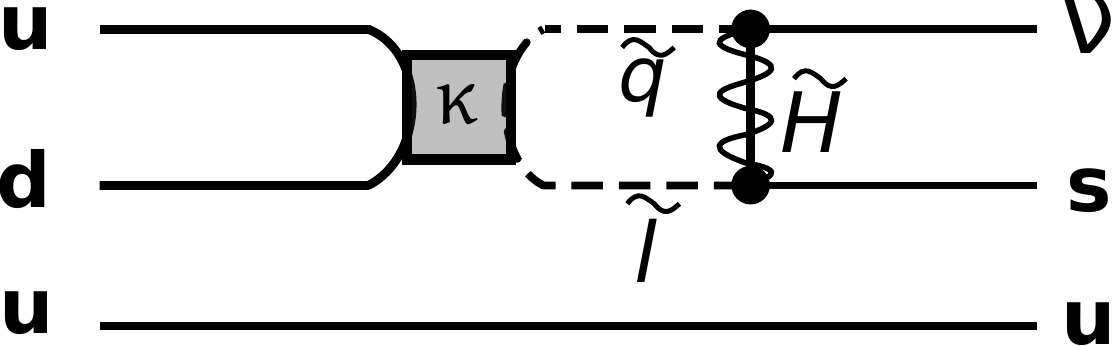}
\caption{The diagrams giving a contribution to proton decay in the usual GUT (left) and in the supersymmetric version (right)}
\label{proton}
\end{figure}

Since in GUTs quarks and leptons belong to the same representation of the gauge group, the interactions with the new gauge bosons leads to the processes where quarks convert into leptons and vice versa, i.e. to the violation of the baryon and lepton numbers, contrary to the SM. The key prediction of GUTs is proton decay. It takes place according to the process shown in Fig.\ref{proton} (left)  with creation of  $\pi^0$ meson and positron. 
The proton life time is proportional to the mass of the heavy $X$ boson $\tau_P\sim M_X^4$ that gives the value bigger than $10^{30}$ years. The modern experimental data give the lower bound $\sim 10^{34}$ years. At the same time, in the supersymmetric case there might be other modes of proton decay with creation of $K^+$ meson and  antineutrino (see Fig.\ref{proton} right). In this case, the decay rate is additionally suppressed due to the loop with superpartners inside. Experimental constraint here is weaker $\sim 10^{33}$ years. The search for the proton decay is continued. The observation of such a decay would be the confirmation of the GUT hypothesis.
\clearpage
\subsection{Extra symmetry factors}

\begin{figure}[htb]
\begin{center}
\leavevmode
\includegraphics[width=0.6\textwidth]{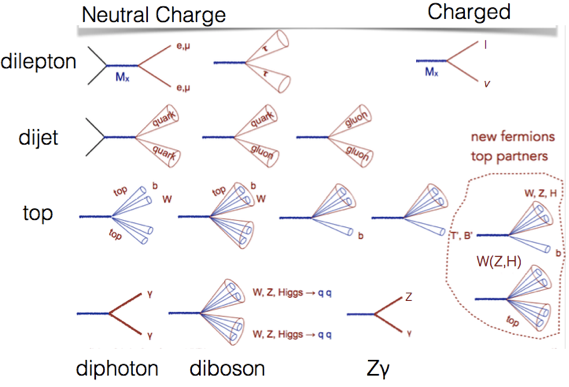}
\end{center}
\caption{Single jet and dijet events with high transverse energy}
\label{dijets}
\end{figure}

A less radical change of the symmetry group of the SM is the presence of additional symmetry factors like $U(1)'$ or $SU(2)' $, etc. These additional factors are typical for the string theory models and might continue the symmetry pattern of the SM. The presence of such factors leads to the existence of additional gauge bosons $A',Z',W'$, etc. At colliders they might appear as characteristic single or double jet events with high energy (see Fig.\ref{dijets}~\cite{Rahatlou}).

\begin{figure}[htb]
\begin{center}
\includegraphics[width=0.65\textwidth]{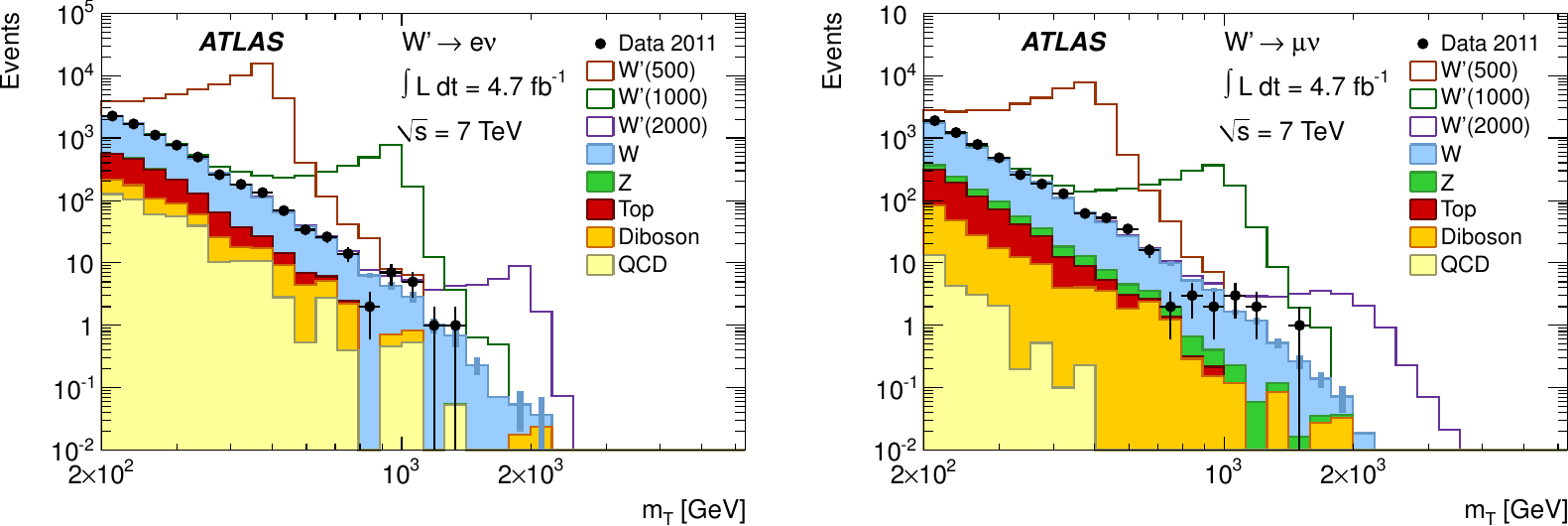}
\includegraphics[width=0.67\textwidth]{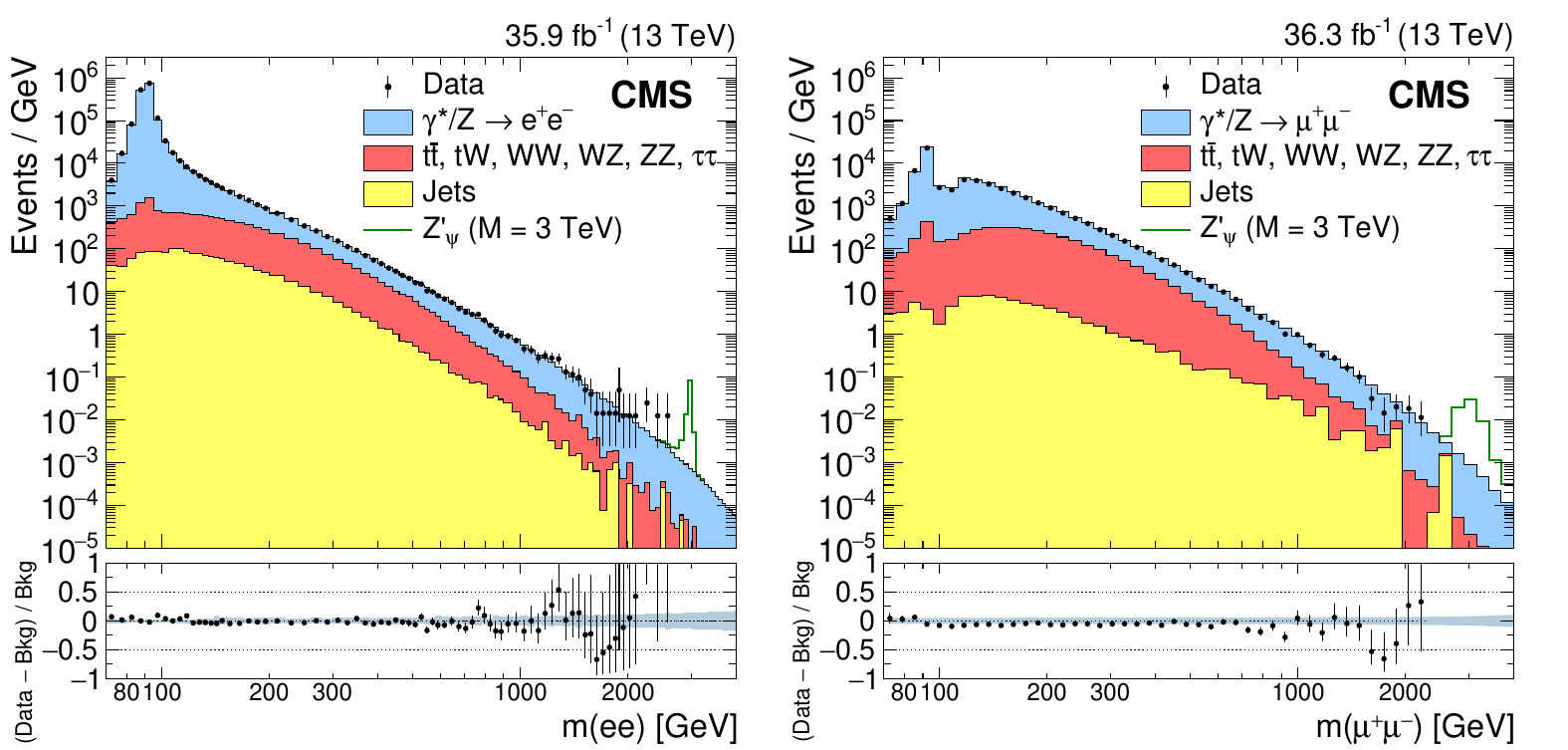}
\end{center}
\caption{Search for additional gauge bosons at the LHC}
\label{extra}
\end{figure}

Experimentally studied are the processes with $Z'$ boson production (dimuon events), $W’$ production (single muon/jets), resonant $t\bar t$ production, diboson events and monojet events with missing energy (see Fig.\ref{extra}~\cite{Search_gauge_bosons}). So far there are no positive signatures and we have just the bounds on the masses of these hypothetical particles of an order of TeV.

The other popular example of hypothetical new symmetries is the additional $U(1)'$ factor associated  with the so-called dark photon. The mixture with the ordinary photon due to the non-diagonal term ${\cal L} \sim F_{\mu\nu}F_{\mu\nu}'$ leads to conversion of the ordinary photon into the dark one that might be observed experimentally. There are already some dedicated experiments. Presumably the dark photon might be the dark matter particle.

\section{New Particles}

The Standard model can be extended introducing new particles as we have seen by example of supersymmetry or additional symmetry factors. However, there are many other possibilities of addition of new particles which are not related to the extension of the symmetry group.

\subsection{Extended Higgs sector}

Possible extension of the Higgs sector of the SM is an actual question which might be answered in the near future. Is the discovered Higgs boson the only one or not? What are the alternatives  to the one Higgs doublet model?  

\begin{figure}[htb]\vspace{0.6cm}
\leavevmode \hspace{1cm}
\includegraphics[width=0.32\textwidth]{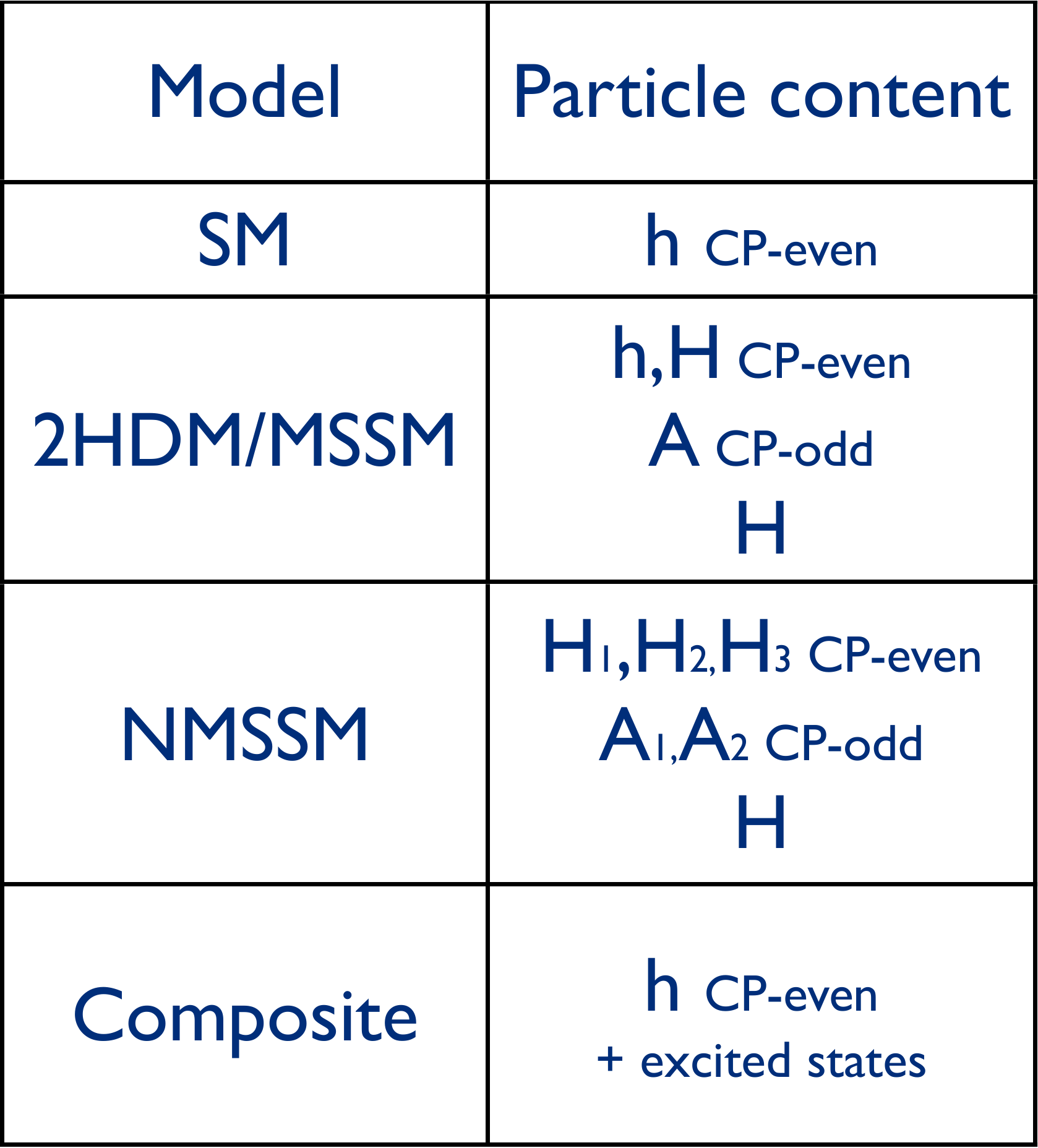} \vspace{-5.6cm}

\hspace{8.4cm}\includegraphics[width=0.32\textwidth,height=6cm]{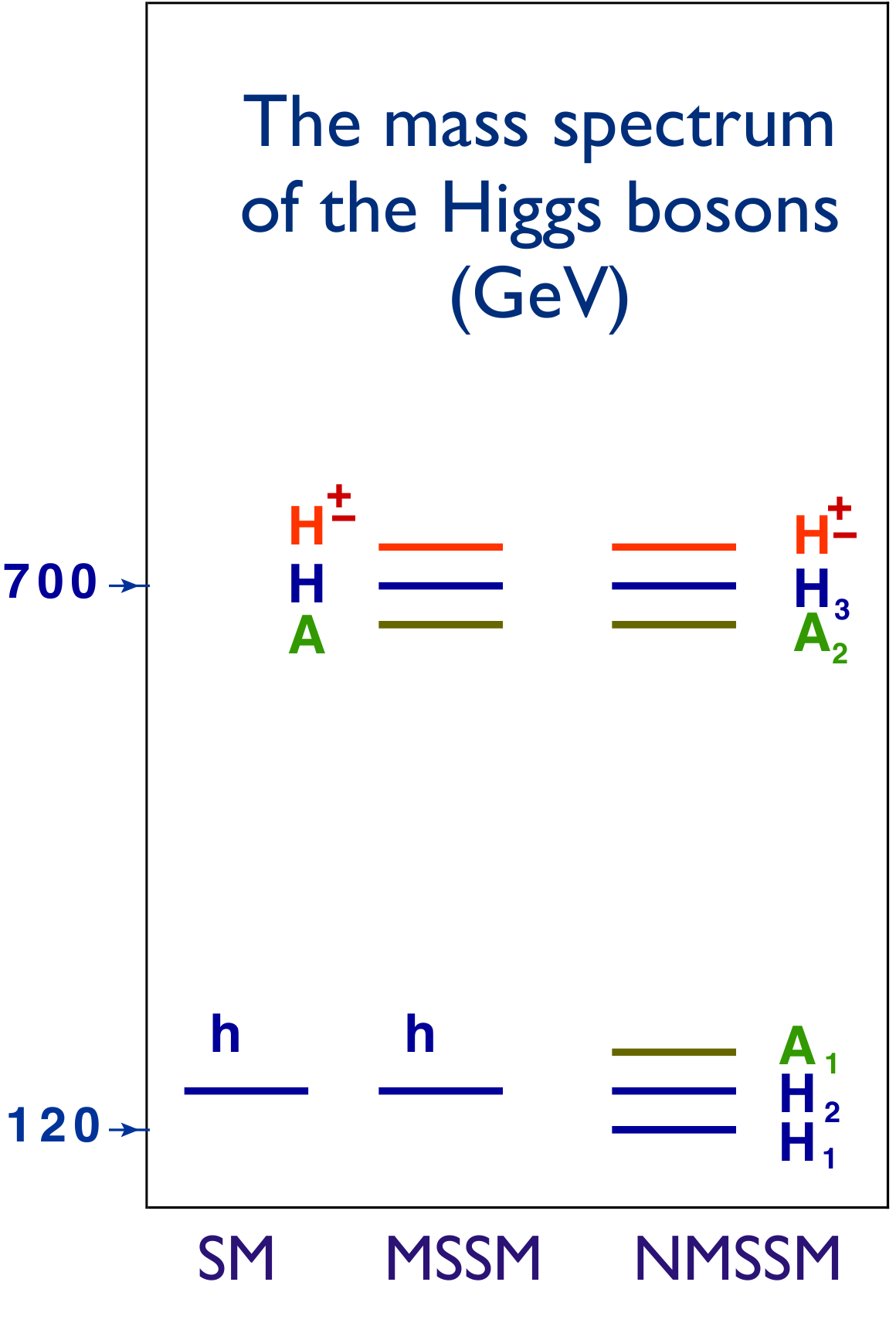}
\caption{The field content and the spectrum in various models of the Higgs sector}
\label{spectr}
\end{figure}

The nearest extension of the SM is the two Higgs doublet model~\cite{THDM}. It is also realized in the case of the Minimal Supersymmetric Standard Model (MSSM)~\cite{MSSM}.  Here the up and down quarks and leptons interact with different doublets  each of which has a vacuum expectation value. In this case, one has 5 Higgs bosons:  two CP-even, one CP-odd and two charged ones (see Fig.\ref{spectr} (left)). 

The next popular step is the introduction of an additional Higgs field which is a singlet with respect to the gauge group of the SM. In the case of supersymmetry, this model is called the NMSSM, the next-to minimal~\cite{NMSSM}.  Here one has already seven Higgs bosons. The sample spectrum of particles for various models is shown in Fig.\ref{spectr}, right. Note that in the case of the NMSSM, one has two light CP-even Higgs  bosons and the discovered particle might correspond to both $H_1$ and to  $H_2$. The reason why we do not see the lightest Higgs boson $H_1$ in the second case is that it has a large admixture of the singlet state and hence very weakly interacts with the SM particles.

\begin{figure}[htb]
\centering
\includegraphics[width=0.35\textwidth]{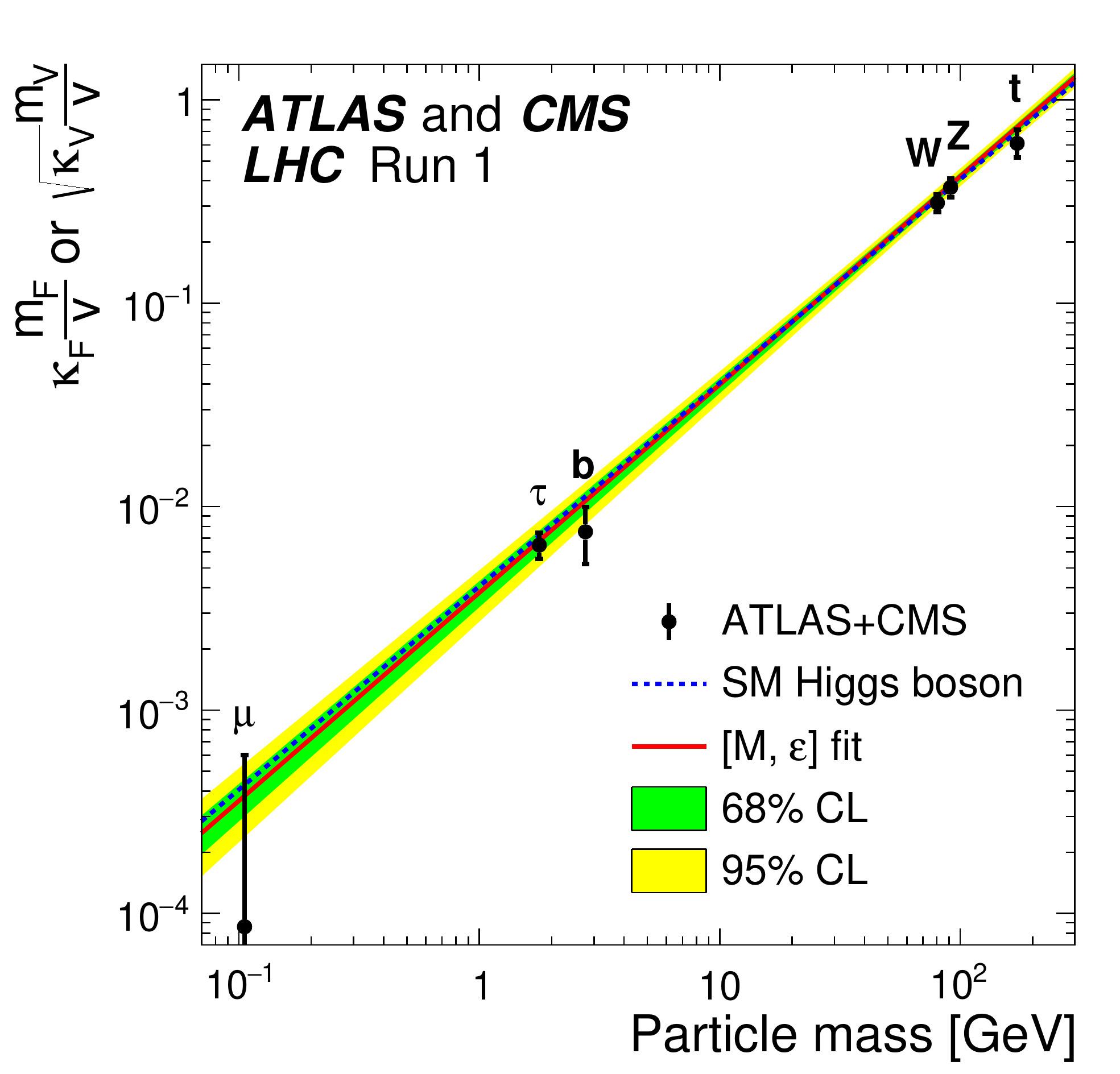}
\caption{Dependence of the Higgs couplings on the masses of quarks, leptons and intermediate gauge bosons}
\label{couplings}
\end{figure}

How to check these options experimentally? There are two methods: to measure the couplings of the 125-GeV Higgs boson with quarks, leptons and intermediate gauge bosons and check whether they deviate from the predictions of the SM. In the latter case they correspond to the straight line in the plot representing the couplings as functions of the masses of particles (see Fig.\ref{couplings}~\cite{ATLAS_Higgs}). Here the name of the game is high precision which can be achieved increasing the luminosity. 

\begin{figure}[htb]
\begin{center}\vspace{-0.2cm}
\leavevmode
\includegraphics[width=0.4\textwidth]{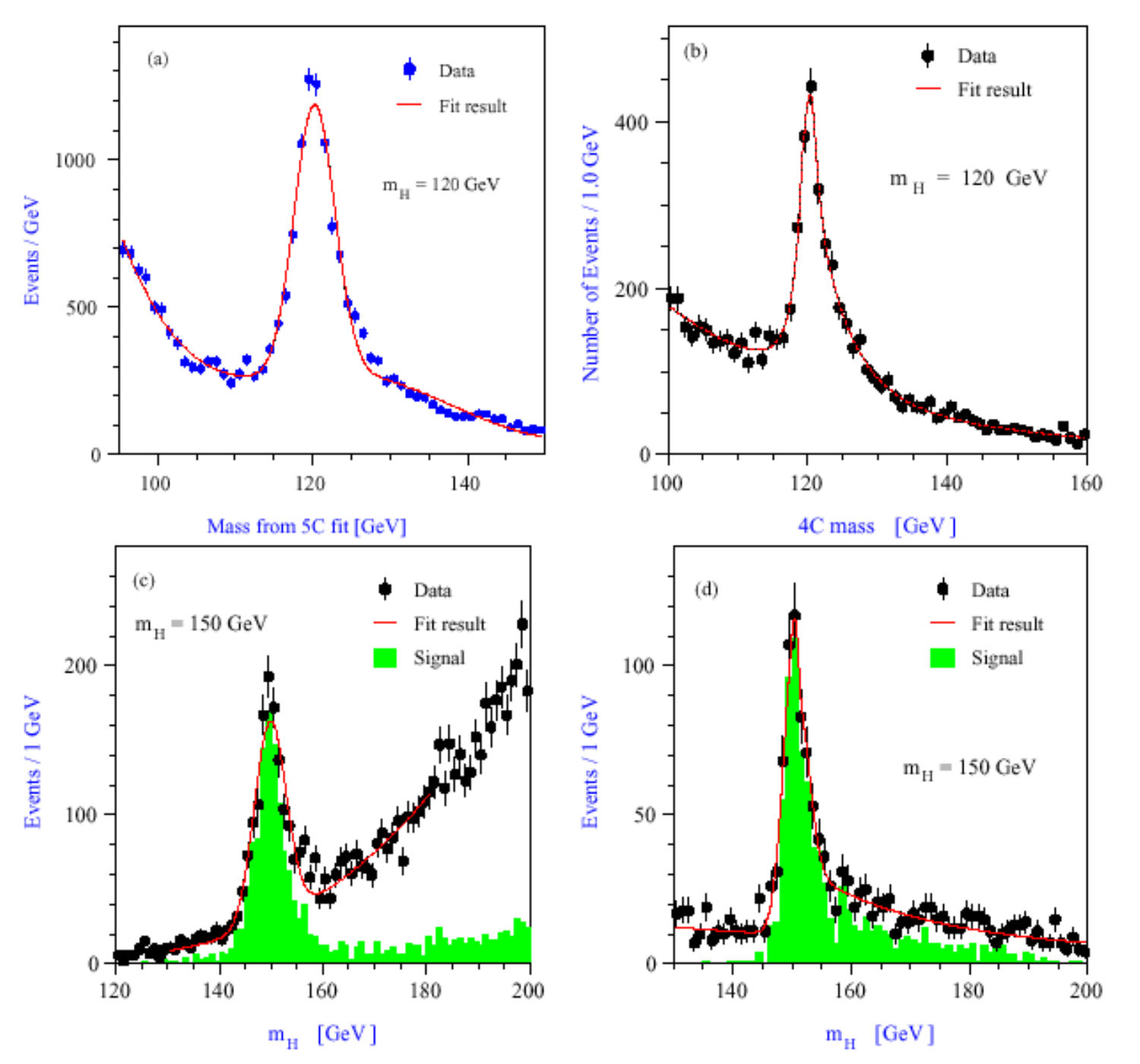}
\end{center}\vspace{-0.2cm}
\caption{The measurement of the mass and the width of the Higgs boson in various channels at the ILC:
$e^+e^-\to HZ\to b\bar b q\bar q,  q\bar q l^+l^-, W^+W^-q\bar q,  W^+W^-l^+l^-$ }
\label{ILC}
\end{figure}
The task for the near future is the precision analysis of the discovered Higgs boson. It is necessary to measure its characteristics  like the mass and the width and also all decay constants with the accuracy  ten times higher than the reached one. Quite possible that this task requires a construction of the electron-positron collider, for instance,  the linear collider ILC.  Figure \ref{ILC} shows the expected results for the Higgs boson mass measurement at the ILC in various channels~\cite{ILC}. 

\begin{figure}[ht]
\begin{center}
\leavevmode
\includegraphics[width=0.315\textwidth]{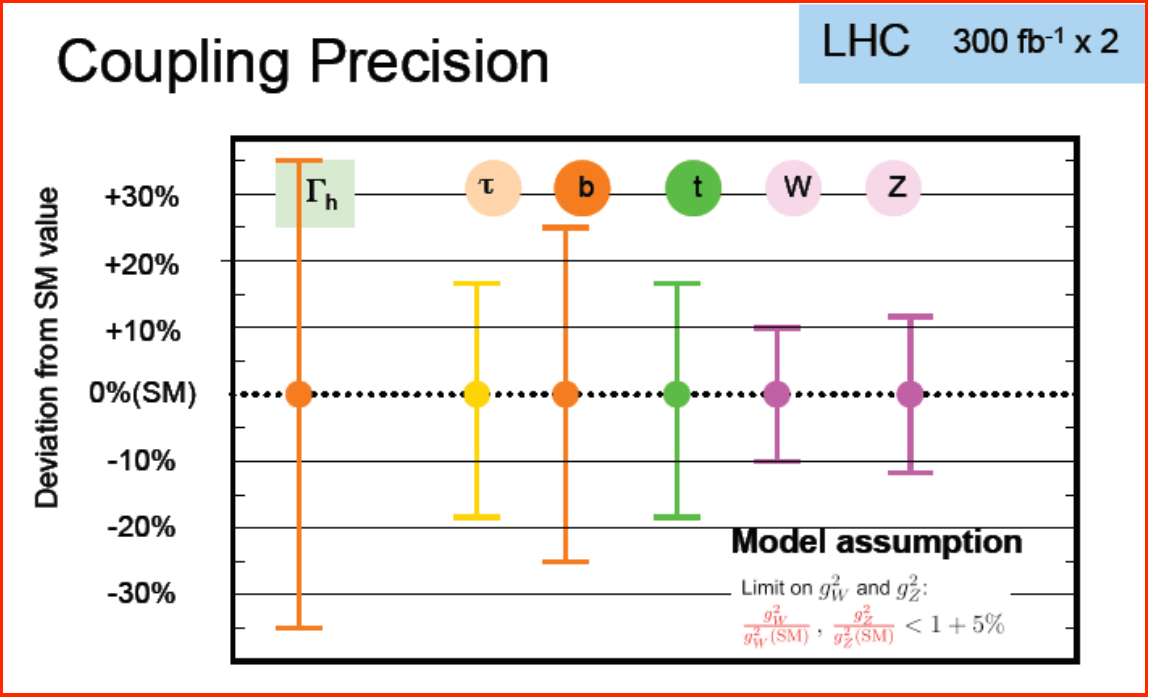}
\includegraphics[width=0.3\textwidth]{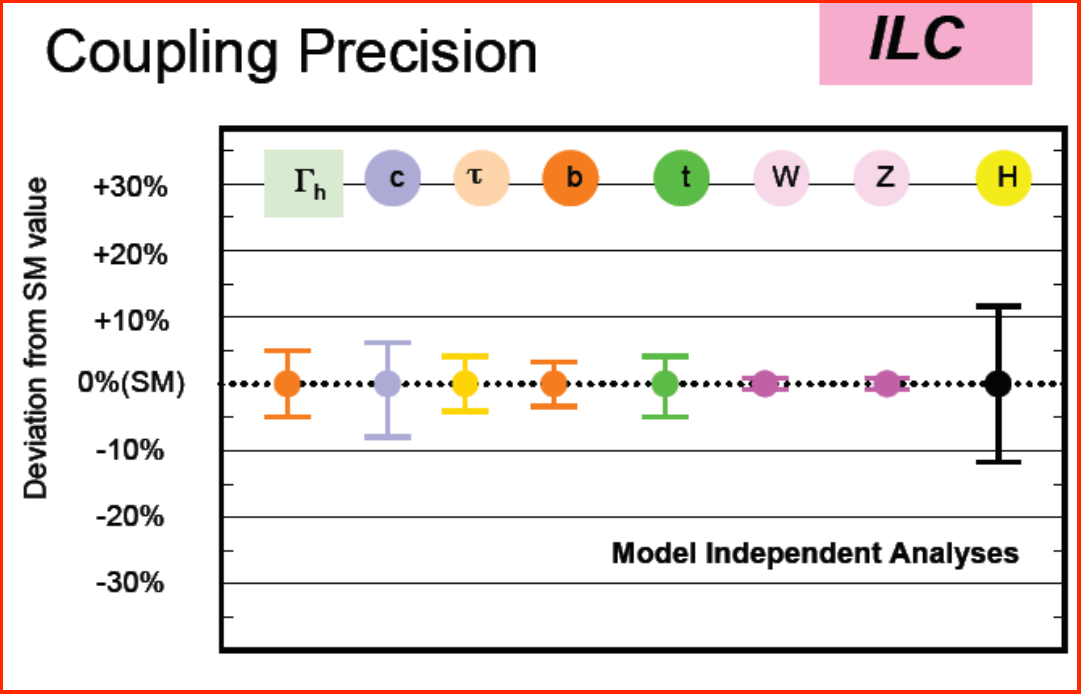}
\includegraphics[width=0.3\textwidth]{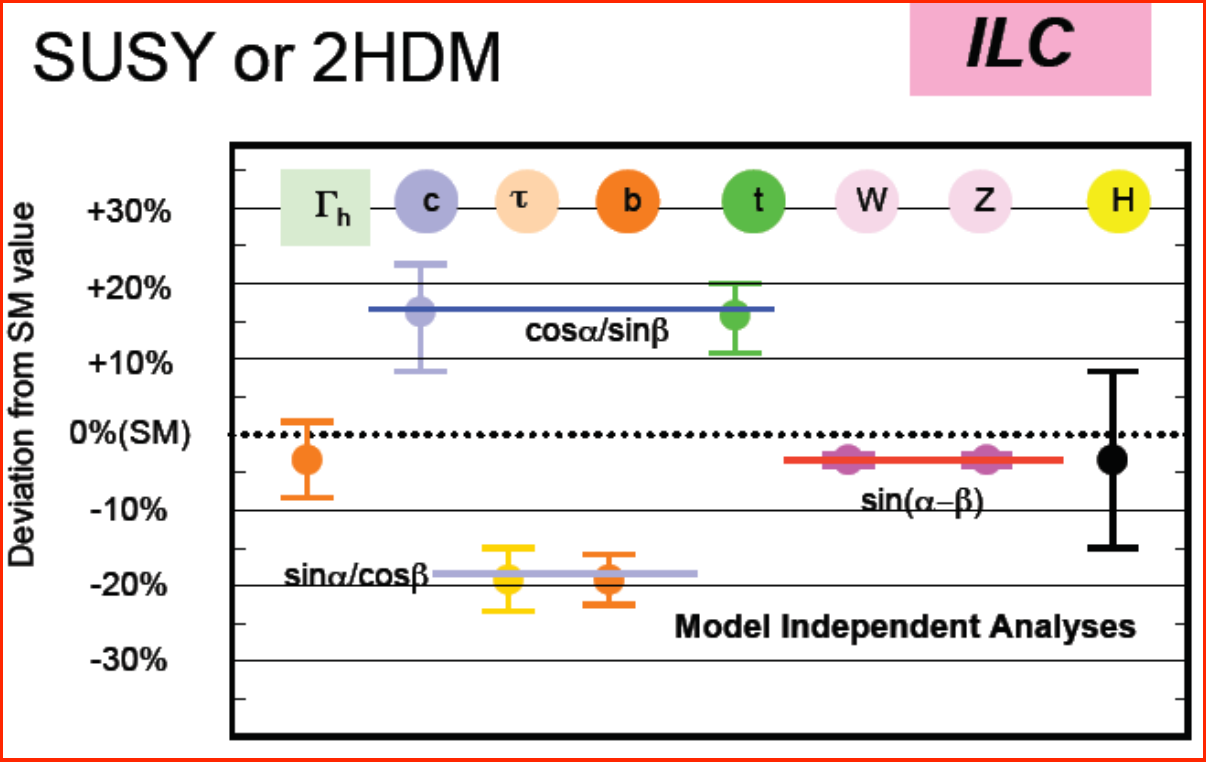}
\end{center}
\caption{The measurement of the Higgs boson couplings at the LHC and ILC~\cite{LHC_ILC}}
\label{ILCH}
\end{figure}

It is planned that the accuracy of the Higgs mass measurement  will achieve $\sim$50  MeV that is 5-7 times higher than the achieved one.  Another task is the accurate determination of the constants of all decays which will possibly allow one to distinguish the one-doublet model from the two-doublet one. Figure \ref{ILCH} shows the planned accuracies of the measurement of the couplings of the Higgs boson with the SM particles at the LHC for the integrated luminosity   of
300 1/fb (left), which is ten times higher than today.  For comparison we also show the same data for the ILC (middle). The accuracy of measurement of the couplings at the ILC will allow one  not only to distinguish different models but also check the predictions of supersymmetric theories (right).

The second way is the direct observation of additional Higgs bosons. 

\begin{figure}[htb]
\begin{center}
\leavevmode
\includegraphics[width=0.3\textwidth]{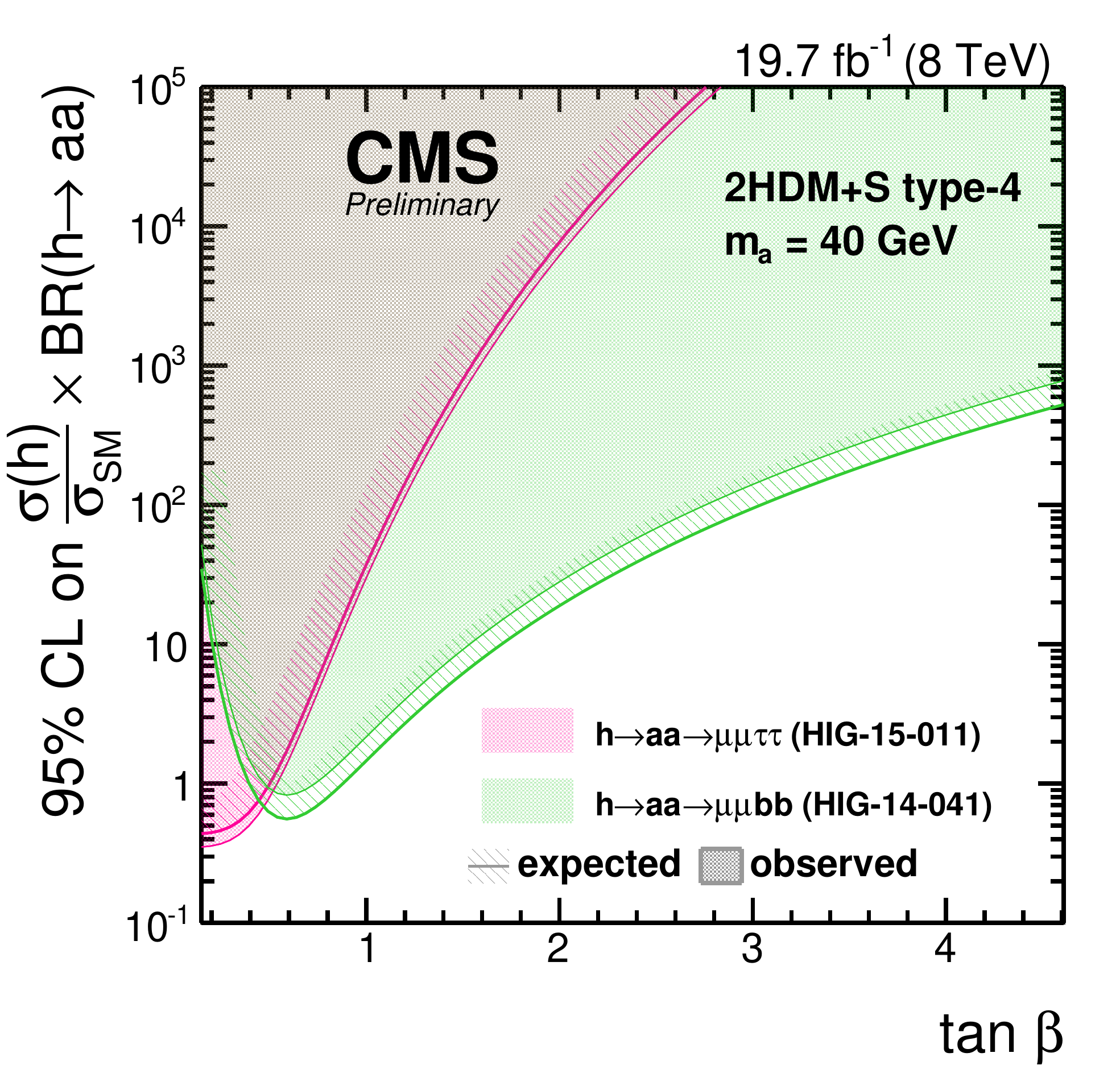}
\includegraphics[width=0.4\textwidth]{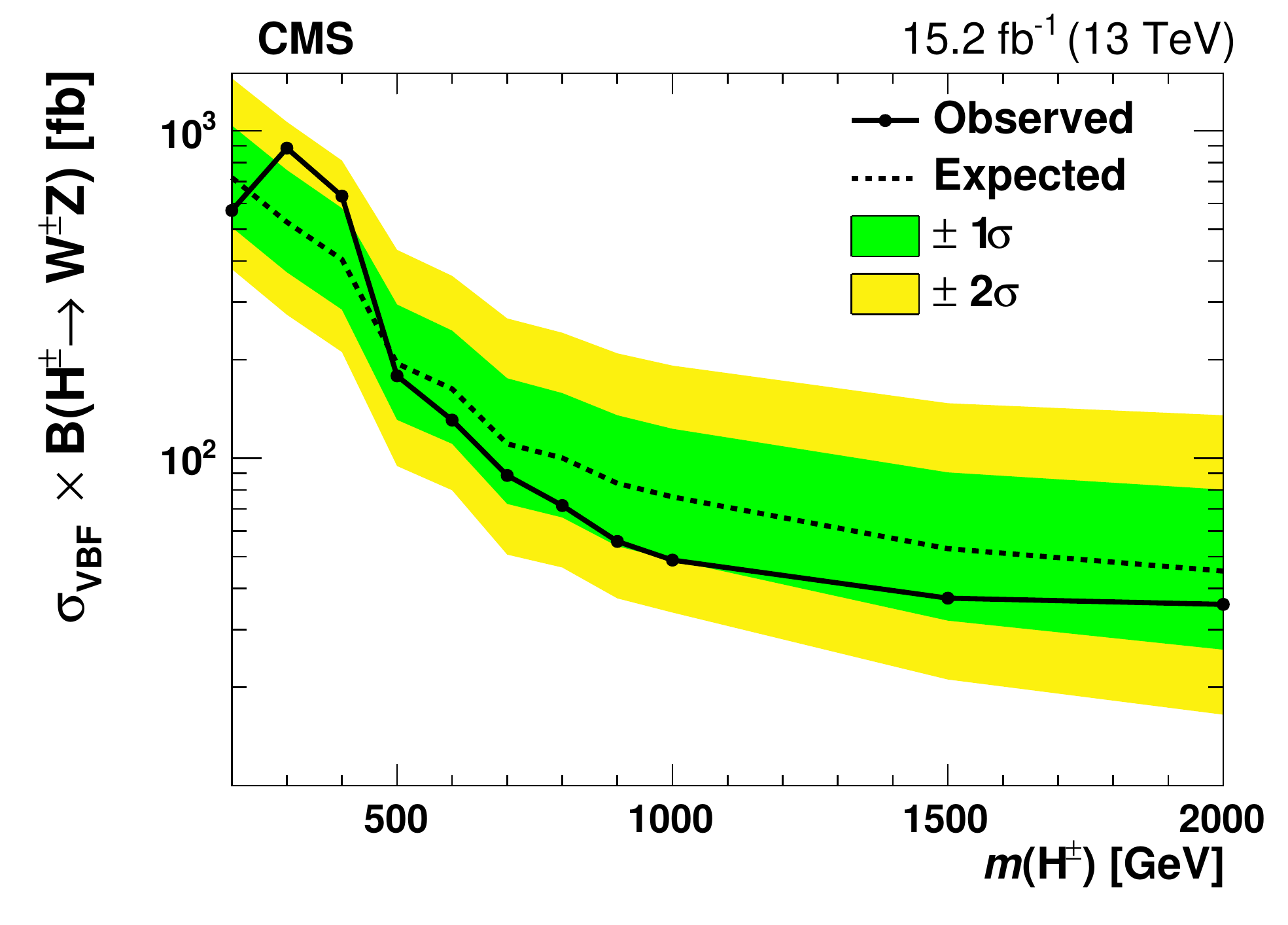}
\includegraphics[width=0.3\textwidth]{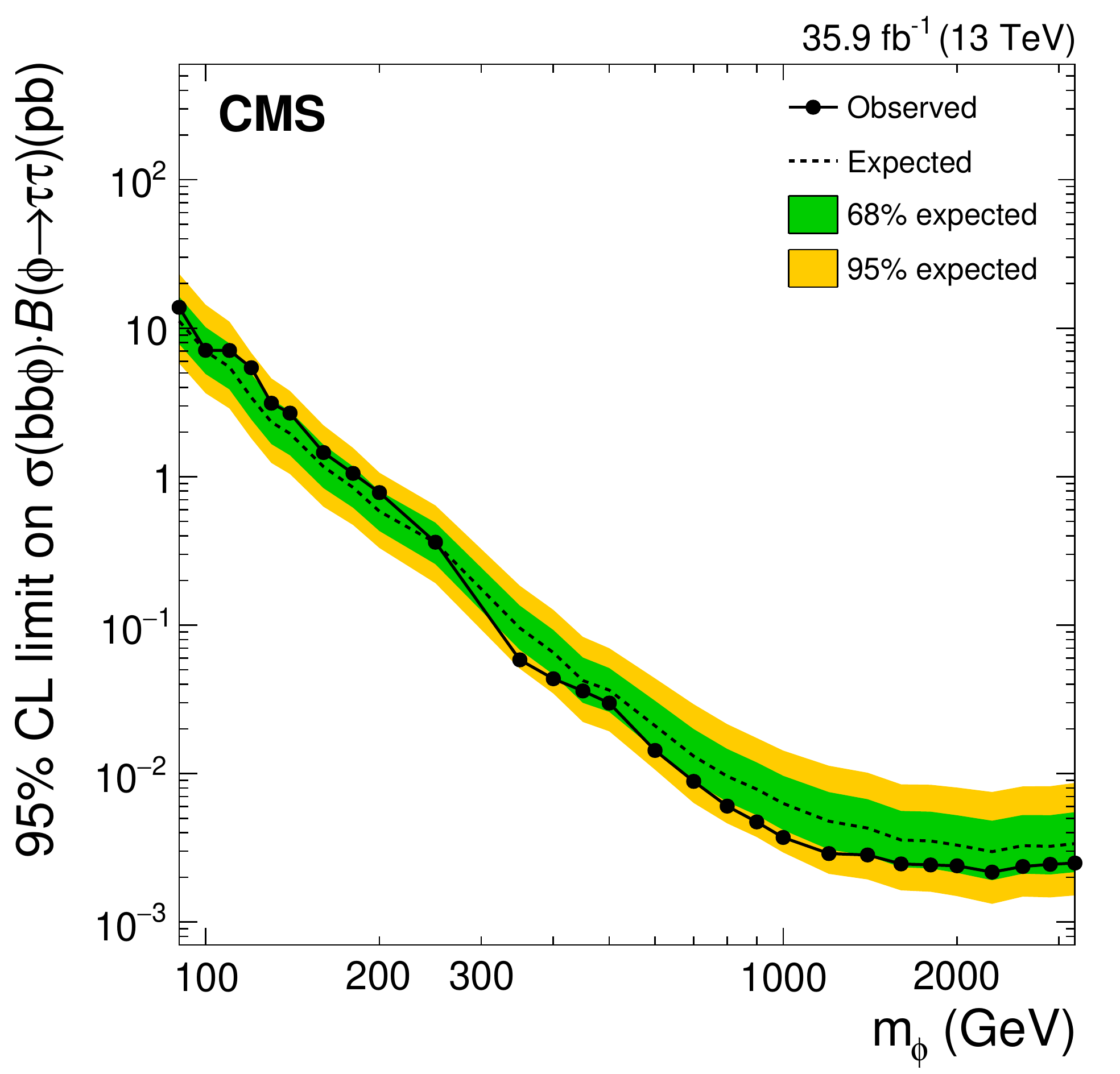}
\includegraphics[width=0.4\textwidth]{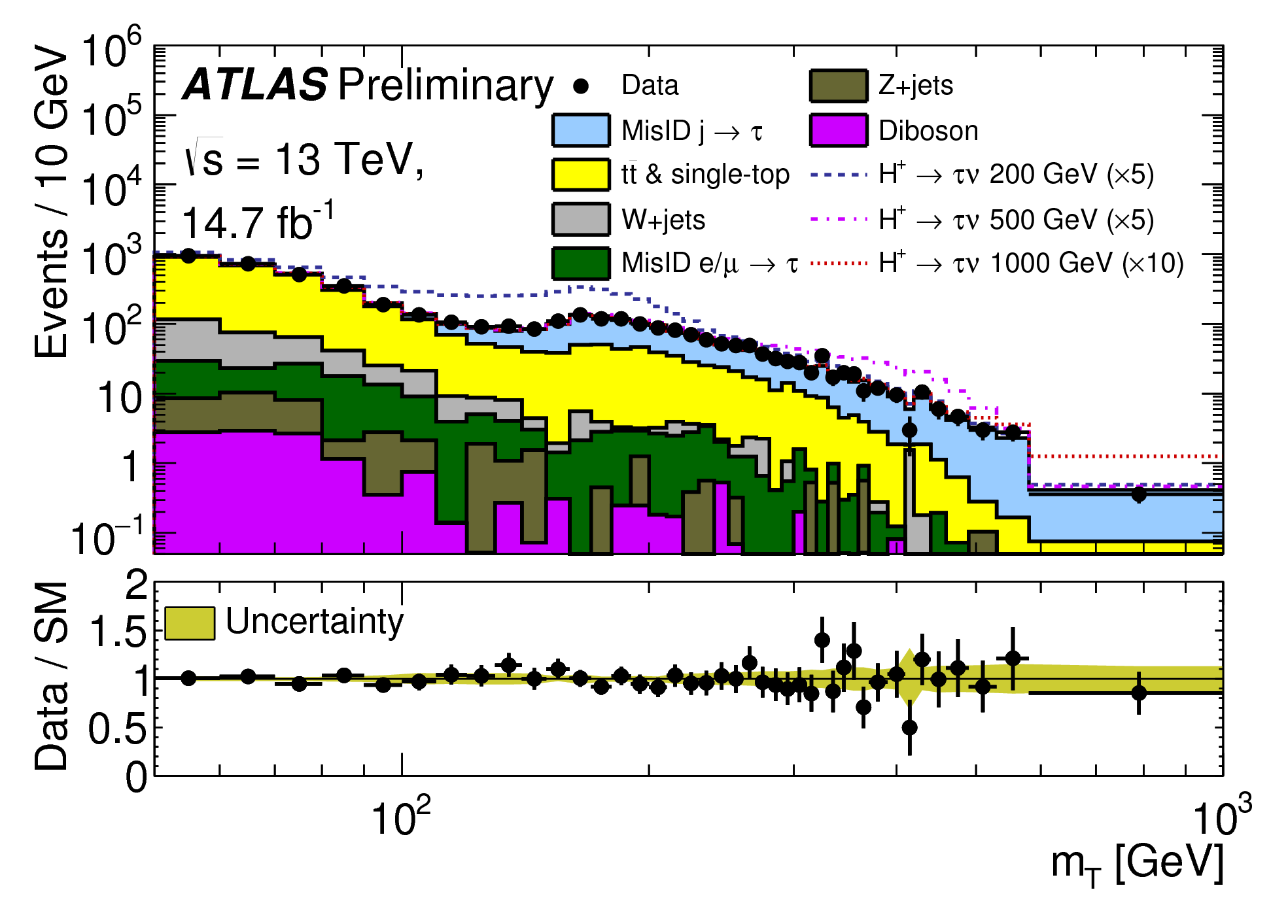}
\end{center}
\caption{The search for the heavy and charged Higgs bosons at the LHC}
\label{chargedH}
\end{figure}

The search for additional Higgs bosons, both the neutral and the charged ones, are performed now at the LHC in various channels. Up to now no signature is seen and we have only the constraints on the masses and parameters of the interaction. Unfortunately, there are no clear predictions for these parameters as it was with the 125-GeV Higgs boson. The results of experimental analysis are shown in Fig.\ref{chargedH}~\cite{chargedH}. The search for additional Higgs bosons in the interval $200 < m_H < 1000$ GeV did not give positive results so far.

\subsection{Axions and axion-like particles}

A completely different type of particles is represented axions and axion-like particles. They are related to the problem of CP-violation in strong interactions. As is well known, in the SM CP-violation is due to the phase factors in the quark and lepton mixing matrices. In the quark sector this phase is very small $\delta_{13}=1.2\pm 0.1$ rad. However, strong interactions due to the axial anomaly produce a new effective interaction $\frac{\alpha_s}{8\pi} G \tilde G \theta_{QCD}$ which has a topological nature and changes the CP-violating phase $\theta=\theta_{QCD}+N_f \delta$. 
$${\cal L}_{SM} \in -\bar q_L\left( \begin{array}{ccc}m_u e^{i\delta/2} & 0& ...\\
0 &m_d e^{i\delta/2} & ...\\
0 & 0& ...\end{array} \right) \left(\begin{array}{c}u \\d\\ ...\end{array}\right)_R-\frac{\alpha_s}{8\pi}G\tilde G \theta_{QCD}$$

The presence of this phase leads to the appearance of  the neutron dipole moment $d_n=-4\times 10^{-3}\times \theta$ [e fm]. At the same time, the experimental bound on the neutral dipole moment is very strict: $|d_n|<3\times 10^{-13}$ [e fm] that gives $\theta <10^{-10}$. Such a small number requires some explanation. And it was found transforming the angle $\theta$ into the dynamic field $a(x)=\theta(x)f_a$ whose vacuum mean value defines the CP-violating phase. This field interacts with gluons 
\begin{equation}
{\cal L}=\frac 12(\partial_\mu a)(\partial^\mu a)-\frac{\alpha_s}{8\pi f_a^2}G_{\mu\nu}^a\tilde G_a^{\mu\nu}a
\end{equation}
and develops a dynamical potential (see Fig.\ref{axion}).
 \begin{figure}[ht!]
\begin{center}
\leavevmode
\includegraphics[width=0.7\textwidth]{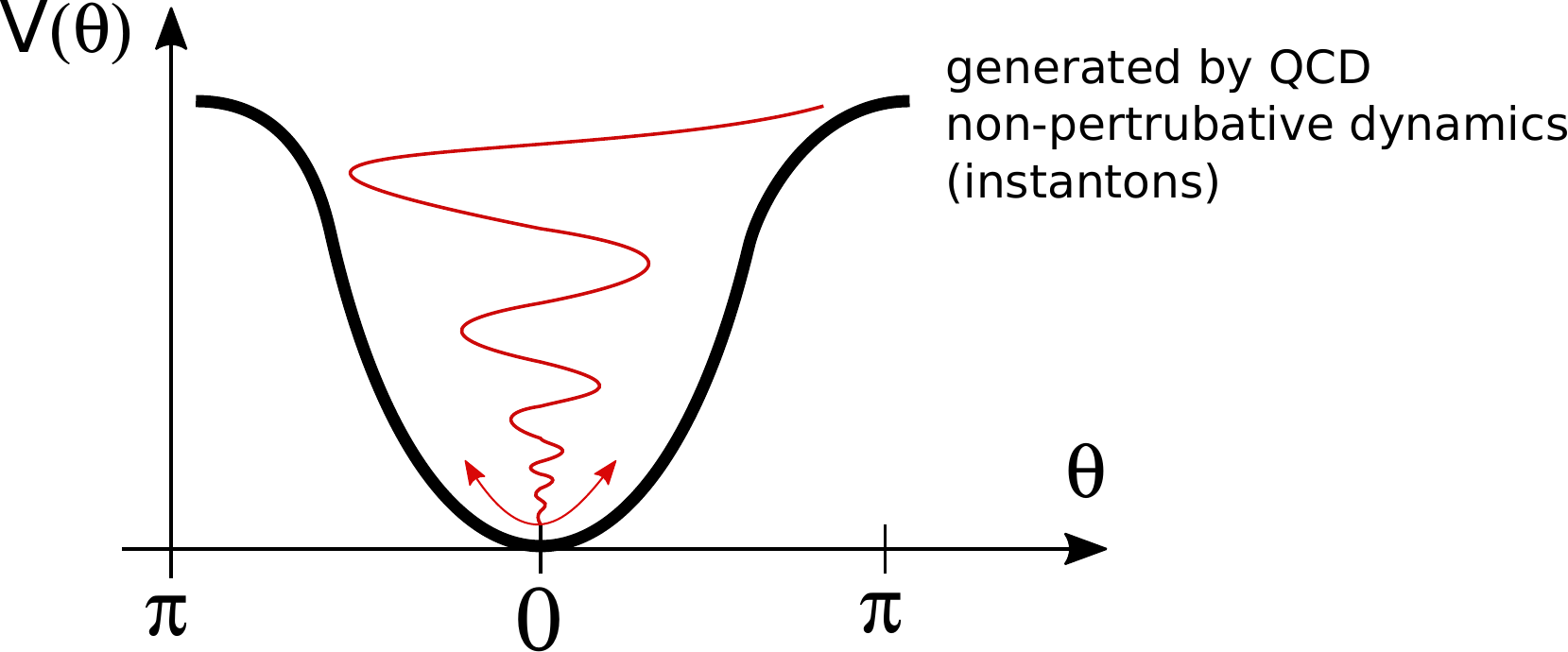}
\end{center}
\caption{The axion potential generated by QCD non-perturbative dynamics}
\label{axion}
\end{figure}
In the minimum of the potential it equals zero and then acquires a small value generated by non-perturbative dynamics. The axial symmetry related to this field is broken spontaneously, which leads to the appearance of a goldstone boson that later obtains a mass. This particle got the name of axion and the mechanism of dynamic suppression of  $\theta$ was called the Peccei-Quinn mechanism~\cite{PQ}.

The axion is characterized by two free parameters, its mass $m_a$ and the interaction with gluons
$1/f_a$. The search for axions has not given a result so far. The allowed regions in parameter space are shown in Fig.\ref{axion}~\cite{PDG_axion}. One can see that the allowed masses are extremely small and the scale of interaction  $f_a$ is very high.
 \begin{figure}[ht!]
\begin{center}
\leavevmode
\includegraphics[width=\textwidth]{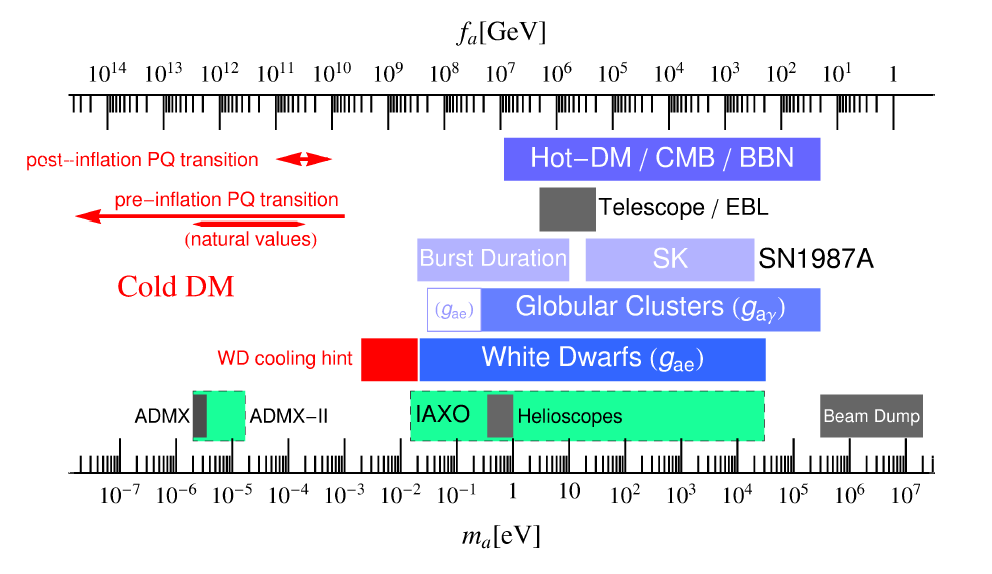}
\end{center}
\caption{The allowed regions for the mass and the coupling of the axion}
\label{axion}
\end{figure}

Later it became clear that coherent oscillations of the axion field ( remind that axion is a boson)
may produce condensate that can be the form of Dark Matter. Despite the small mass of the axion, the axion Dark matter might be cold since it is not in the state of thermal equilibrium. Therefore, if the axion exists, some amount of Dark Matter of the axion type is inevitable. 

\subsection{Neutrinos}
We know now  3 generations of matter particles. At the moment, there is no theoretical answer to the question of this fact.  We have only the experimental data that can be interpreted  as an indication of the existence of three generations. They assume the presence of the quark-lepton symmetry since refer to the number of light neutrinos and, due to this symmetry, to the number of generations. 

\begin{figure}[ht]
\begin{center}
\leavevmode
\includegraphics[width=0.35\textwidth]{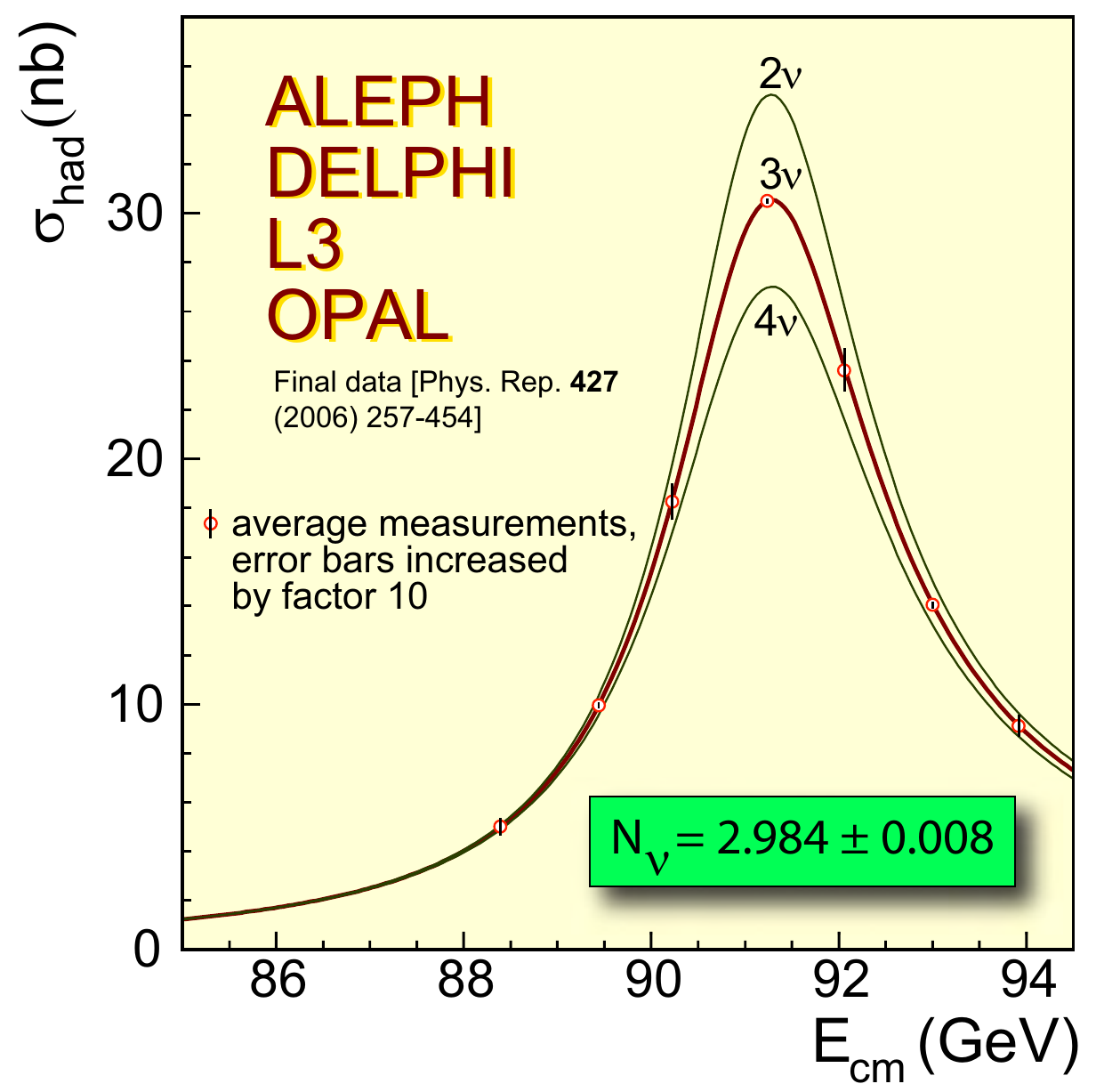}\hspace{1cm}
\includegraphics[width=0.40\textwidth]{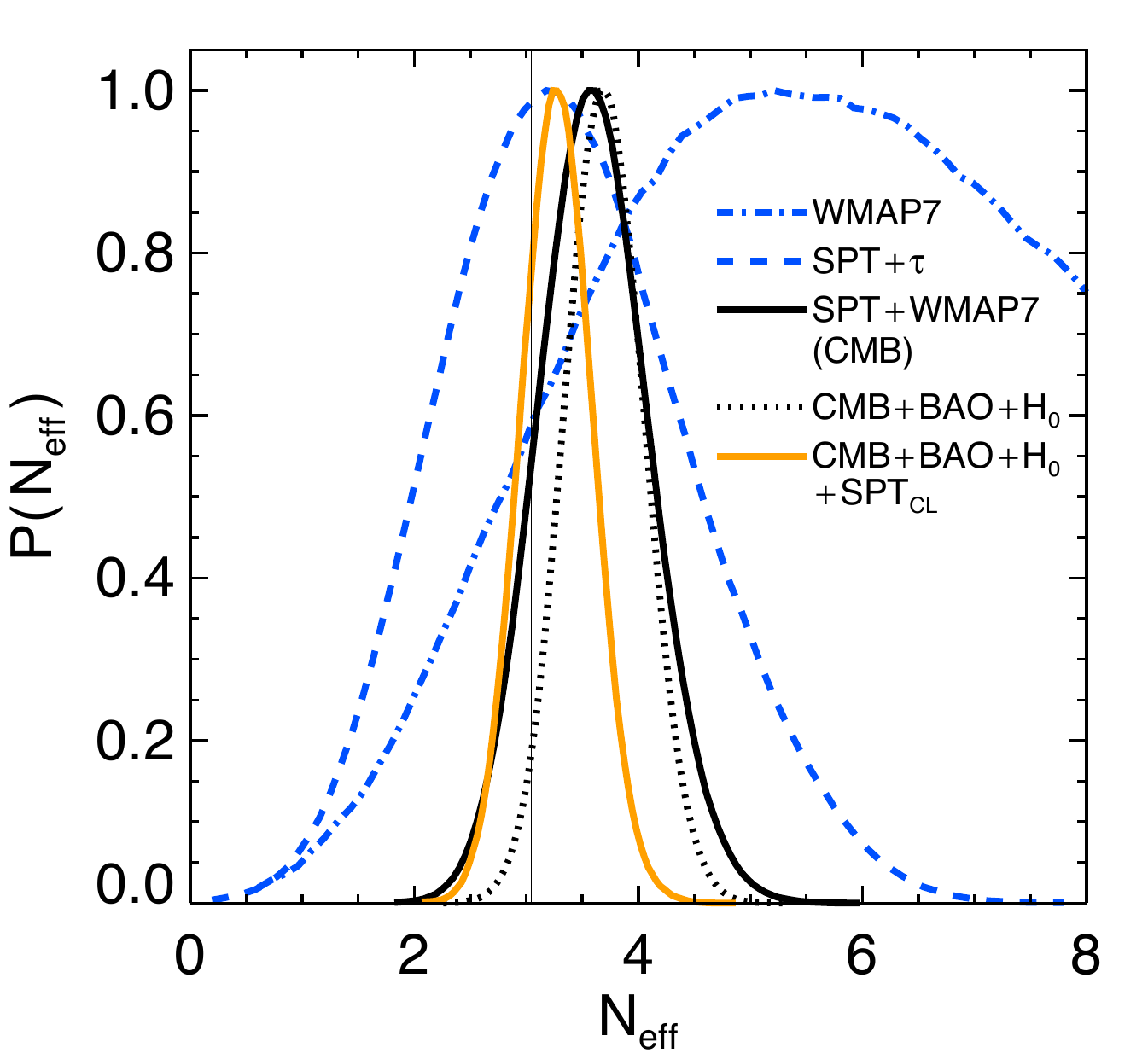}
\end{center}\caption{Experimentally measured profile of  the Z-boson, and the number of light  neutrinos (left) and the fit of the number of light neutrinos   from the temperature fluctuations of CMB (right)}
\label{width}
\end{figure}

The first fact is the measurement at the electron-positron collider LEP of the profile and width of  the Z-boson. The Z-boson can decay into quarks, leptons and neutrinos with the total mass less than  its own mass and measuring the width of the Z-boson, one can find out the number of light neutrinos.  This is not true for neutrinos with the mass bigger than 45 GeV.  The fit to the data corresponds to the number of neutrinos equal to $N_\nu=2.984\pm0.008$, i.e. 3 (see  Fig.\ref{width} left)~\cite{Z}.

The same conclusion follows from the fit of the spectrum of thermal fluctuations of the cosmic microwave background (CMB). The number of light neutrinos as well as the spectra of their masses are reliably defined from the CMB shape  (see Fig.\ref{width} right). The obtained number is:  $N_\nu= < 3.30\pm0.27$~\cite{CMBNeutrino2}, i.e. is also  consistent with 3 but still leaves some space for an additional sterile neutrino. 

\begin{figure}[ht]
\begin{center}
\leavevmode
\includegraphics[width=0.35\textwidth]{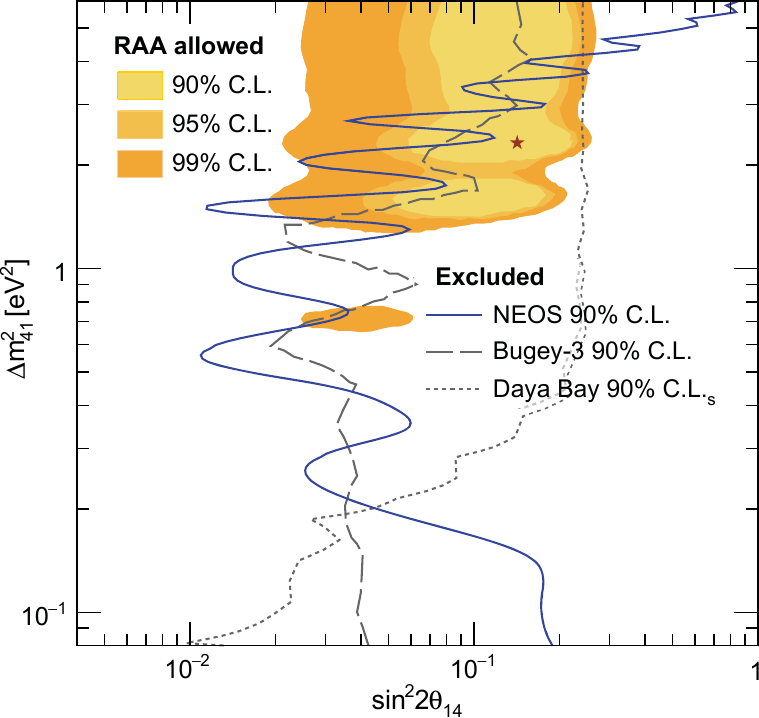}
\end{center}\caption{Constraints on the mass and mixing papameters of  a sterile neutrino}
\label{sterile}
\end{figure}

The search for  a sterile neutrino is on the way. Its existence may eliminate some tension in neutrino oscillation data coming from the LSND and MiniBoone experiments due to an admixture of the fourth component, which gives additional contribution to neutrino transformation probabilities
$$\begin{array}{l}
P_{\nu_e\to\nu_e}\approx 1-2|U_{e4}|^2(1-|U_{e4}|^2)\\
P_{\nu_\mu\to\nu_\mu}\approx 1-2|U_{\mu4}|^2(1-|U_{\mu4}|^2)\\
P_{\nu_\mu\to\nu_e}\approx 2|U_{e4}|^2 |U_{e4}|^2
\end{array}$$
for $4\pi E/\Delta m^2_{41}<<L<<4\pi E/\Delta m^2_{31}$.
Nevertheless, a recent direct search for a sterile neutrino gave negative results and imposed constraints on the mass and the mixing of the fourth neutrino (see Fig.\ref{sterile}~\cite{sterile}).

At last, there are complimentary data on precision measurements of the probabilities of rare decays where hypothetical additional heavy quark generations might contribute. According to these data, the fourth generation is excluded at the 90\% confidence level~\cite{4gen}.

A natural question arises:  Why does dNature need 3 copies of quarks and leptons?  All what we see around us is made of protons, neutrons and electrons, i.e. of  $u$ and  $d$ quarks and electrons - particles of the first generation. The particles made of the quarks of the next two generations and heavy leptons, copies of the electron, quickly decay and are observed only in cosmic rays or accelerators. Why do we need them? 

Possibly, the answer to this question is concealed not in the SM but in the properties of the Universe.  The point is that for the existence of baryon asymmetry of the Universe, which is the necessary condition for the existence of a stable matter, one needs the CP-violation~\cite{Sakharov}. This requirement in its turn is achieved in the SM due to the nonzero phase  in the mixing matrices of quarks and leptons.The nonzero phase appears only when the number of generations  $N_g \geq 3$.

With the discovery of neutrino oscillations neutrino physics has entered the new phase: the mass differences of different neutrino  types and the mixing angles were measured.  At last, the answer to the question of neutrino mass was obtained. Now we know that neutrinos are massive. This way, the lepton sector of the SM took the form identical to the  quark one and it was confirmed that the SM possesses the quark-lepton symmetry.  Nevertheless, the reason for such symmetry remains unclear, it might well be that it is a consequence of the Grand unification of interactions. However, the answer to this question lies beyond the SM. 

\begin{figure}[ht]
\begin{center}
\leavevmode
\includegraphics[width=0.8\textwidth]{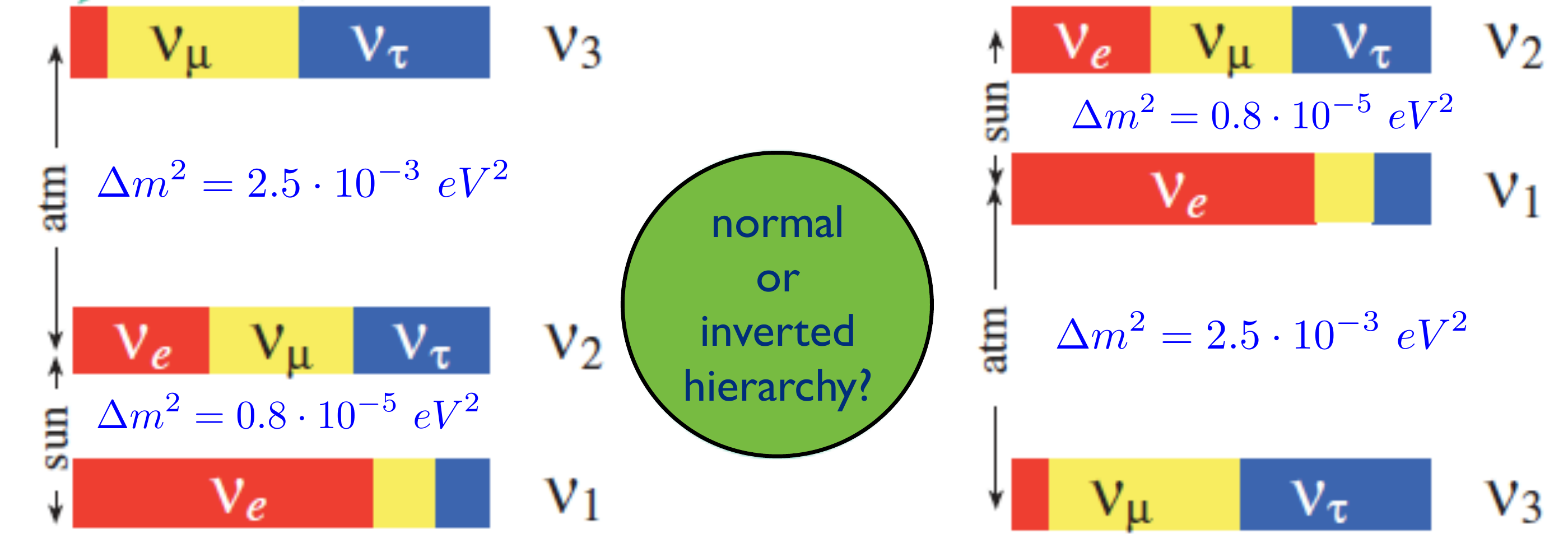}
\end{center}
\caption{Normal and inverse hierarchy of neutrino masses}
\label{neutrino}
\end{figure}

At the same time, the neutrino sector of the SM is still not fully understood.  First of all, this concerns the mass spectrum.  Neutrino oscillations allow one to determine only the squares of the mass difference for various neutrinos. The obtained picture is shown in Fig.\ref{neutrino}~\cite{neutrino_hierarchy}.  The color pattern shows the fraction of various types of neutrino in mass eigenstates. 

Besides the hierarchy problem (normal or inverted) there is also an unclear question of the absolute scale of neutrino masses.  One may hope to get an answer to this question in two ways. The first one is a direct measurement of the electron neutrino mass in the $\beta$-decay experiment. According to the Troitsk-Mainz experiment, the upper bound on the neutrino mass today is $m_{\nu_e}< 2$ eV ~\cite{TroitskMainz}. The upcoming experiment KATRIN~\cite{KATRIN} will be able to move this bound up to
$<0.2$ eV. However, this might not be enough if one believes in astrophysical data. The determination of the sum of neutrino masses from the spectrum of the cosmic microwave background  is an indirect but rather an accurate way to find the absolute mass scale. At the early stage of the Universe during the fast cooling
process particles  fell out of the thermodynamic  equilibrium  at the temperature proportional to their masses and their abundance ``froze down" influencing the spectrum. Hence, fitting the spectrum of the CMB fluctuations one can determine the number of neutrino species and the sum of their masses.  The result of the latest space mission PLANK~\cite{Plank} looks like $\sum m_\nu <0.23$ eV. This number is still much bigger than the neutrino mass difference shown in Fig.\ref{neutrino}.
Thus, the absolute scale of neutrino masses is still an open question.

Another unsolved problem of the neutrino sector is the nature of neutrino: Is it a Majorana particle or a Dirac one, is it an antiparticle  to itself or not?  Remind that  particles with spin 1/2  are described by the Dirac equation, the solutions being the bispinors. They can be divided into two parts corresponding to the left or right polarization
\begin{equation}
\nu_D=\left(\begin{array}{c} \nu_L \\ 0 \end{array}\right)+\left(\begin{array}{c} 0 \\ \nu_R \end{array}\right), \ \ \ \nu_L \neq \nu_R^*, \ \ \  m_L=m_R.
\end{equation}
Both parts have the same mass since this is just one particle with two polarization states. At the same time, in the case of a neutral particle the Dirac  bispinor can be split into two real parts  
\begin{equation}
\nu_D=\left(\begin{array}{c} \xi _1\\ \xi_1^* \end{array}\right)+ \left(\begin{array}{c} \xi_2 \\ \xi_2^* \end{array}\right), \ \ \  m_{\xi_1}\neq m_{\xi_2}.
\end{equation}
each of these parts is a Majorana spinor obeying the condition  $\nu_M=\nu_M^*$,
i.e. if the neutrino is a Majorana spinor, then it is an antiparticle to itself. These two Majorana spinors can have different masses. Hence, if this possibility is realized in Nature, we have just discovered the light neutrino and the heavy ones can have much bigger masses.

An argument in favour of the Majorana neutrino  is  the smallness of their masses. If one gets them through the usual Brout-Englert-Higgs mechanism, the corresponding Yukawa couplings are extremely small of an order of $10^{-12}$. In the case of the Majorana neutrino one can avoid it using the see-saw mechanism~\cite{seesaw}:  The small masses of light neutrinos appear due to the heaviness of the Majorana mass 
\begin{eqnarray}
&&\hspace{1.3cm} L \ \ \ \ R \nonumber\\
M_\nu&=&\begin{array}{c} L \\ R \end{array}\left( \begin{array}{cc} 0 & m_D \\ m_D & M \end{array}\right), \ \ \  m_1=\frac{m_D^2}{M}, \ m_2= M.
\end{eqnarray}
Thus, the neutrino Yukawa coupling  may have the usual lepton value and the Majorana mass  $M$ might be of the order of the Grand Unification scale. In this case, one also has the maximal mixing in the neutrino sector.

\begin{figure}[ht]
\begin{center}
\leavevmode
\centering
\includegraphics[width=0.27\textwidth]{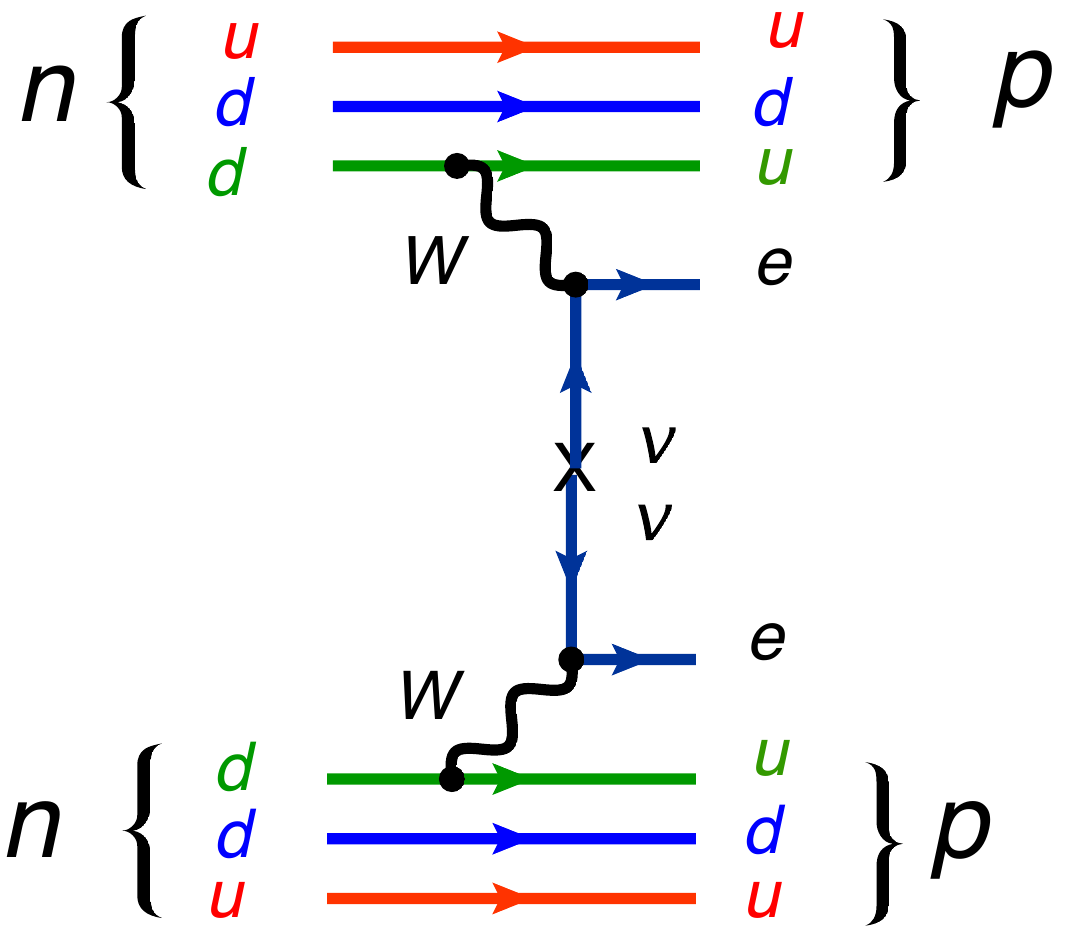}
\includegraphics[width=0.35\textwidth]{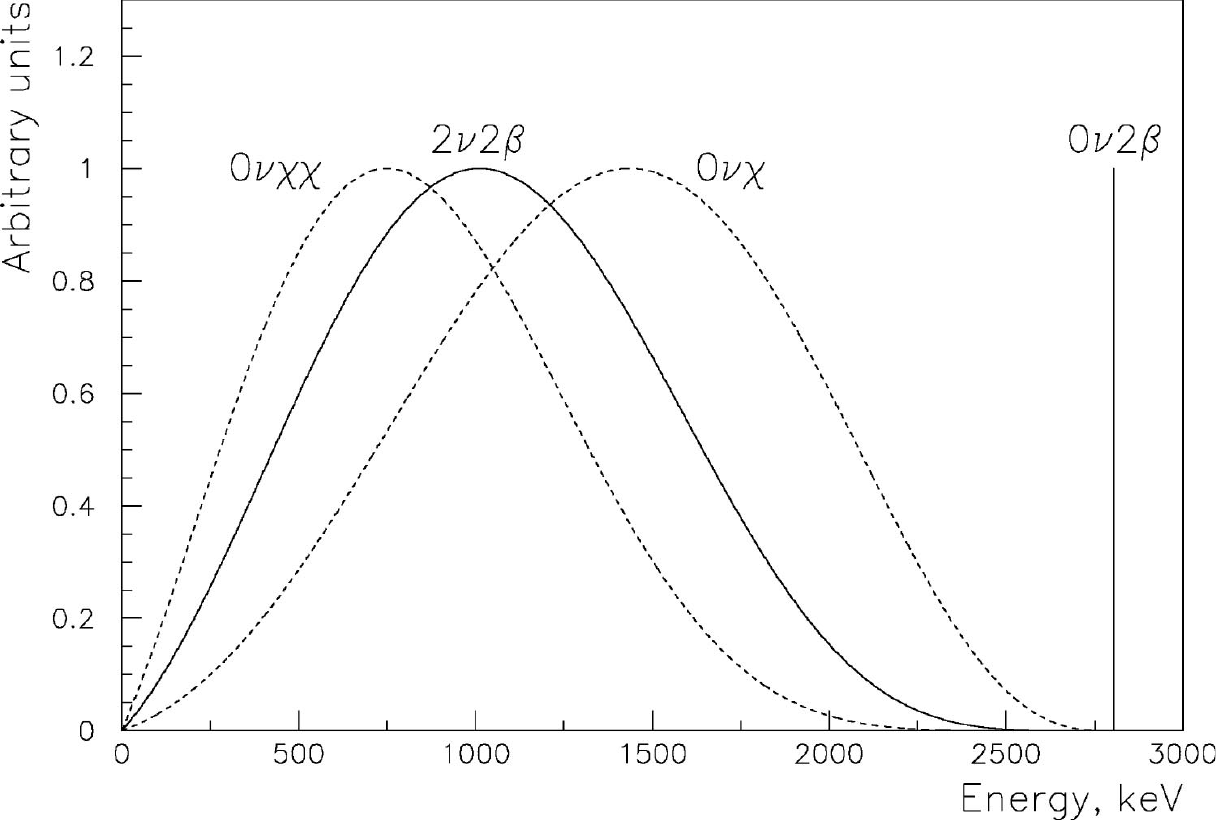}
\includegraphics[clip, trim = 0 0 2.2cm 2.3cm, width=0.36\textwidth]{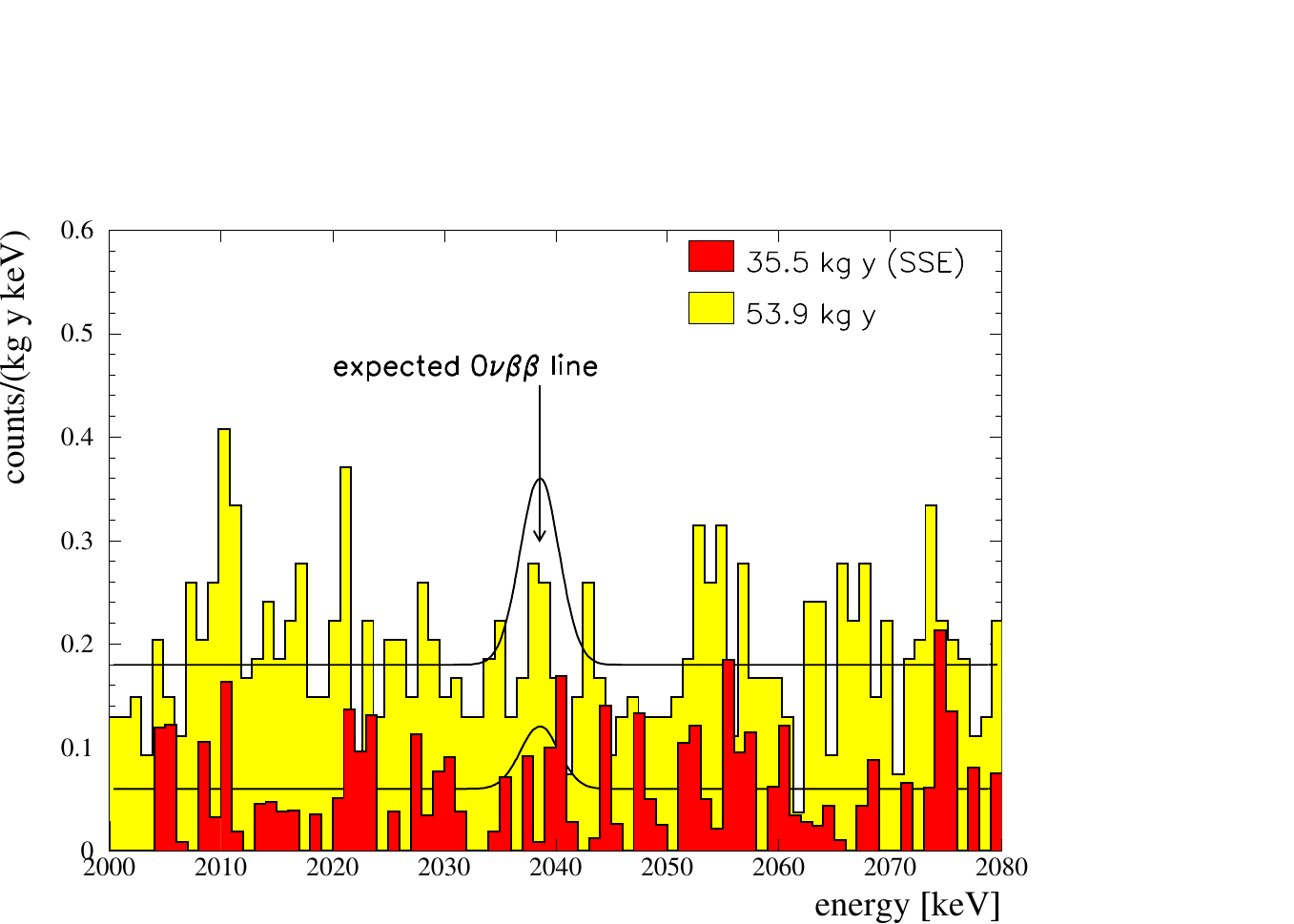}
\end{center}\caption{Neutrinoless double $\beta$- decay (left) and the energy spectrum of electrons in the case of a usual and neutrinoless decay of the isotope $^{76}Ge$ (center).  The experimentally measured spectrum of electrons is shown on the right~\cite{02bb_exp}}
\label{02bb}
\end{figure}

One can find out the nature of the neutrino  studying the double $\beta$-decay. If the neutrinoless double $\beta$-decay is possible, then the neutrino is a Majorana since for the Dirac neutrino it is forbidden. The corresponding Feynman diagram is shown in Fig.\ref{02bb}. It also shows the energy spectrum of electrons in the case of the usual and neutrinoless  $\beta$-decay~\cite{02bb}. 
As one can see, two types of spectrum are easily distinguishable. However, practical observation is rather 
cumbersome.  The histogram shown in Fig.\ref{02bb} (right) is the experimentally measured electron spectrum of the double $\beta$-decay. The solid line shows the expected position of the maximum in the spectrum of two electrons corresponding to the double neutrinoless $\beta$-decay.

\begin{figure}[ht]
\begin{center}
\includegraphics[width=0.45\textwidth]{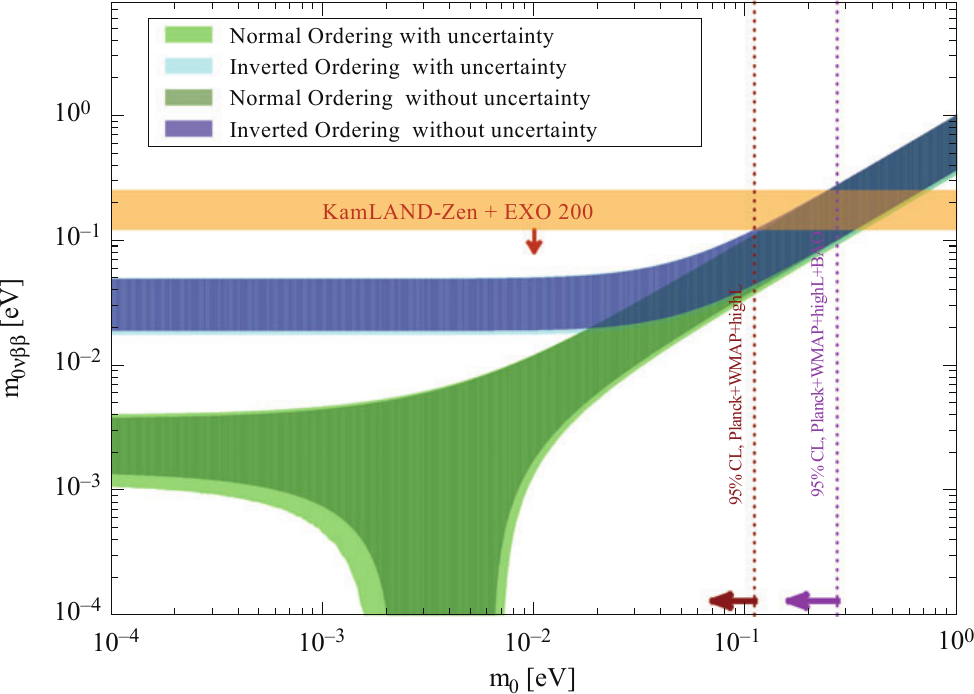}
\end{center}\caption{Effective neutrino mass in neutrinoless double beta decay. Green and pink areas correspond to the inverse and normal hierarchy of neutrino masses, respectively}
\label{effectivemass}
\end{figure}

As a result, today there are no clear indications of the existence of the double neutrinoless  $\beta$-decay.  The experiments are carried out on the isotopes  $ ^{48}Ca, ^{76}Ge, ^{82}Se$, $^{130}Te, ^{136}Xe, ^{150}Nd.$  Modern estimates of the lifetime are~\cite{halflife}
\begin{eqnarray*}
T_{1/2}2\nu\beta\beta (^{136}Xe) &\times& 10^{21} \ yr = 2.23\pm 0.017\  stat \pm 0.22 \ sys ,  \\
T_{1/2}0\nu\beta\beta (^{136}Xe) &\times& 10^{25} \ yr > 1.6 \ (90\% \ CL).
\end{eqnarray*} 
It is an interesting question whether it will be possible to find the neutrinoless double beta decay
increasing the accuracy of the observation in principle since the effective coupling might be very small. It so happens that the answer to this question depends on the hierarchy of neutrino masses:
for the inverse hierarchy the situation is optimistic and there is a lower limit on effective mass while for the normal hierarchy the lower limit is absent and the effective mass can be unlimitedly small. The situation is illustrated in Fig.\ref{effectivemass}~\cite{effectivemass}.
Thus, the nature of the neutrino remains an open problem of the SM.

\subsection{Dark Matter}
\begin{figure}[ht]
\begin{center}
\leavevmode
\includegraphics[width=0.56\textwidth, height=6.0cm]{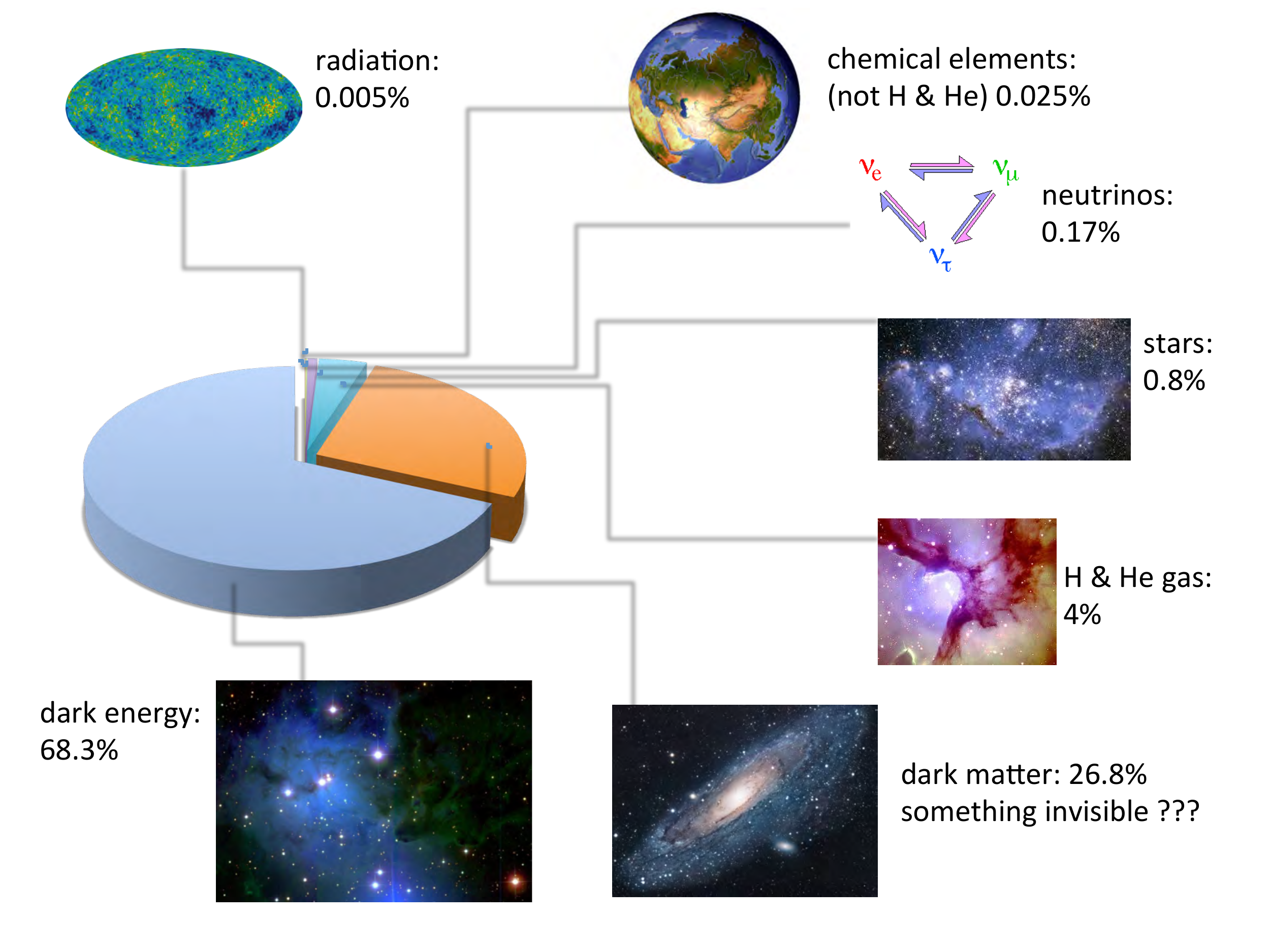}\hspace{0.6cm}
\includegraphics[width=0.38\textwidth, height=4.6cm]{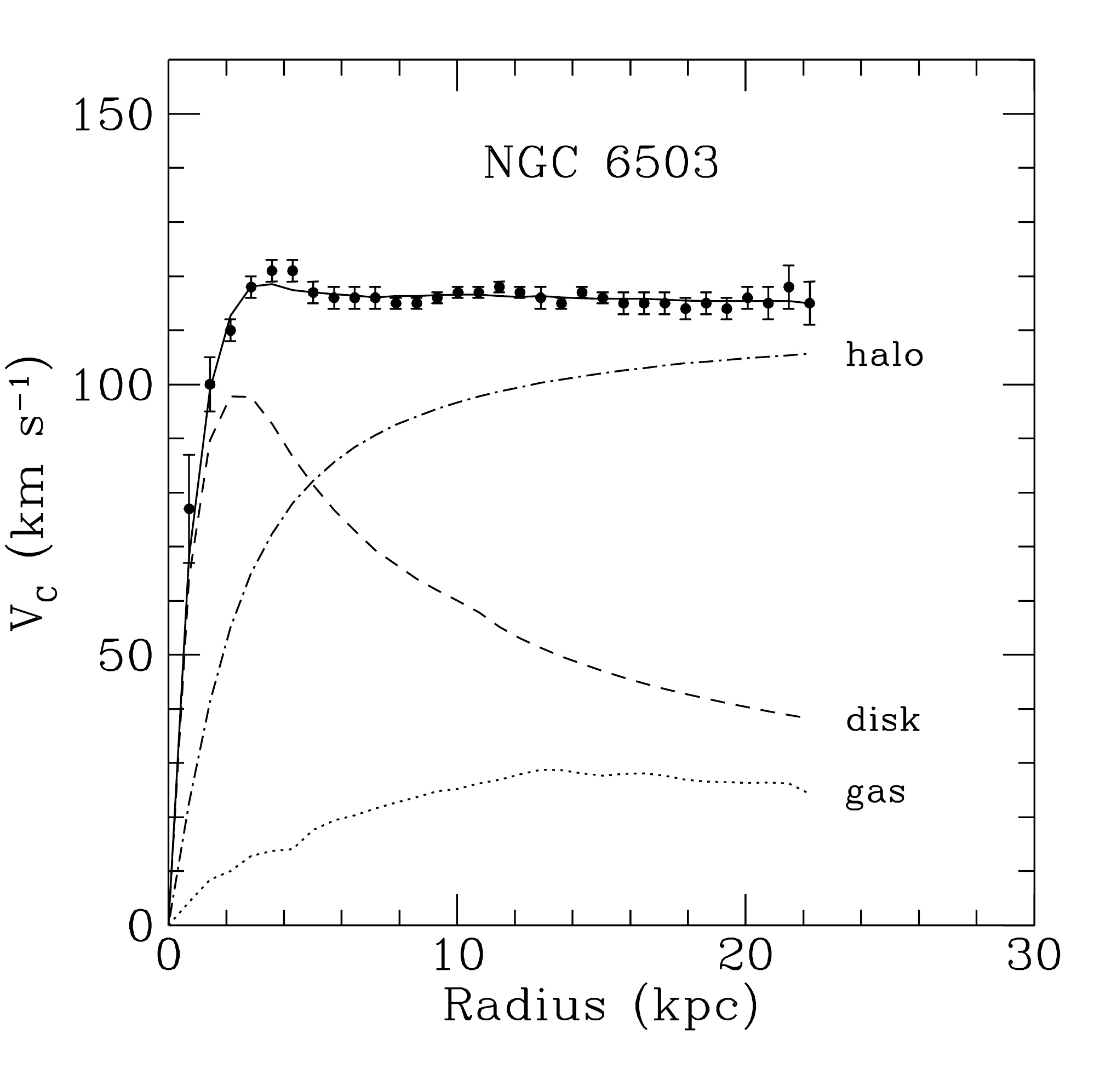}
\end{center}\caption{The energy balance of the Universe (left)~\cite{balance} and rotation curves of stars in the spiral galaxy (right)~\cite{rotation}}
\label{balance}
\end{figure}

The existence of Dark Matter is known since the 30s of the last century. However, the situation has changed when the energy balance of the Universe was obtained and became clear that there is 6 times as much of  Dark Matter  than ordinary matter (see Fig.\ref{balance}, left)~\cite{WMAP}. The existence of  Dark Matter, which is known so far due to its gravitational influence, is supported by the rotational curves of the stars, galaxies and clusters of galaxies (see Fig.\ref{balance} right), the gravitational lenses, and the large scale structure of the Universe~\cite{Kolb}. 
Therefore, the question appears: What is the dark matter made of, can it be some non-shining macro objects like the extinct stars, molecular clouds, etc., or these are micro particles? In the last case Dark Matter becomes the object of particle physics. 

According to the last astronomical data, at least in our galaxy, there is no evidence of the existence of macro objects, the so called MACHOs. At the same time, Dark Matter is required for a correct description of the star rotation. Therefore,  the hypothesis of the microscopic nature of the Dark matter is the dominant one. 
In this case, in order to form the large scale structure of the Universe, Dark Matter has to be cold, i.e. nonrelativistic; hence, DM particles have to be heavy. According to the estimates, their mass has to be above a few dozens of keV~\cite{coldDM}. Besides, DM particles have to be stable or long-lived to survive since the Big Bang. Thus, one needs a neutral, stable and relatively heavy particle.

If one looks at the SM, the only stable neutral particle is the neutrino. However, if the neutrino is the Dirac particle, its mass is too small to form Dark Matter. Therefore, within the SM the only possibility to describe Dark Matter is the existence of heavy Majorana neutrinos. Otherwise, one needs to assume some new physics beyond the SM. The possible candidates are: 
neutralino, sneutrino and gravitino in the case of supersymmetric extension of the SM~\cite{abun},  and also a new heavy neutrino~\cite{Canetti}, a heavy photon, a sterile Higgs boson, etc.~\cite{Feng}. An alternative way to form Dark Matter is the axion field, the hypothetical light strongly interacting particle~\cite{axionDM}. In this case, Dark Matter differs by its properties. 

The dominant hypothesis is that Dark Matter is made of weakly interacting massive particles - WIMPs. 
This hypothesis is supported by the following fact: the concentration of Dark Matter after the moment when a particle fell down from the thermal equilibrium is given by the Boltzmann equation~\cite{abun}
\begin{equation}
\frac{dn_\chi}{dt}+ 3 H n_\chi = - < \sigma v > ( n^2_\chi- n^2_{\chi,eq}),	
\end{equation}
where $H = \dot{R}/ R$  is the Hubble constant, $n_{\chi,eq}$ is the concentration in the equilibrium, and $\sigma$  is the Dark matter annihilation cross-section.The relic density is expressed through the concentration   $n_\chi$  in the following way:
\begin{equation}
\Omega_\chi h^2 =\frac{m_\chi n_\chi}{\rho_c}\approx \frac{2\cdot 10^{27}\ cm^3\ sec^{-1}}{<\sigma v>}.
\end{equation}
Having in mind that $\Omega_\chi h^2 \approx 0.113\pm 0.009$ and $v\sim 300$ km/sec,  one gets for the cross-section
\begin{equation}
\sigma\approx 10^{-34}\ cm^2 = 100\ pb,
\end{equation}
that is a typical cross-section for a weakly interacting particle with the mass of the order of the Z-boson mass.

These particles presumably  form an almost spherical galactic halo with the radius a few times bigger than the size of the shining matter. The DM particles cannot leave the halo being gravitationally bounded and cannot stop since they cannot drop down the energy emitting photons like the charged particles.  In the Milky Way, in the region of the Sun the density of Dark Matter should be   $\sim$ 0.3 GeV/sm$^3$ in order to get the observed rotation velocity of the Sun around the center of the galaxy $\sim$ 220 km/sec.

\begin{figure}[h]
\begin{center}
\leavevmode
\includegraphics[width=0.6\textwidth]{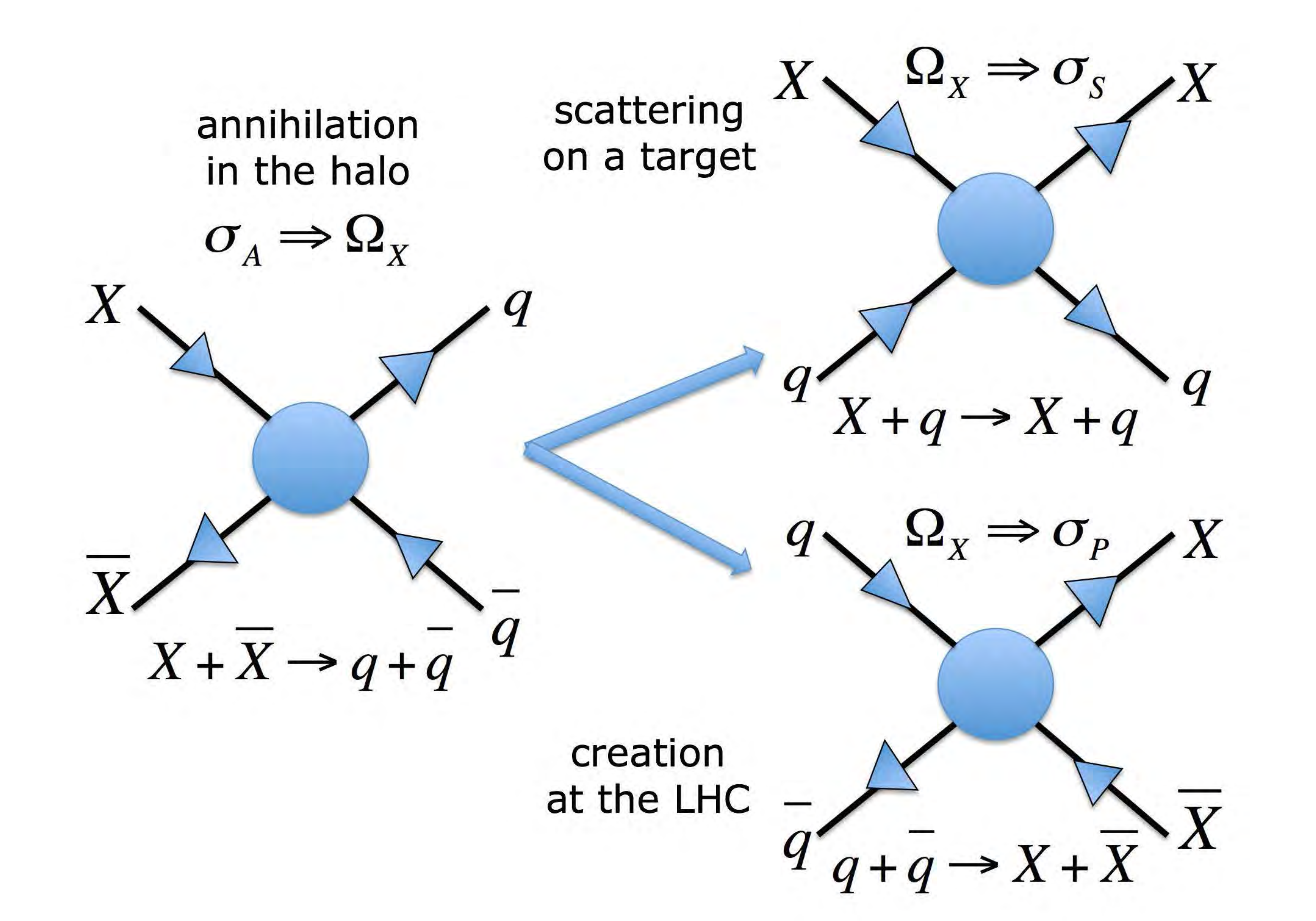}
\end{center}\vspace{-0.5cm}
\caption{The search for Dark Matter in three cross related channels}
\label{DMsearch}
\end{figure}

The search for Dark Matter particles is based on three reactions the cross-sections of which are related by the crossing symmetry (see Fig.\ref{DMsearch})~\cite{Kolb}.

This is, first of all, the annihilation of Dark Matter in the galactic halo that leads to the creation of ordinary particles and should appear as the ``knee" in the spectrum of the cosmic rays for diffused gamma rays, antiprotons and positrons.  Secondly, this is the scattering of DM on the target which should lead to a recoil  of the nucleus of the target when hit by a particle with the mass of the order of the Z-boson mass. And, third, this is a direct creation of DM particles at the LHC which, due to their neutrality, should manifest themselves in the form of missing energy and transverse momentum.

\begin{figure}[htb]
\begin{center}
\leavevmode
\includegraphics[width=0.35\textwidth]{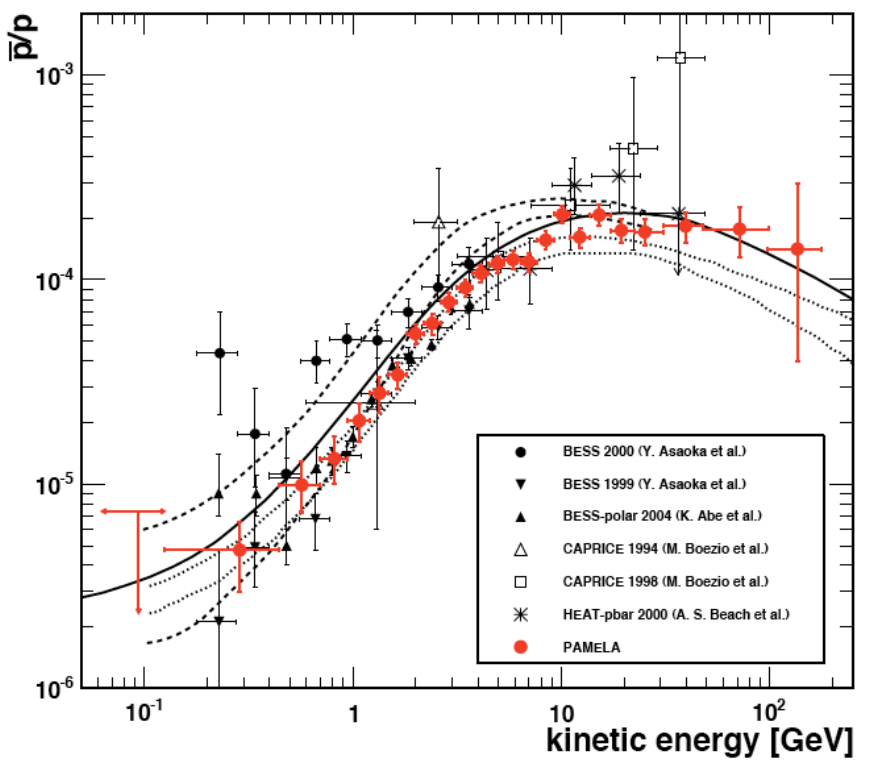}
\includegraphics[width=0.45\textwidth]{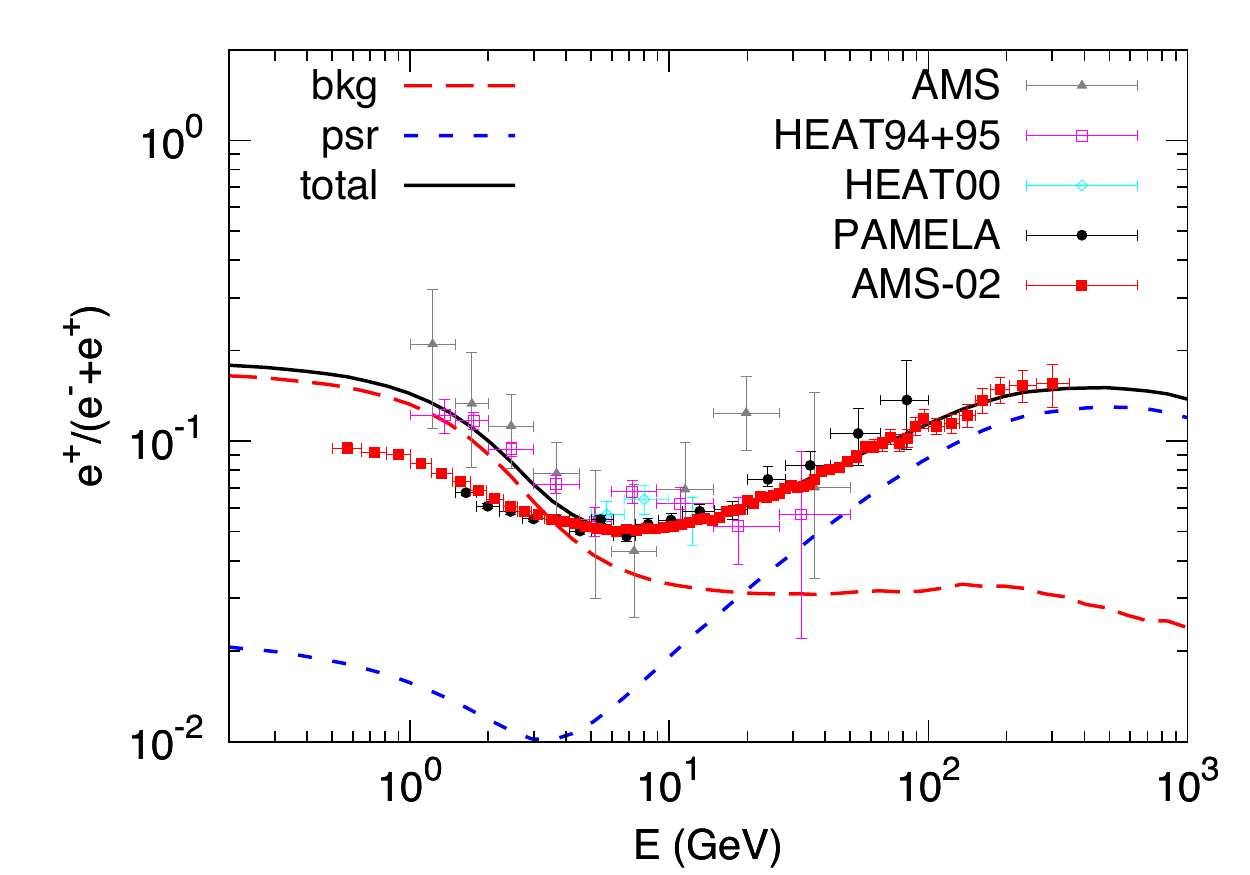}
\includegraphics[clip, trim = 5cm 20cm 5cm 3cm, width=0.50\textwidth]{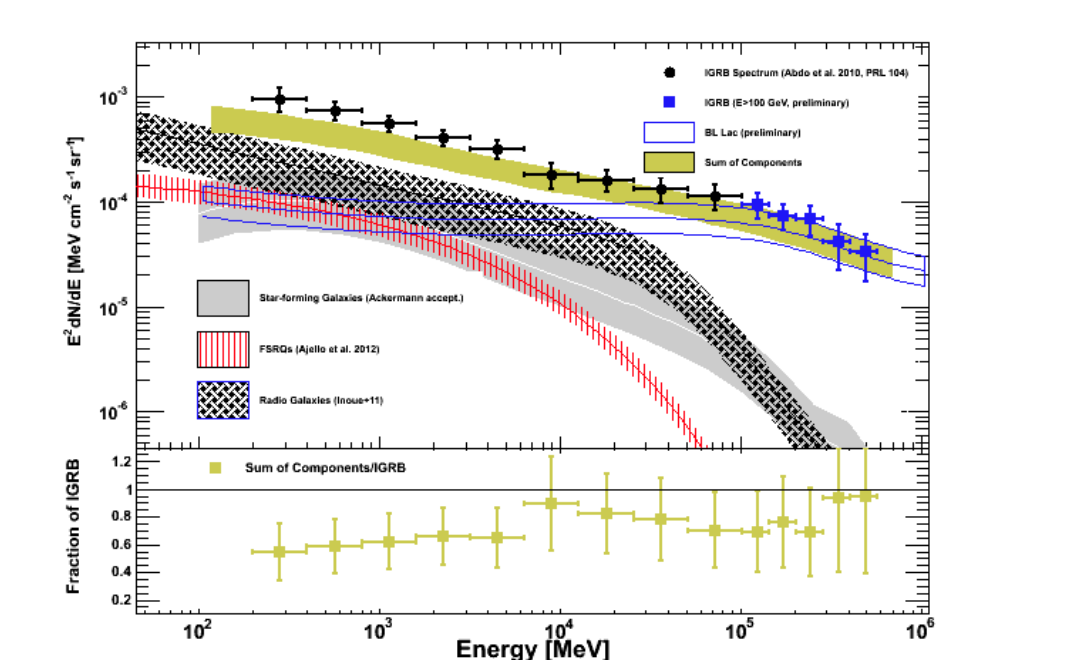}
\end{center}\caption{Indirect search for Dark Matter: antiproton~\cite{Antiproton}, positron ~\cite{positron}, and diffuse $\gamma$ ray~\cite{fermi-lat} data}
\label{indirect}
\end{figure}

\begin{figure}[htb]
\begin{center}\vspace{-1cm}
\leavevmode
\includegraphics[clip, trim = 10.5cm 21.5cm 2cm 3cm, width=0.44\textwidth]{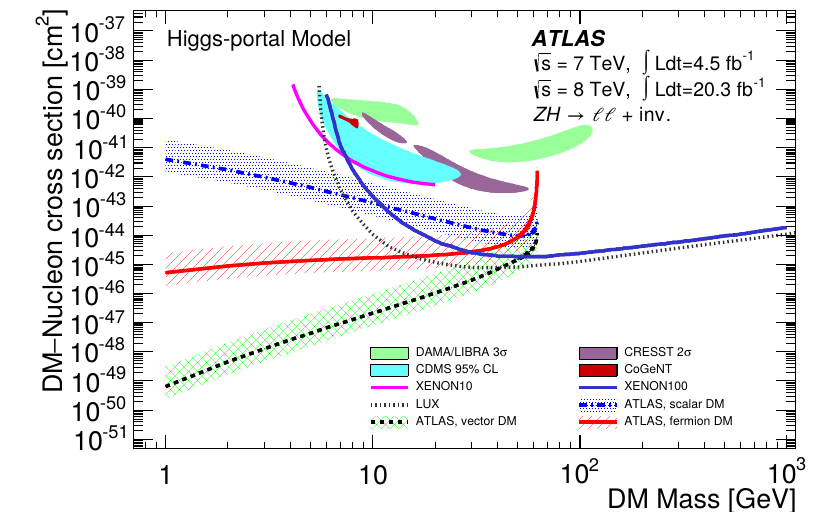}
\includegraphics[width=0.44\textwidth]{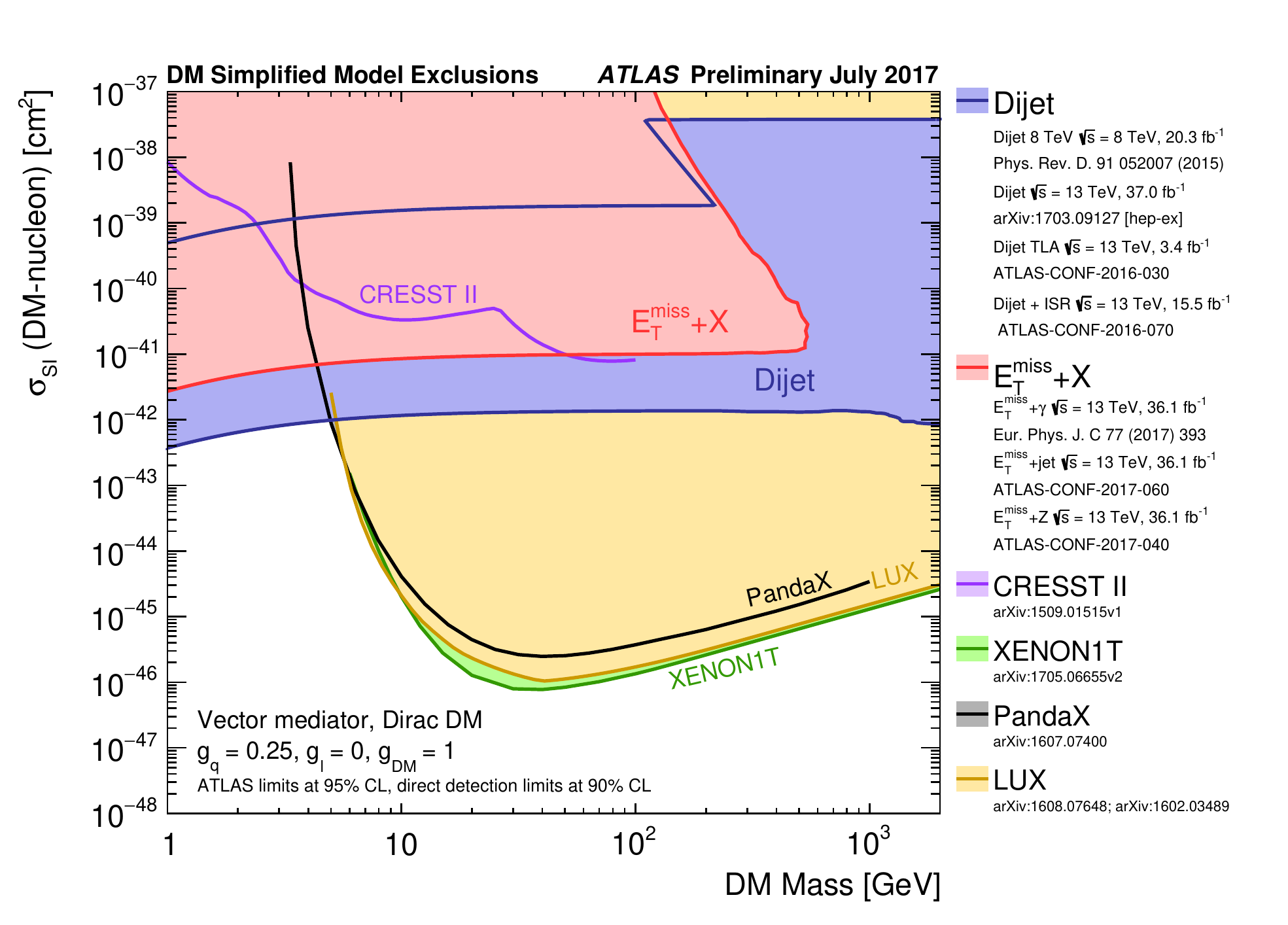}
\includegraphics[width=0.8\textwidth]{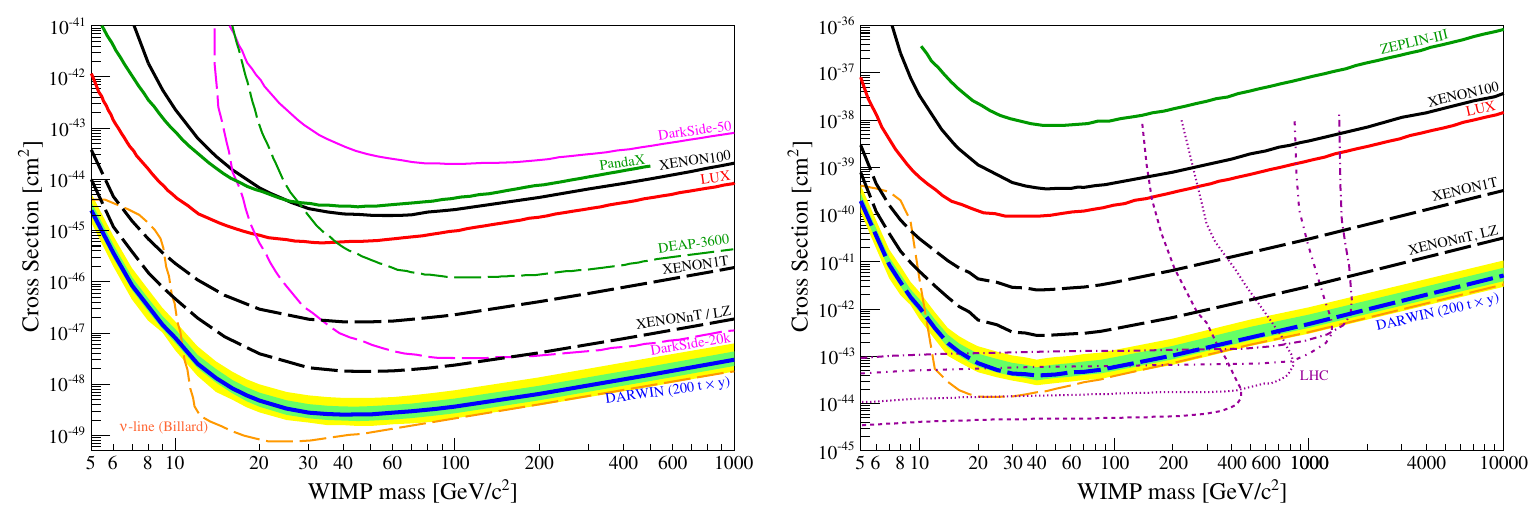}
\end{center}\vspace{-0.5cm}
\caption{Direct search for the Dark matter at accelerators~\cite{atlas_dm} and underground experiments~\cite{direct}}
\label{direct}
\end{figure}

In all these directions  there is an intensive search for a signal of the DM. The results of this search for all three cases are shown in Figs.\ref{indirect}, \ref{direct}.
As one can see from the cosmic ray data (Fig.\ref{indirect}), in the antiproton sector there is no any statistically significant excess above the background~\cite{Antiproton}. In the positron data there exists some confirmed increase; however,  its origin is usually connected not with the DM annihilation but with the new astronomical source~\cite{positron}. The spectrum of diffused gamma rays like antiprotons is consistent with the background within the uncertainties.

As for the direct detection of Dark Matter, there is  no any positive signal so far. The results of the search are presented in the plane  mass--cross-section. One can see from Fig.\ref{direct}~\cite{direct} that today the cross-sections up to $10^{-45}$ sm$^2$ are reached for the mass near 100 GeV.  In the near future it is planned to advance  two orders of magnitude.

The results of the DM search at the LHC are also shown in the plane mass--cross-section~\cite{LHCDM}. Here the signal of the DM creation is also absent. As it follows from the plot, the achieved bound of possible cross-sections at the LHC  is worse than in  the underground experiments for all mass regions except for the small masses $<10$ GeV where the accelerator is more efficient. Note, however, that the interpretation of the LHC data as the registration of DM particles is ambiguous and definite conclusions can be made only together with the data from the cosmic rays and direct detection of the scattering of  DM.

All available experimental data combined (LHC,LUX,Planck) are still consistent with even the simplest versions of SUSY (cNMSSM, NUHM).  The remaining parameter space is directly probed by direct WIMP searches with tonne scale detectors: DEAP-3600, XENON1T, LUX/LZ.  Complimentarity with the LHC (cMSSM, NUHM are mostly out of reach of the 14 TeV run!)

\begin{figure}[htb]
\begin{center}
\leavevmode
\includegraphics[width=0.42\textwidth,height=5.5cm]{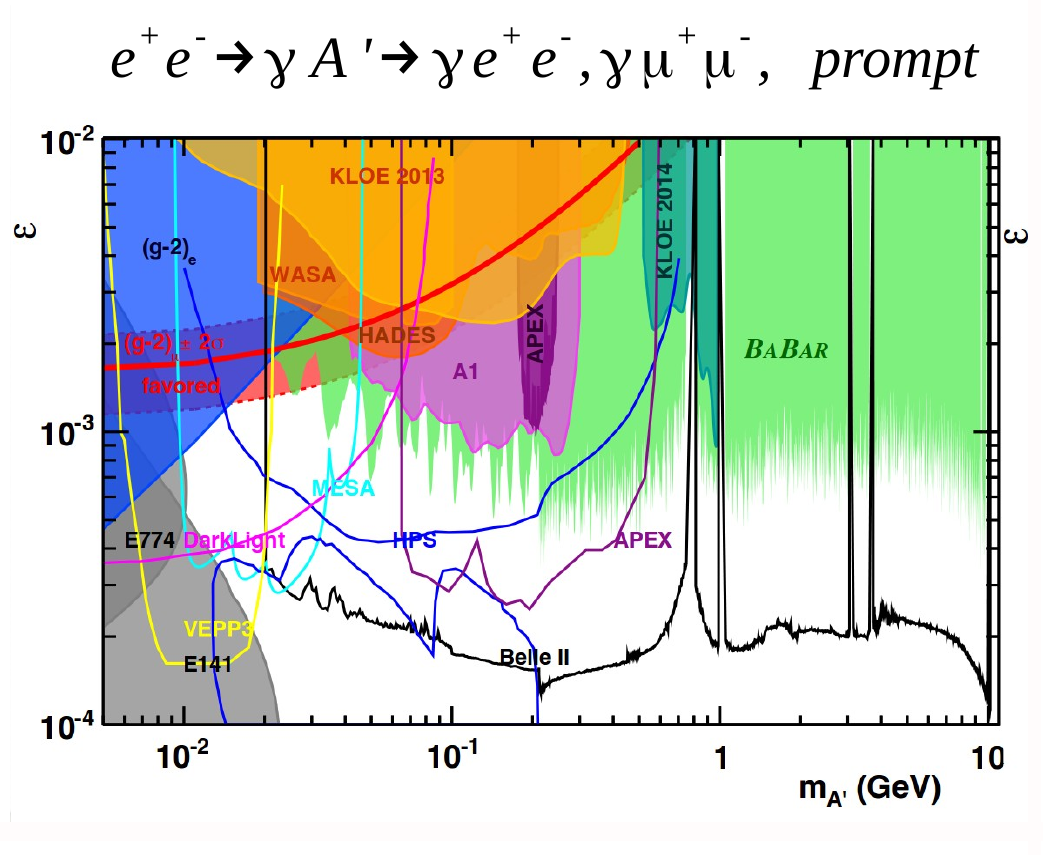}\hspace{1cm}
\includegraphics[width=0.38\textwidth, height=5.5cm]{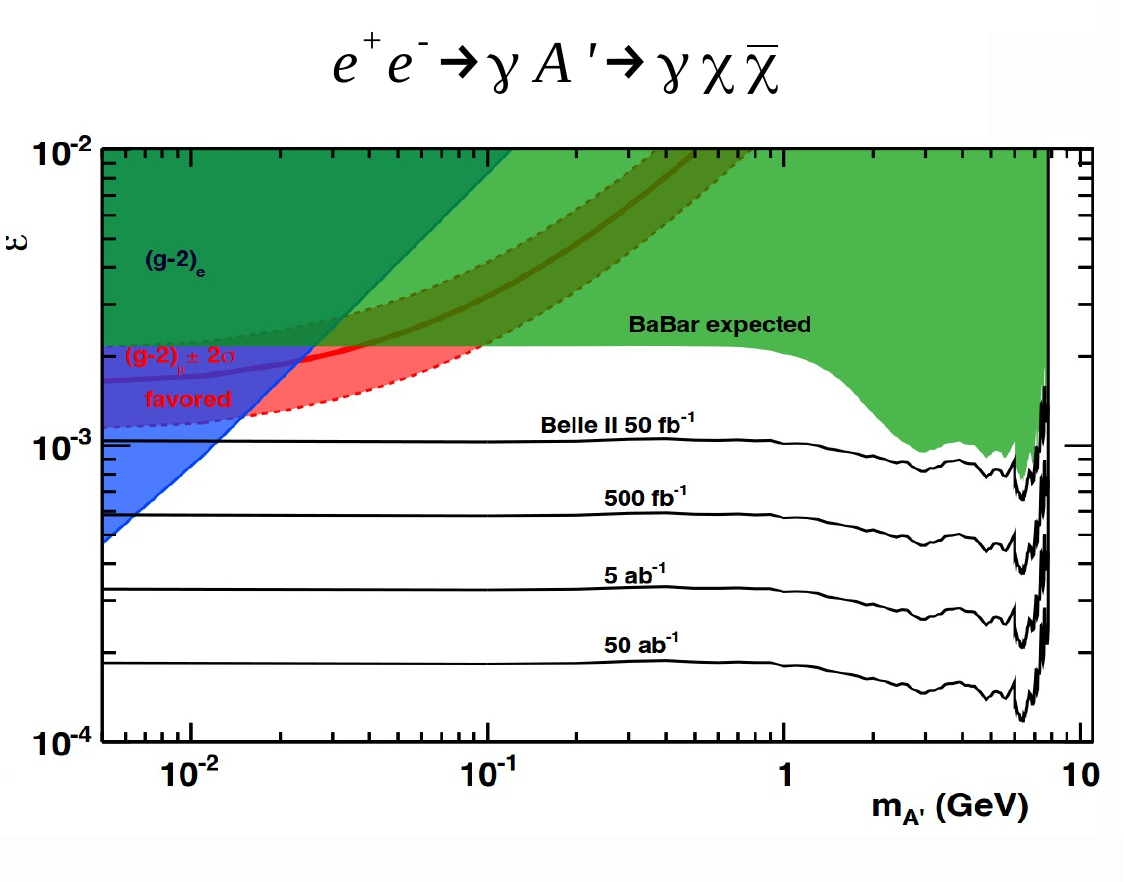}
\end{center}\caption{The search for Dark photon}
\label{dphoton}
\end{figure}

The other possibility mentioned already is the dark photon. In the process of annihilation one may produce the dark photon together with the ordinary one. It will decay later producing the  pair of charged particles which may be detected or invisible matter in the form of neutralino. The search for such decays is running and new dedicated experiments are in progress. The results are presented in the plane  of the dark photon mass versus the mixing with ordinary photon (see Fig.\ref{dphoton}~\cite{1705}).

\section{New Dimensions}

The paradoxical  idea of extra dimensions attracted  considerable interest in recent years
despite the absence of any experimental confirmation. This is 
mainly due to unusual possibilities and intriguing effects even in
classical physics  (For review see, e.g. Refs.\cite{ED} ), and  the
requirement from  the  string theory which  allows for consistent formulation  in the critical dimension
equal to 26 for the bosonic and 10 for the fermionic string~\cite{string}. This way the string theory stimulated the study of ED theories. 

\begin{figure}[htb]
\begin{center}
\leavevmode
\includegraphics[width=0.42\textwidth]{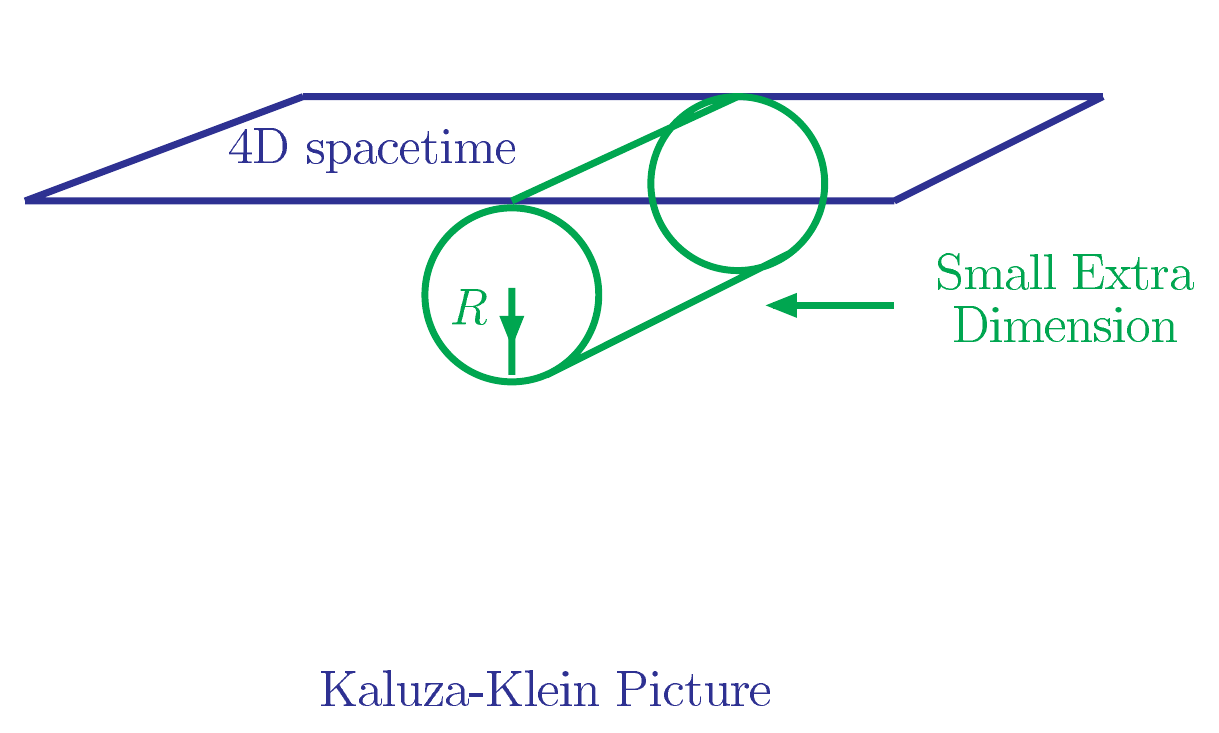}\hspace{1cm}
\includegraphics[width=0.38\textwidth]{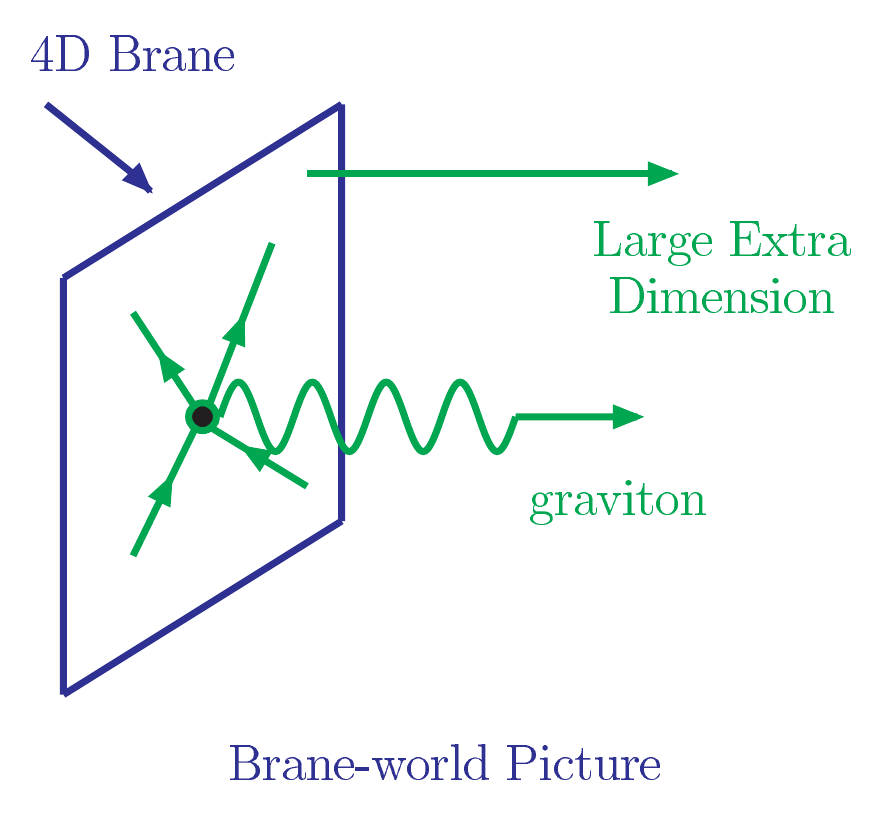}
\end{center}\caption{Compact (left) and large (right) extra space dimensions}
\label{large}
\end{figure}

The natural question arises: why don't we  see these extra space dimensions? There are two possibilities: compact ED of small radius  and localization of observables on a 4-dimensional hyper surface (brane) (see Fig.\ref{large}).

\subsection{Compact Extra Dimensions}
The idea of compact extra dimensions goes back to the
so-called Kaluza-Klein theories~\cite{KK}.
We do not see ED because their radius  is too
small for the present energies, say, equal to the Planck length,
$10^{-33}$ cm. 
The KK approach is based on the hypothesis that the space-time is
a (4+d)-dimensional pseudo Euclidean space~\cite{KK-review}
 $$E_{4+d} = M_{4} \times K_{d},$$
  where $M_{4}$ is the
four-dimensional space-time and $K_{d}$ is the $d$-dimensional compact
space of characteristic size (scale) $R$.  In accordance with the
direct product structure of the space-time, the metric is usually
chosen to be
\begin{equation}
ds^{2} = \hat{G}_{MN}(\hat{x}) d\hat{x}^{M} d\hat{x}^{N} = g_{\mu
\nu}(x) dx^{\mu} dx^{\nu} + \gamma_{mn}(x,y) dy^{m} dy^{n}.
\end{equation}
To interpret the theory as an effective four-dimensional one, the
field $\hat{\phi}(x,y)$ depending on both coordinates is expanded
in a Fourier series over the compact space
\begin{equation}
\hat{\phi} (x,y) = \sum_{n} \phi^{(n)}(x) Y_{n}(y), \label{KKexp}
\end{equation}
 where $Y_{n}(y)$ are orthogonal normalized
eigenfunctions of the Laplace operator $\Delta_{K_{d}}$ on the
internal space $K_{d}$,
\begin{equation}\label{Y-ef}
  \Delta_{K_{d}} Y_{n}(y) = \frac{\lambda_{n}}{R^{2}} Y_{n}(y).
\end{equation}

The coefficients $\phi^{(n)}(x)$ of the Fourier expansion
(\ref{KKexp}) are called the Kaluza-Klein modes and play the role
of fields of the effective four-dimensional theory. Their masses
are given by
\begin{equation}\label{mass}
  m_{n}^{2} = m^{2} + \frac{\lambda_{n}}{R^{2}},
\end{equation}
where $R$ is the radius of the compact dimension.

 The coupling constant $g_{(4)}$ of the 4-dimensional theory
is related to the coupling constant $g_{(4+d)}$ of the initial
(4+d)-dimensional one by
\begin{equation}\label{coupl}
  g_{(4)} = \frac{g_{(4+d)}}{V_{(d)}},
\end{equation}
$V_{(d)}\propto R^{d}$ being the volume of the space of extra
dimensions.

\underline{Low scale gravity}

Consider now the Einstein $(4+d)$-dimensional gravity with the
action
\[
S_{E} = \int d^{4+d}\hat{x} \sqrt{-\hat{G}} \frac{1}{16 \pi
G_{N(4+d)}} {\cal R}^{(4+d)} [\hat{G}_{MN}],
\]
where the scalar curvature ${\cal R}^{(4+d)} [\hat{G}_{MN}]$ is
calculated using the metric $\hat{G}_{MN}$. Performing the mode
expansion and integrating over $K_{d}$, one arrives at the
four-dimensional action
\[
S_{E} = \int d^{4}x \sqrt{-g} \left\{ \frac{1}{16 \pi G_{N(4)}}
{\cal R}^{(4)} [g^{(0)}_{MN}] + \mbox{non-zero KK modes} \right\},
\]
 Similar to eq.(\ref{coupl}), the relation between the 4-dimensional and
$(4+d)$-dimensional gravitational (Newton) constants is given by
\begin{equation}\label{G-rel}
  G_{N(4)} = \frac{1}{V_{(d)}} G_{N(4+d)}.
\end{equation}
One can rewrite this relation in terms of the 4-dimensional Planck
mass $M_{Pl} = (G_{N(4)})^{-1/2} = 1.2 \cdot 10^{19}\; \mbox{GeV}$
and a fundamental mass scale of the $(4+d)$-dimensional theory $M
\equiv (G_{N(4+d)})^{-\frac{1}{d+2}}$. One gets
\begin{equation}\label{M-rel}
  M_{Pl}^{2} = V_{(d)} M^{d+2}.
\end{equation}
This formula is often referred to as the reduction formula.

\begin{figure}[htb]
\begin{center}
\leavevmode
\includegraphics[width=0.33\textwidth]{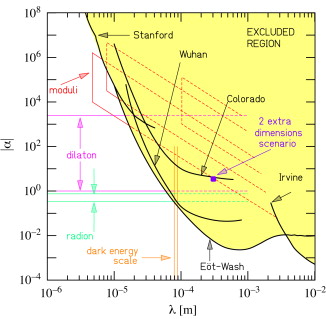}
\end{center}\caption{The allowed region in parameter space  for the modified Newton potential}
\label{newforce}
\end{figure}

The presence of ED leads to the modification of classical gravity.
The Newton potential between two test masses $m_{1}$ and $m_{2}$,
separated by a distance $r$, is in this case equal to
\[
V(r) = G_{N(4)} m_{1}m_{2} \sum_{n} \frac{1}{r} e^{-m_{n}r} =
G_{N(4)} m_{1}m_{2} \left( \frac{1}{r} +
   \sum_{n \neq 0} \frac{1}{r} e^{-|n|r/R}  \right).
\]
The first term in the last bracket is the contribution of the
usual massless graviton (zero mode) and the second term is the
contribution of the massive gravitons. For the size $R$ large
enough (i.e. for the spacing between the modes small enough) this
sum can be replaced by the integral and one gets \cite{ADD}
\begin{equation}
  V(r)  =   G_{N(4)} \frac{m_{1}m_{2}}{r} \left[
1\! + \!S_{d-1} \int_{1}^{\infty}\!\! \!\!\!e^{-mr/R} m^{d-1} dm \right]
    =   G_{N(4)} \frac{m_{1}m_{2}}{r} \left[
1\! +\! S_{d-1} \left( \frac{R}{r} \right)^{d} \int_{r/R}^{\infty}\!\! \!\!\!
e^{-z} z^{d-1} dz \right], \label{Npot}
\end{equation}
 where $S_{d-1}$ is the area of the $(d-1)$-dimensional sphere
 of the unit radius. This leads to the following
behaviour of the potential at short and long distances
\begin{equation}\label{pot}
  V\approx \left\{ \begin{array}{ll} G_{N(4)} \frac{\displaystyle m_{1}
  m_{2}}{\displaystyle r}& r \gg R,\\
G_{N(4)} \frac{\displaystyle m_{1} m_{2}}{\displaystyle r} S_{d-1}
     \left( \frac{\displaystyle R}{\displaystyle r} \right)^{d} \Gamma (d) =
     G_{N(4+d)} \frac{\displaystyle m_{1} m_{2}}{\displaystyle r^{d+1}}
      S_{d-1} \Gamma (d) & r \ll R,
     \end{array} \right.
\end{equation}
The attempts to observe the modification of the Newton law did not come out with a positive result but the accuracy was increased by two orders of magnitude. In Fig.\ref{newforce}~\cite{newton} we show the 
allowed regions in parameter space for the modified potential of the  form $V=-G\frac{m_1m_2}{r}(1+\alpha e^{-r/\lambda})$.

\underline{The ADD model}

The ADD model was proposed  by N. Arkani-Hamed, S. Dimopoulos and
G. Dvali in Ref. \cite{ADD}. The model includes the SM localized
on a 3-brane embedded into the $(4+d)$-dimensional space-time with
compact extra dimensions. The gravitational field is the only
field which propagates in the bulk.

 To analyze the field content of the effective
(dimensionally reduced) four-dimensional model, consider the field
$\hat{h}_{MN}(x,y)$ describing the linear deviation of the metric
around the $(4+d)$-dimensional Minkowski background $\eta_{MN}$
\begin{equation}\label{ADD:Gh}
  \hat{G}_{MN} (x,y) = \eta_{MN} +
\frac{2}{M^{1+d/2}} \hat{h}_{MN}(x,y)
\end{equation}
Let us assume, for simplicity, that the space of extra dimensions
is the $d$-dimensional torus. Performing the KK mode expansion
\begin{equation}\label{ADD-exp}
  \hat{h}_{MN} (x,y) =
\sum_{n} h^{(n)}_{MN} (x) \frac{1}{\sqrt{V_{(d)}}}
\exp(-i\frac{n_{m}y^{m}}{R}),
\end{equation}
where $V_{(d)}$ is the volume of the space of extra dimensions, we
obtain the KK tower of states $h^{(n)}_{MN} (x)$ with masses
\begin{equation}\label{ADD-mass}
m_{n} = \frac{1}{R} \sqrt{n_{1}^{2}+n_{2}^{2}+ \ldots + n_{d}^{2}}
\equiv \frac{|n|}{R},
\end{equation}
 so that the mass splitting  is $\Delta m \propto 1/R.$

The interaction of the KK modes $h^{(n)}_{MN}(x)$ with fields on
the brane is determined by the universal minimal coupling of the
$(4+d)$-dimensional theory
\[
S_{int} = \int d^{4+d}\hat{x} \sqrt{-\hat{G}} \hat{T}_{MN}
\hat{h}^{MN} (x,y),
\]
where the energy-momentum tensor of the matter  localized on the
brane at $y=0$ has the form
\[
\hat{T}_{MN}(x,y) = \delta_{M}^{\mu} \delta_{N}^{\nu} T_{\mu
\nu}(x) \delta^{(d)}(y).
\]
Using the reduction formula (\ref{M-rel}) and the KK expansion
(\ref{ADD-exp}), one obtains that
\begin{equation}
  S_{int} =   \int d^{4}x
T_{\mu \nu} \sum_{n} \frac{1}{M^{1+d/2} \sqrt{V_{(d)}}} h^{(n)\mu
\nu} (x) =  \sum_{n} \int d^{4}x \frac{1}{M_{Pl}} T^{\mu \nu}(x)
h^{(n)}_{\mu \nu}(x), \label{ADD-int1}
\end{equation}
which is the usual interaction of matter with gravity suppressed
by $M_{Pl}$.

The degrees of freedom of the four-dimensional theory, which
emerge from the multidimensional metric, include~\cite{GRW,Hew}
\begin{enumerate}
  \item the massless graviton and the massive KK gravitons $h^{(n)}_{\mu\nu}$
(spin-2 fields) with masses given by eq.(\ref{ADD-mass});
  \item $(d-1)$ KK towers of spin-1 fields which do not couple to $T_{\mu \nu}$;
  \item $(d^{2}-d-2)/2$ KK towers of real scalar fields (for $d \geq 2$), they
do not couple to $T_{\mu \nu}$ either;
\item  a KK tower of scalar fields coupled to the trace of the
energy-momentum tensor $T_{\mu}^{\mu}$, its zero mode is called
radion and describes fluctuations of the volume of extra
dimensions.
\end{enumerate}
Alternatively, one can consider the $(4+d)$-dimensional theory
with the $(4+d)$-dimensional massless graviton $\hat{h}_{MN}(x,y)$
interacting with the SM fields with couplings $\sim 1/M^{1+d/2}$.

In the 4-dimensional  picture the coupling of each individual
graviton (both massless and massive) to the SM fields is small
$\sim 1/M_{Pl}$. However, the smallness of the coupling constant
is compensated by the high multiplicity of states with the same
mass. Indeed, the number $d{\cal N}(|n|)$ of modes with the
modulus $|n|$ of the quantum number being in the interval
$(|n|,|n|+d|n|)$ is equal to
\begin{equation}\label{ADD-dN}
  d{\cal N}(|n|) = S_{d-1} |n|^{d-1}d|n| = S_{d-1} R^{d}
m^{d-1}dm \sim S_{d-1} \frac{M_{Pl}}{M^{d+2}} m^{d-1}dm,
\end{equation}
 where we used the mass formula $m=|n|/R$
and the reduction formula (\ref{M-rel}). The number of KK
gravitons $h^{(n)}$ with masses $m_{n} \leq E < M$ is equal to
\[
{\cal N}(E) \sim \int_{0}^{ER} d{\cal N}(|n|) \sim S_{d-1}
\frac{M^{2}_{Pl}}{M^{d+2}} \int_{0}^{E} m^{d-1}dm =
\frac{S_{d-1}}{d} \frac{M^{2}_{Pl}}{M^{d+2}} E^{d} \sim R^{d}
E^{d}.
\]
 One can see that for $E \gg R^{-1}$ the multiplicity of states which
can be produced is large. Hence, despite the fact that due to
eq.(\ref{ADD-int1}) the amplitude of emission of the mode $n$ is
${\cal A} \sim 1/M_{Pl}$, the total combined rate of emission of
the KK gravitons with masses $m_{n} \leq E$ is
\begin{equation}\label{ADD-rate}
  \sim \frac{1}{M_{Pl}^{2}} {\cal N}(E) \sim
\frac{E^{d}}{M^{d+2}}.
\end{equation}
 We can see that there is a considerable
enhancement of the effective coupling due to the large phase space
of KK modes or due to the large volume of the space of extra
dimensions. Because of this enhancement the cross-sections of
processes involving the production of KK gravitons may turn out to
be quite noticeable at future colliders.

\underline{HEP phenomenology}

There are two types of processes at high energies in which the
effect of the KK modes of the graviton can be observed in running
or planned experiments. These are the graviton emission and
virtual graviton exchange processes~\cite{GRW}-\cite{ChK}.

We start with the graviton emission, i.e., the reactions where the
KK gravitons are created as final state particles. These particles
escape from the detector so that a characteristic signature of
such processes is missing energy. Though the rate of production of
each individual mode is suppressed by the Planck mass, due to the
high multiplicity of KK states the magnitude of the total rate of
production is determined by the TeV scale (see
eq.(\ref{ADD-rate})). Taking eq.(\ref{ADD-dN}) into account, the
relevant differential cross section \cite{GRW} is
\begin{equation}\label{ADD-sigma}
  \frac{d^{2}\sigma}{dt dm} \sim  S_{d-1}
\frac{M^{2}_{Pl}}{M^{d+2}} m^{d-1} \frac{d \sigma_{m}}{dt} \sim
\frac{1}{M^{d+2}},
\end{equation}
 where $d \sigma_{m}/dt $ is the differential cross section of
the production of a single KK mode with mass $m$.

\begin{figure}[htb]
\begin{center}
\leavevmode
\includegraphics[width=0.40\textwidth,height=5.5cm]{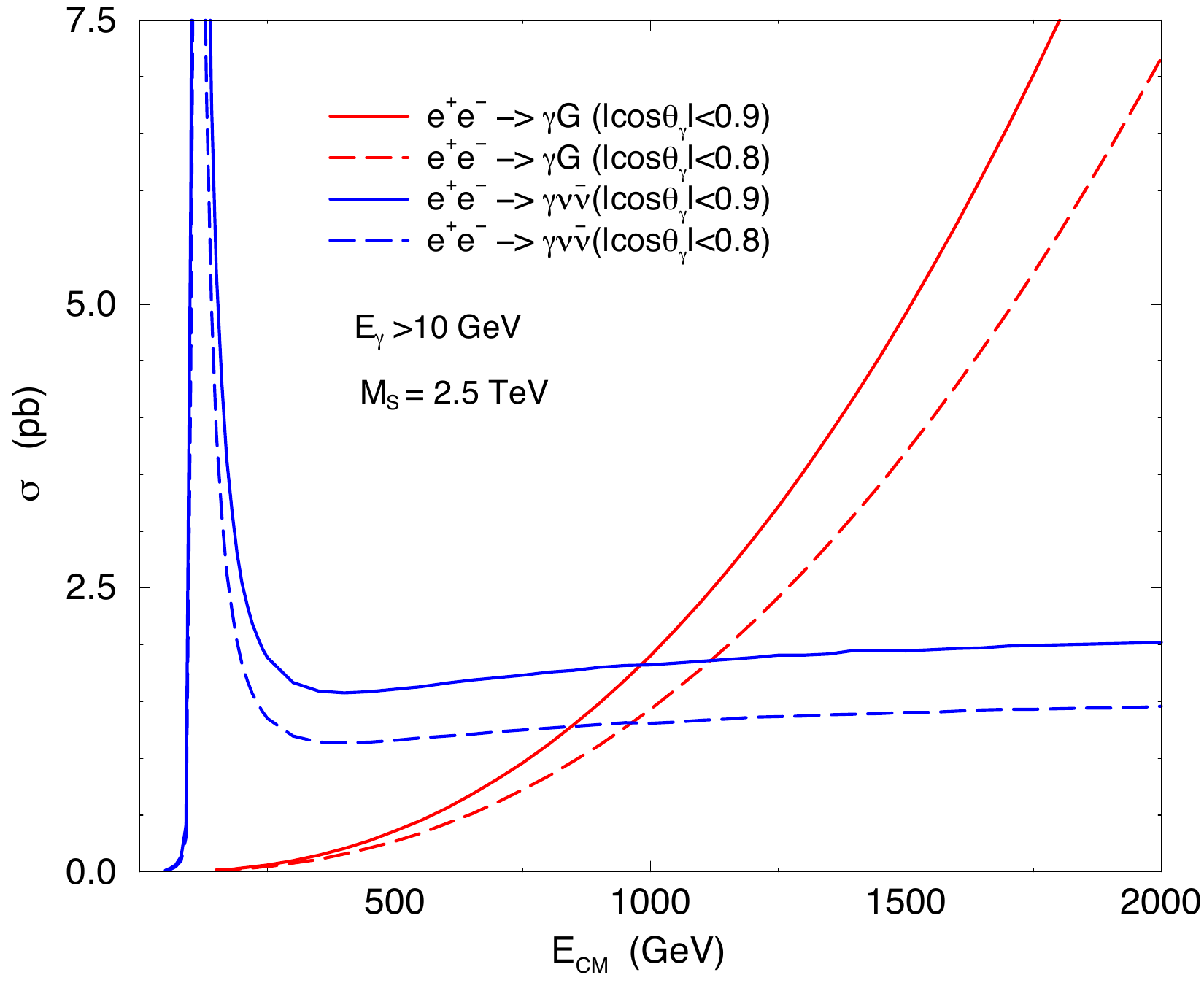}\hspace{1cm}
\includegraphics[width=0.42\textwidth, height=5.5cm]{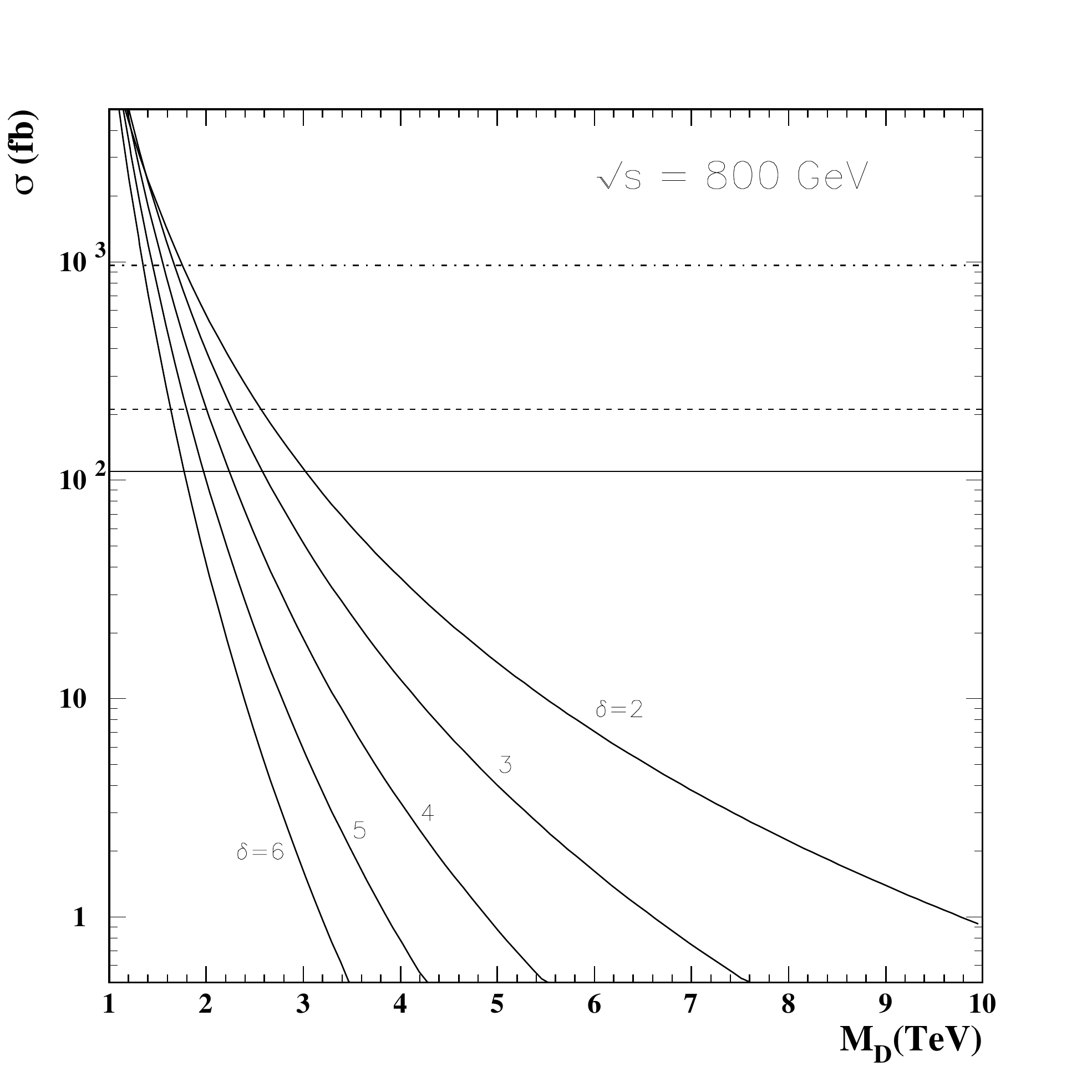}
\end{center}
\caption{The total cross sections for $e^{+}e^{-} \rightarrow
\gamma \nu_{i} \bar{\nu}_{i}$ ($i=e,\mu,\tau$)  and $e^{+}e^{-}
\rightarrow \gamma h$ (faster growing curves) for $d=2$ and $M =
2.5$ TeV \cite{ChK} (left) and the  cross section for $e^{+}e^{-}
\rightarrow \gamma h^{(n)}$ at $\sqrt{s}=800$ GeV as a function of
the scale $M$  for A different number  $\delta$ of extra
dimensions (right). Horizontal lines indicate the background.
\cite{TDR-Wil}} \label{fig:ADD-ee}
\end{figure}

At $e^{+}e^{-}$ colliders the main contribution comes from the
$e^{+}e^{-} \rightarrow \gamma h^{(n)}$ process. The main
background comes from the process $e^{+}e^{-} \rightarrow \nu
\bar{\nu} \gamma$ and can be effectively suppressed by using
polarized beams. Figure \ref{fig:ADD-ee} shows the total cross
section of the graviton production in electron-positron collisions
\cite{ChK}. To the right is  the same cross section as a function
of $M$ for $\sqrt{s} = 800 \; \mbox{GeV}$ \cite{TDR-Wil}.

Effects due to gravitons can also be observed at hadron colliders.
A characteristic process at the LHC would be $pp \rightarrow
(\mbox{jet} + \mbox{missing} \; E)$. The subprocess that gives the
largest contribution is the quark-gluon collision $qg \rightarrow
qh^{(n)}$. Other subprocesses are $q\bar{q} \rightarrow gh^{(n)}$
and $gg \rightarrow gh^{(n)}$.

\begin{figure}[htb]
\begin{center}
\leavevmode
\includegraphics[width=0.38\textwidth]{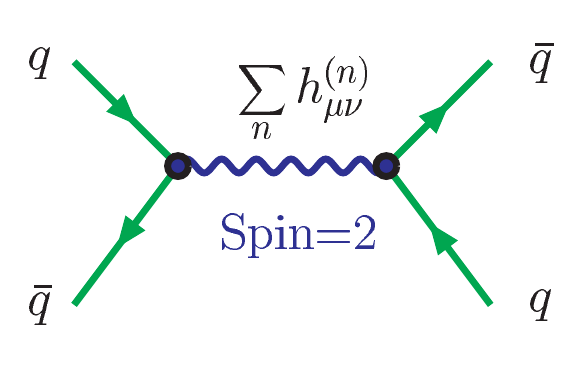}\hspace{1cm}
\includegraphics[width=0.40\textwidth]{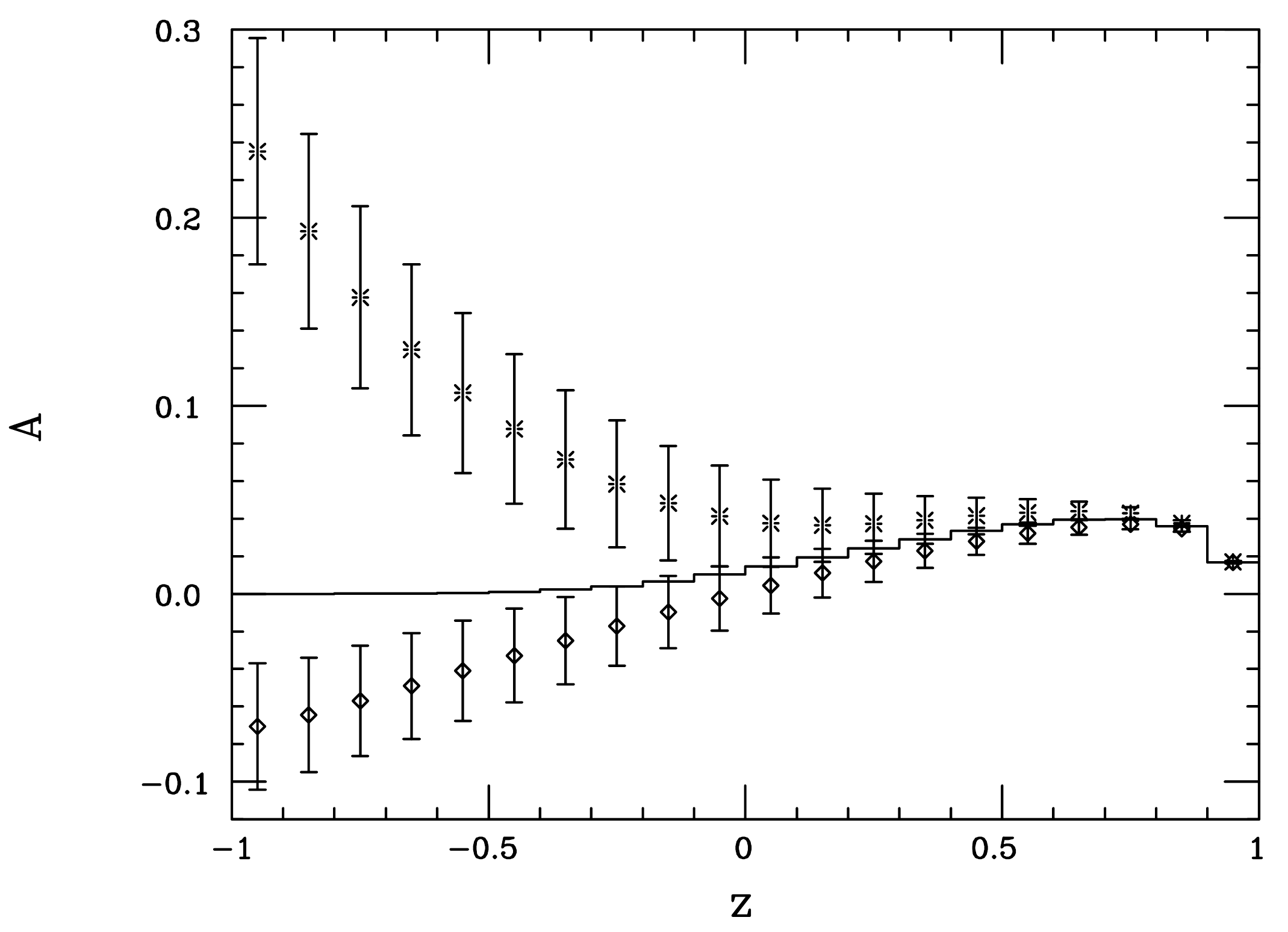}
\end{center}
\caption{
The Feynman diagram for the virtual graviton exchange
(left) and deviation from the expectations of the SM (histogram)
for the Bhabha scattering at a $500 \; \mbox{GeV}$ $e^{+}e^{-}$
collider for the Left-Right polarization asymmetry as a function
of $z=\cos \theta$ for $M = 1.5 \; \mbox{TeV}$ and the integrated
luminosity ${\cal L}=75 \; \mbox{fb}^{-1}$ (right) \cite{Ri99}.} \label{grav}
\end{figure}

Processes of another type, in which the effects of extra
dimensions can be observed, are exchanges of virtual KK modes, in
particular, the virtual graviton exchanges. Contributions to the
cross section from these additional channels lead to  deviation
from the behaviour expected in the 4-dimensional model.  An
example is $e^{+}e^{-} \rightarrow f \bar{f}$ with $h^{(n)}$ being
the intermediate state (see Fig.\ref{grav}). Moreover, gravitons
can mediate processes absent in the SM at the tree-level, for
example, $e^{+}e^{-} \rightarrow HH$, $e^{+}e^{-} \rightarrow gg$.
Detection of such events with  large  cross sections may serve as
an indication of the existence of extra dimensions.

The $s$-channel amplitude of a graviton-mediated scattering
process is given by
\begin{equation}\label{amp}
{\cal A} = \frac{1}{M_{Pl}^{2}} \sum_{n} \left\{ T_{\mu \nu}
\frac{P^{\mu \nu} P^{\rho \sigma}}{s-m_{n}^{2}} T_{\rho \sigma} +
\sqrt{\frac{3(d-1)}{d+2}} \frac{T^{\mu}_{\mu}
T^{\nu}_{\nu}}{s-m_{n}^{2}} \right\},
\end{equation}
where $P_{\mu \nu}$ is the polarization factor coming from the
propagator of the massive graviton and $T_{\mu \nu}$ is the
energy-momentum tensor \cite{GRW}. It contains  a kinematic factor
\begin{eqnarray}
  {\cal S} & = &  \frac{1}{M_{Pl}^{2}}
\sum_{n}\frac{1}{s-m_{n}^{2}} \approx \frac{1}{M_{Pl}^{2}} S_{d-1}
\frac{M_{Pl}^{2}}{M^{d+2}}
\int^{\Lambda} \frac{m^{d-1}dm}{s-m^{2}}  \nonumber \\
   & = &  \frac{S_{d-1}}{2M^{4}}
\left\{ i\pi \left( \frac{s}{M^{2}} \right)^{d/2-1}  +
\sum_{k=1}^{[(d-1)/2]} c_{k} \left( \frac{s}{M^{2}} \right)^{k-1}
\left( \frac{\Lambda}{M} \right)^{d-2k} \right\}.   \label{kin}
\end{eqnarray}
Since the integrals are divergent for $d \geq 2$, the cutoff
$\Lambda$ was introduced. It sets the limit of applicability of
the effective theory. Because of the cutoff,the amplitude cannot
be calculated explicitly without the knowledge of a full
fundamental theory. Usually, in the literature it is assumed that
the amplitude is dominated by the lowest-dimensional local
operator (see \cite{GRW}).

The characteristic feature of expression (\ref{kin}) different
from the 4-dimensional model is the increase of the cross section
with energy. This is a consequence of the exchange of the infinite
tower of the KK modes. Note, however, that this result is based on
a tree-level amplitude, while the radiative corrections in this
case are power-like and may well change this behaviour.

Typical processes, in which the virtual exchange via massive
gravitons can be observed, are: (a) $e^{+}e^{-} \rightarrow \gamma
\gamma$; (b) $e^{+}e^{-} \rightarrow f \bar{f}$, for example the
Bhabha scattering $e^{+}e^{-} \rightarrow e^{+}e^{-}$ or M\"oller
scattering $e^{-}e^{-} \rightarrow e^{-}e^{-}$; (c) graviton
exchange contribution to the Drell-Yang production. A signal of
the KK graviton mediated processes is the deviation in the number
of events and in the left-right polarization asymmetry from those
predicted by the SM (see Figs.~\ref{grav}) \cite{Ri99}.

\subsection{Large Extra Dimensions}

The alternative to compact ED are the large ones which we do not see for the reason that observables are localized on a 4-dimensional hyper surface called {\it brane}. The particles can be 
pressed to the brane by some force, and to leave the brane, they have to gain  high energy.

\begin{figure}[htb]
\begin{center}
\leavevmode
\includegraphics[width=0.30\textwidth]{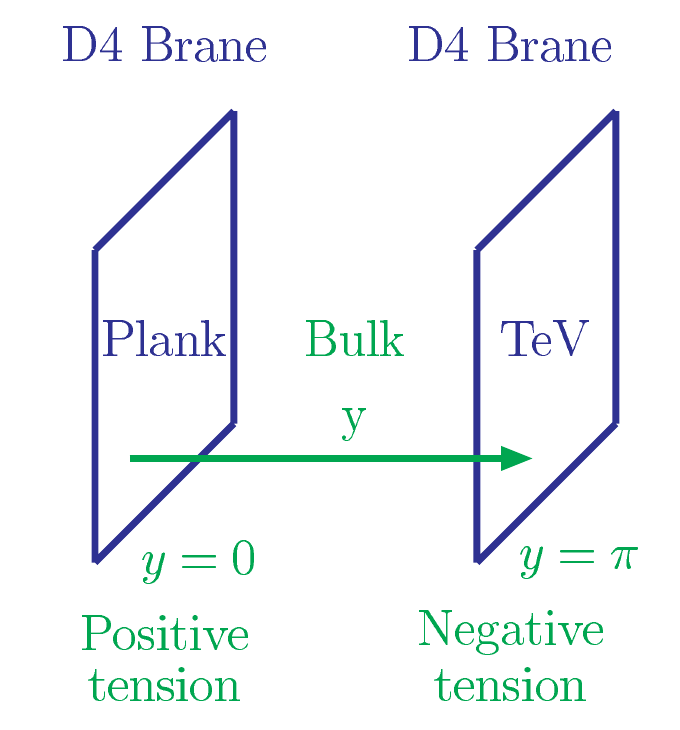}
\end{center}
\caption{The Randall-Sundrum construction of the extra-dimensional
space} \label{ransun}
\end{figure}

The Randal-Sundrum model~\cite{RS} is a model of Einstein gravity in the
five-dimensional Anti-de Sitter space-time with extra dimension
being compactified to the orbifold $S^{1}/Z_{2}$. There are two
3-branes in the model located at the fixed points $y=0$ and $y=\pi
R$ of the orbifold, where $R$ is the radius of the circle $S^{1}$.
The brane at $y=0$ is usually referred to as A Planck brane,
whereas the brane at $y=\pi R$ is called A TeV brane (see
Fig.\ref{ransun}). The SM fields are constrained to the TeV brane,
while gravity propagates in additional dimension.

The action of the model is given by
\begin{eqnarray}
  S & = &  \int d^{4}x \int_{-\pi R}^{\pi R} dy \sqrt{-\hat{G}}
\left\{ 2 M^{3} {\cal R}^{(5)} \left[\hat{G}_{MN}\right] + \Lambda
\right\} \nonumber \\
   & +&  \int_{B_{1}} d^{4}x \sqrt{-g^{(1)}} \left( L_{1} - \tau_{1}
\right) + \int_{B_{2}} d^{4}x \sqrt{-g^{(2)}} \left( L_{2} -
\tau_{2} \right),
           \label{RS:L}
\end{eqnarray}
where ${\cal R}^{(5)}$ is the five-dimensional scalar curvature,
$M$ is the mass scale (the five-dimensional "Planck mass") and
$\Lambda$ is the cosmological constant; $L_{j}$ is a matter
Lagrangian and $\tau_{j}$ is a constant vacuum energy on brane $j$
$(j=1,2)$.

The RS  solution describes the space-time with nonfactorizable
geometry with the metric  given by
\begin{equation}\label{RS:RS1}
ds^{2} = e^{-2\sigma (y)} \eta_{\mu \nu} dx^{\mu} dx^{\nu} +
dy^{2}.
\end{equation}
The additional coordinate changes inside the  interval $-\pi R < y
\leq \pi R$ and the function $\sigma (y)$ in the warp factor $\exp
(-2\sigma)$ is equal to
\begin{equation}\label{RS:sigma}
  \sigma (y) = k |y|, \; \; \; (k > 0).
\end{equation}
For the solution to exist the parameters must be fine-tuned to
satisfy the relations
\[
\tau_{1} = - \tau_{2} = 24 M^{3} k, \; \; \; \Lambda = 24 M^{3}
k^{2}.
\]
Here $k$ is a dimensional parameter which was introduced for
convenience. This fine-tuning is equivalent to the usual
cosmological constant problem. If $k > 0$, then  the tension on
brane 1 is positive, whereas the tension $\tau_{2}$ on brane 2 is
negative.

For a certain choice of the gauge the most general perturbed
metric is given by
\[
ds^{2} = e^{-2k|y|} \left( \eta_{\mu \nu} + \tilde{h}_{\mu
\nu}(x,y) \right) dx^{\mu} dx^{\nu} + (1 + \phi (x) ) dy^{2}.
\]
and describes the  graviton field $\tilde{h}_{\mu \nu}(x,y)$ and
the radion field $\phi (x)$~\cite{ADM}.

As the next step, the field $h_{\mu \nu}(x,y)$ is decomposed over an
appropriate system of orthogonal and normalized functions:
\begin{equation}\label{RS:decomp}
  h_{\mu \nu} (x,y) = \sum_{n=0}^{\infty} h_{\mu \nu}^{(n)}(x)
  \frac{\chi_{n}(y)}{R}.
\end{equation}
The particles localized on the branes are:\vspace{0.3cm}

\begin{minipage}[h]{7cm}
 \underline{Brane 1 (Planck):}
\begin{itemize}
  \item massless graviton $h_{\mu \nu}^{(0)}(x)$,
  \item massive KK gravitons $h_{\mu \nu}^{(n)}(x)$ with masses
  $m_{n}=\beta_{n} k e^{-\pi k R}$, where
  $\beta_{n} = 3.83, 7.02, 10.17, 13.32, \ldots$
  are the roots of the Bessel function,
  \item massless radion $\phi (x)$.
\end{itemize}
\end{minipage}\vspace{-4.3cm}

\hspace*{8cm}\begin{minipage}[h]{7cm}
 \underline{Brane 2 (TeV):}
\begin{itemize}
  \item massless graviton $h_{\mu \nu}^{(0)}(x)$,
  \item massive KK gravitons $h_{\mu \nu}^{(n)}(x)$ with masses
  $m_{n}=\beta_{n} k $,
  \item massless radion $\phi (x)$.
\end{itemize}
\end{minipage}\vspace{1cm}

The brane 2  is most interesting from the point of view of high
energy physics phenomenology.  Because of the nontrivial warp
factor $e^{-2\sigma (\pi R)}$, the Planck mass here is related to
the fundamental 5-dimensional scale $M$ by
\begin{equation}\label{RS:M-Pl}
  M_{Pl}^{2} =
e^{2k\pi R} \int_{-\pi R}^{\pi R} dy e^{-2k|y|} = \frac{M^{3}}{k}
\left( e^{2k \pi R} - 1 \right).
\end{equation}
This way one obtains the solution of the hierarchy problem. The
large value of the 4-dimensional Planck mass is explained by an
exponential wrap factor of geometrical origin, while the scale $M$
stays small.

The general form of the interaction of the fields, emerging from
the five-dimensional metric,  with the matter localized on the
branes is given by the expression:
\[
\frac{1}{2 M^{3/2}} \int_{B_{1}} d^{4}x\ h_{\mu \nu}(x,0)
T^{(1)}_{\mu \nu} + \frac{1}{2 M^{3/2}} \int_{B_{2}} d^{4}x\
h_{\mu \nu}(x,0) T^{(2)}_{\mu \nu} \sqrt{-\det \gamma_{\mu \nu}
(\pi R)}
\]
Decomposing the field $h_{\mu\nu}(x,y)$ according to
(\ref{RS:decomp}) we can write  the interaction Lagrangian  as
\begin{equation}
  \frac{1}{2} \int_{B_{2}} d^{4}z \left[ \frac{1}{M_{Pl}}
h^{(0)}_{\mu \nu}(z) T^{(2)\mu \nu} - \sum_{n=1}^{\infty}
\frac{w_{n}}{\Lambda_{\pi}} h^{(n)}_{\mu \nu} T^{(2)\mu \nu} -
\frac{1}{\Lambda_{\pi}\sqrt{3}}T^{(2)\mu}_{\mu} \right],
\label{RS:int2}
\end{equation}
where $\Lambda_{\pi} = M_{Pl} e^{-k \pi R} \approx \sqrt{M^{3}/k}$
and $M_{Pl}$ is given by eq.(\ref{RS:M-Pl}) .

The massless graviton, as in the standard gravity, interacts with
matter with the coupling $M_{Pl}^{-1}$. The interaction of the
massive gravitons and radion is considerably stronger: their
couplings are $\propto \Lambda_{\pi}^{-1} \sim 1 \;
\mbox{TeV}^{-1}$. If the first few massive KK gravitons have masses
$M_{n} \sim 1$TeV, then this leads to new effects which in
principle can be seen at future colliders. To have this situation,
the fundamental mass scale $M$ and the parameter $k$ are taken to
be $M \sim k \sim 1$TeV.

\underline{HEP phenomenology}

With the mass of the first KK mode $M_{1} \sim 1 \; \mbox{TeV}$
direct searches for the first KK graviton $h^{(1)}$ in the
resonance production at future colliders become quite possible.
Signals of the graviton detection can be~\cite{DHR}\vspace{-0.5cm}

$\bullet$ an excess in the Drell-Yan processes \ \ \ \
$\begin{array}{l}  \\ q \bar{q}\rightarrow h^{(1)} \rightarrow
l^{+}l^{-},\\
gg \rightarrow h^{(1)} \rightarrow l^{+}l^{-}
\end{array}$

 $\bullet$ an excess in the dijet channel \hspace{1.5cm}  $
q \bar{q}, gg \rightarrow h^{(1)} \rightarrow q \bar{q},
gg.$\vspace{0.3cm}

\begin{figure}[htb]
\begin{center}
\leavevmode
\includegraphics[clip, trim = 2.5cm 2.5cm 2.5cm 2.5cm,width=0.45\textwidth]{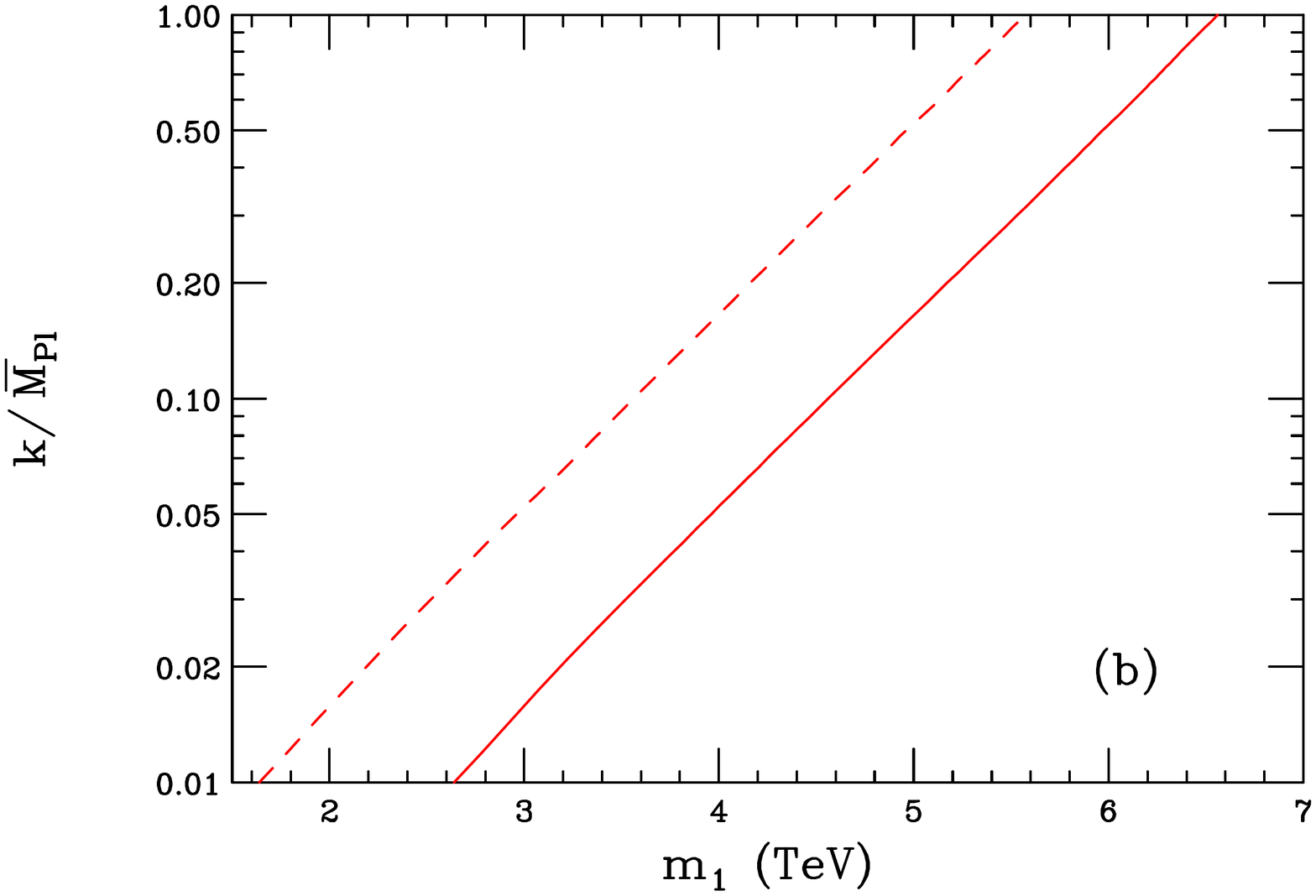}
\includegraphics[width=0.45\textwidth]{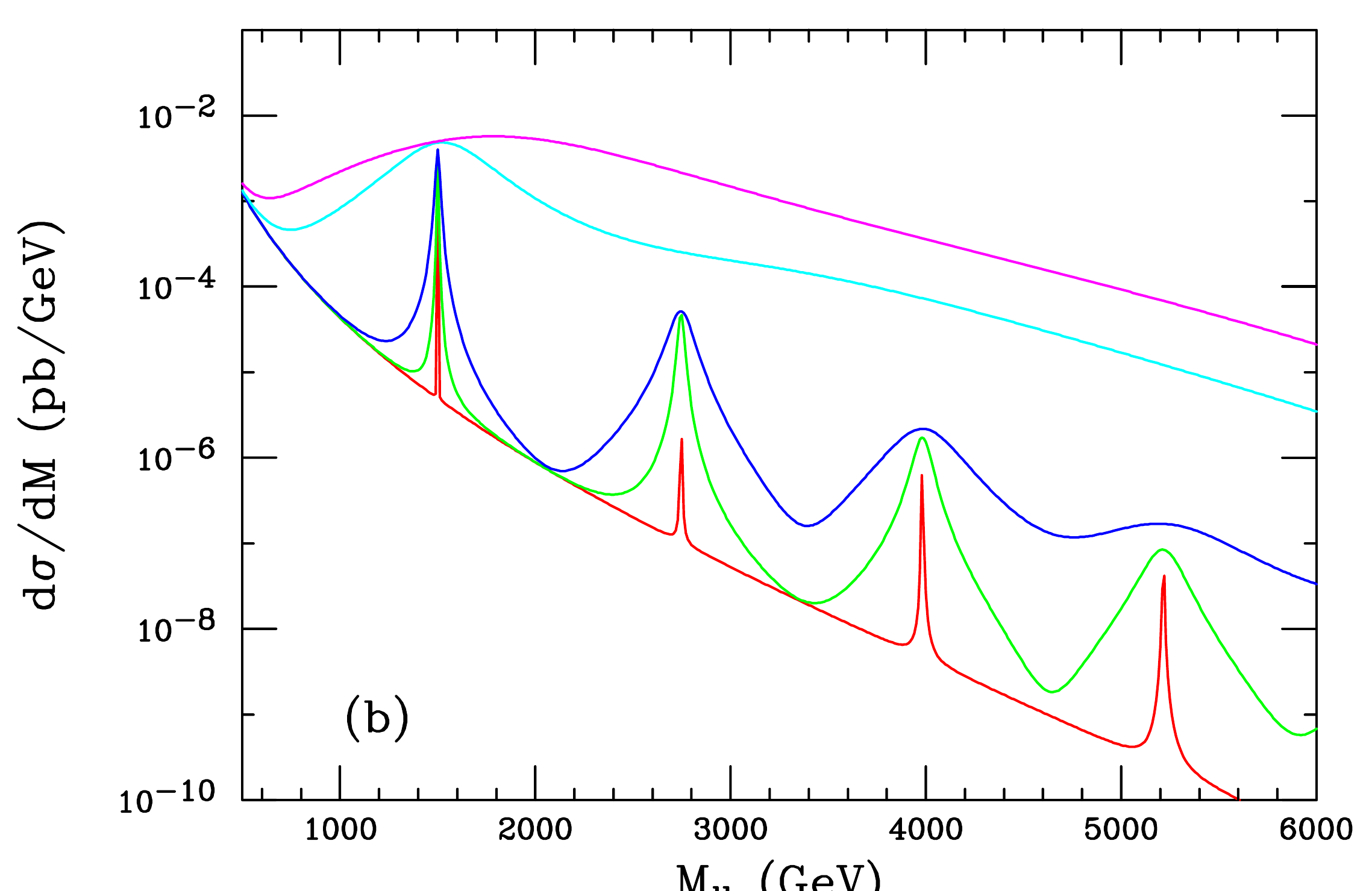}
\end{center}
\caption{Exclusion region for resonance production of the first KK
graviton excitation in the Drell-Yan (corresponding to the
diagonal lines) and dijet (represented by the bumpy curves)
channels at the 
LHC. The dashed and solid
curves correspond to 10, 100  fb$^{-1}$ of integrated luminosity,
respectively (left). Drell-Yan production of the KK graviton  for the LHC (right) for $M_1 = 1500 \text{GeV}$ and its subsequent tower states (right)
} \label{RS-DY}
\end{figure}

\noindent The plots of the exclusion regions for the LHC~\cite{DHR} are presented in Fig.~\ref{RS-DY}.\\

 They show the exclusion region for resonance
production of the first KK graviton excitation in the Drell-Yan
and dijet channels. The excluded region lies above and to the left
of the curves.

The next plots present the behaviour of the cross-section of the
Drell-Yan process as a function of the invariant mass of the final
leptons. It is shown for two values of $M_{1}=1500$  GeV for the LHC in  Fig.~\ref{RS-DY}~\cite{DHR}. One can see the
characteristic peaks in the cross section for one or a series of
massive graviton modes.

\begin{figure}[htb]
\begin{center}
\leavevmode
\includegraphics[width=0.35\textwidth]{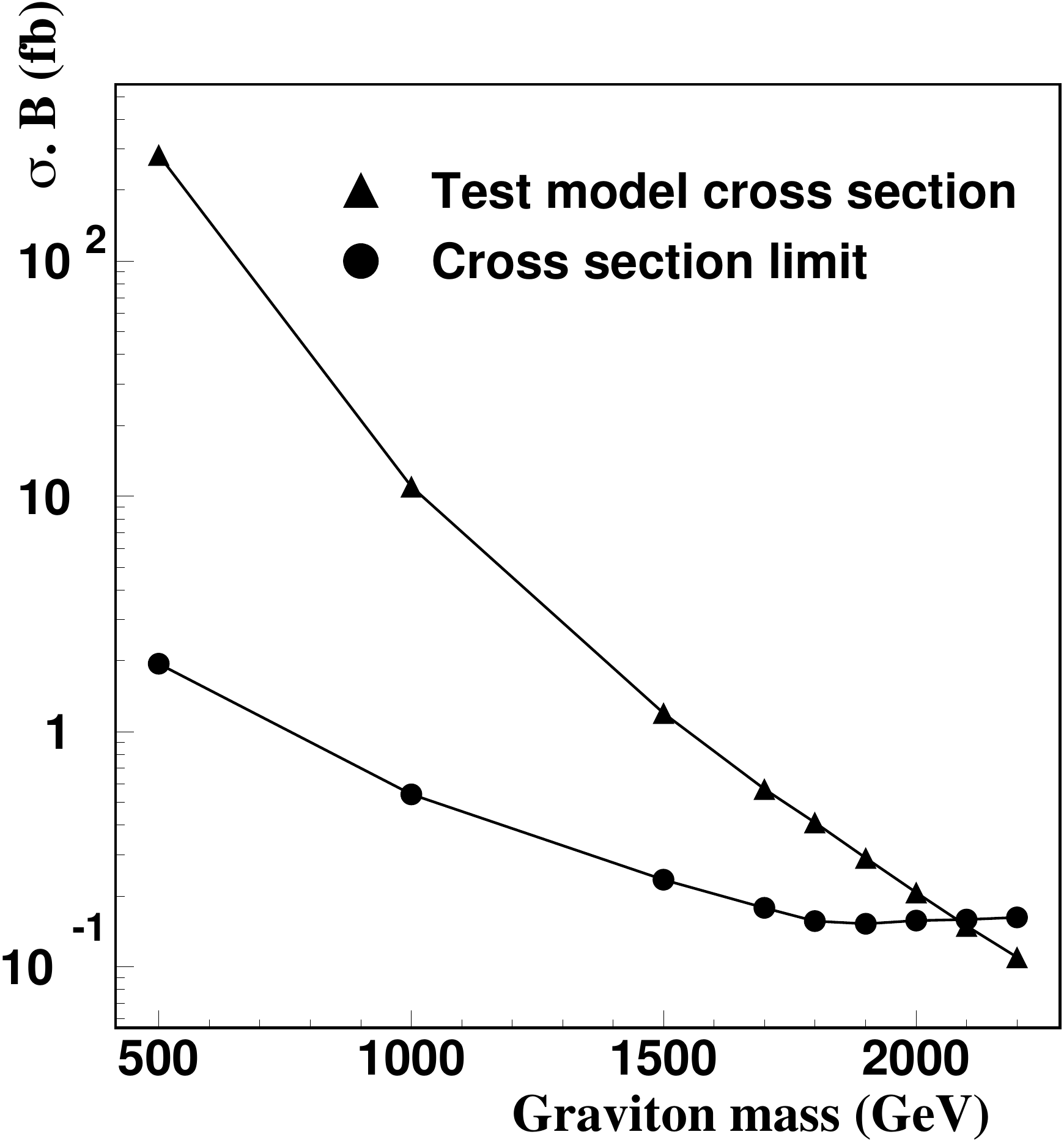}
\includegraphics[width=0.50\textwidth]{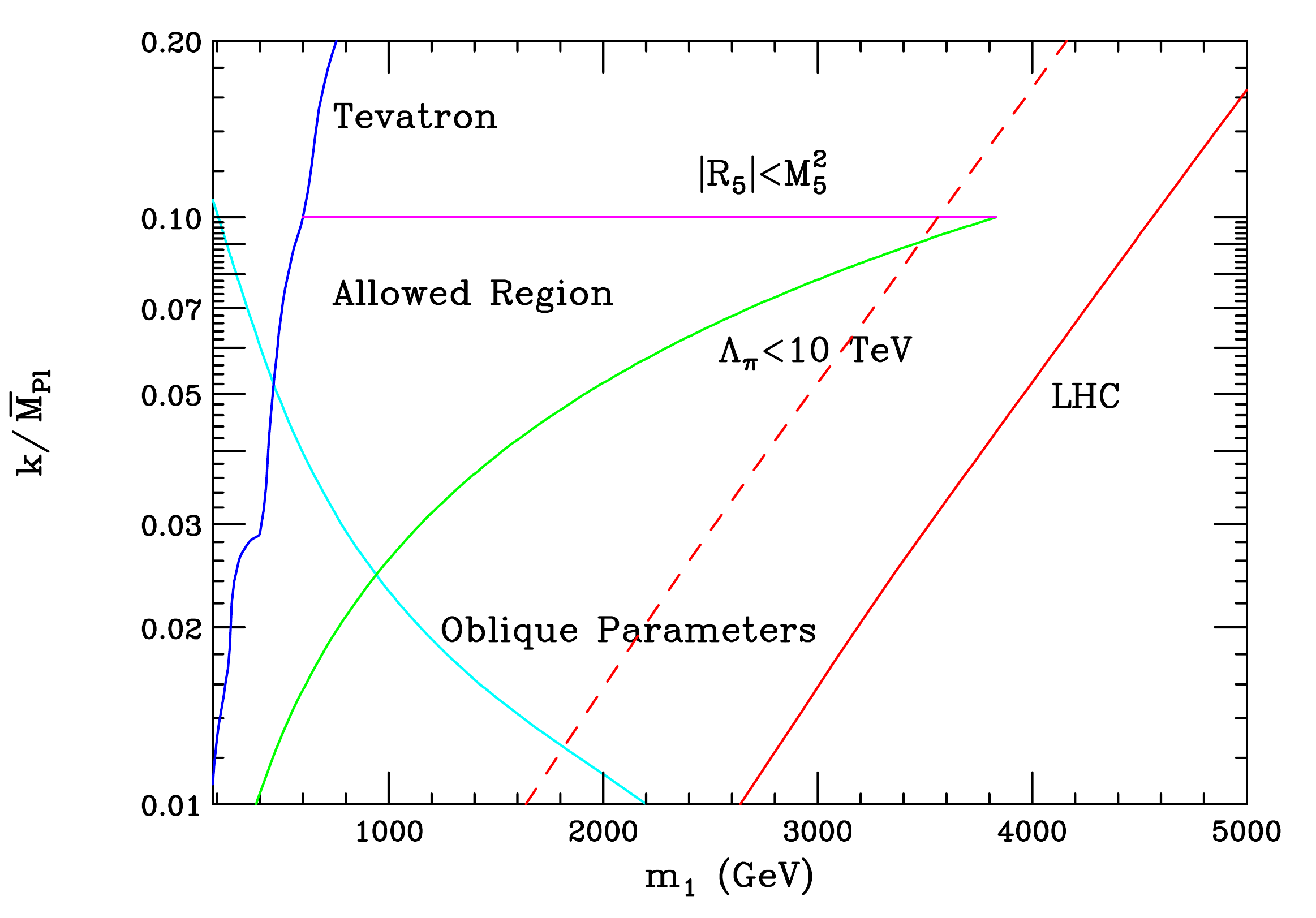}
\end{center}
\caption{The cross-section times branching ratio, $\sigma \cdot
B$, for $h^{(1)} \rightarrow e^{+}e^{-}$ in the RS model and the
smallest detectable cross-section times the branching ratio,
$(\sigma \cdot B)^{min}$ \cite{AOPW} (left)
 and the summary of
experimental and theoretical constraints on the parameters $M_{1}$
and $\eta = (k/M_{Pl}) e^{k\pi R}$ (right) \cite{DHR}. The allowed
region lies as indicated. The LHC sensitivity to graviton
resonances in the Drell-Yan channel is represented by diagonal
dashed and solid curves, corresponding to 10 and 100 fb$^{-1}$ of
integrated luminosity, respectively} \label{RS:G}
\end{figure}

The possibility to detect  the resonance production of the first
massive graviton in the proton - proton collisions $p p
\rightarrow h^{(1)} \rightarrow e^{+}e^{-}$ at the LHC  depends on
the cross section. The main background processes are $p p
\rightarrow Z/\gamma^{*} \rightarrow e^{+}e^{-}$. The estimated
cross section of the process $h^{(1)} \rightarrow e^{+}e^{-}$ as a
function of $M_{1}$ in the RS model is shown in
Fig.~\ref{RS:G}~\cite{AOPW}. One can see that the detection might
be possible if $M_{1} \leq 2080 \; \mbox{GeV}$ .

 To be able to conclude that the observed resonance is a
graviton and not, for example, a spin-1 $Z'$ resonance or a
similar particle, it is necessary to check that it is produced by
a spin-2 intermediate state. The spin of the intermediate state
can be determined from the analysis of the angular distribution
function $f(\theta)$ of the process, where $\theta$ is the angle
between the initial and final beams. This function is
\begin{eqnarray*}
  Spin\ 0 & => &  f(\theta)=1 ,\\
  Spin\ 1 & => &  f(\theta)=1 + \cos^{2} \theta, \\
  Spin\ 2 & => & \left\{ \begin{array}{ll}
q\bar{q} \rightarrow h^{(1)} \rightarrow e^{+}e^{-} & f(\theta) =
1 - 3 \cos^{2} \theta + 4\cos^{4} \theta ,\\
gg \rightarrow h^{(1)} \rightarrow e^{+}e^{-} & f(\theta) = 1 -
\cos^{4} \theta. \end{array} \right.
\end{eqnarray*}
 The analysis, carried out in
Ref.~\cite{AOPW}, shows that angular distributions allow one to
determine the spin of the intermediate state with 90\% C.L. for
$M_{1} \leq 1720$ GeV.

As the next step, it would be important to check the universality of
the coupling of the first massive graviton $h^{(1)}$ by studying
various processes, e.g. $pp \rightarrow h^{(1)} \rightarrow
l^{+}l^{-}, \; \mbox{jets}, \; \gamma \gamma, W^{+}W^{-}, HH$,
etc. If it is kinematically feasible to produce higher KK modes,
measuring the spacings of the spectrum will be another strong
indication in favour of the RS model.

The conclusion is~\cite{DHR}  that with the integrated luminosity
${\cal L} = 100 \; \mbox{fb}^{-1}$ the LHC will be able to cover
the natural region of parameters $(M_{1},\eta = (k/M_{Pl}) e^{k\pi
R})$ and, therefore, discover or exclude the RS model. This is
illustrated in the r.h.s. of  Fig.~\ref{RS:G}.

We finish  with a short summary of the main features of the ADD
and RS models.

\underline{ADD Model}.

\begin{enumerate}
\item The ADD model removes the $M_{EW}/M_{Pl}$ hierarchy, but replaces
it by the hierarchy
$
\frac{R^{-1}}{M} \sim \left( \frac{M}{M_{Pl}} \right)^{2/d} \sim
10^{-\frac{30}{d}}.
$
For $d=2$ this relation gives $R^{-1}/M \sim 10^{-15}$. This
hierarchy is of a different type and might be easier to understand
or explain, perhaps with no need for SUSY;
\item The model predicts the modification of the Newton law at
short distances, which may be checked in precision experiments;
\item For $M$ small enough high-energy physics effects,
predicted by the model, can be discovered at future collider
experiments.
\end{enumerate}

\underline{RS model}

\begin{enumerate}
\item The model solves the $M_{EW}/M_{Pl}$ hierarchy problem
without generating a new hierarchy.

\item A large part of the allowed range of parameters
of the RS model will be studied in future collider experiments,
which will either discover new phenomena or exclude the most
"natural" region of its parameter space.

\item With a mechanism of radion stabilization added the model is quite
viable. In this case, cosmological scenarios, based on the RS
model, are consistent without additional fine-tuning of parameters
(except the cosmological constant problem)~\cite{Cosmoed}.
\end{enumerate}

\section{New Paradigm}
The most radical way out of the SM is the change of the paradigm of local quantum field theory and transition to non-local theories. And the first attempt of this kind is the string theory - the theory of one-dimensional  extended objects~\cite{string}. The natural development of this idea  is the consideration of the objects of an arbitrary dimension  which are called branes (from membrane - two-dimensional surface). The theory of these objects is in progress but some qualitative features are widely discussed.

\subsection{String Theory}
The string theory describes one-dimensional extended objects which in their motion swap a two-dimensional world surface. 
\begin{figure}[htb]\vspace{0.3cm}
\begin{center}
\leavevmode
\includegraphics[width=0.7\textwidth]{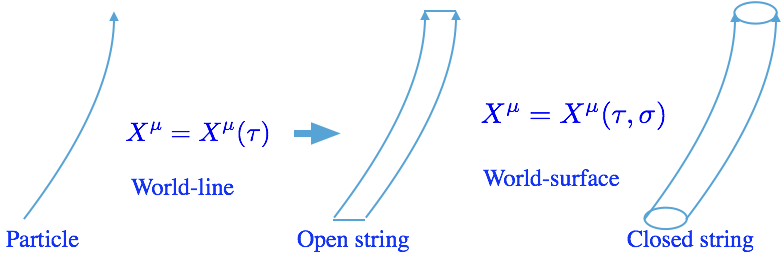}
\end{center}
\caption{From a point-like particle to a one-dimensional string}
\end{figure}
The action for such objects is the straightforward generalization of the action for a point-like particle
\beq
S=-m\int\ d\tau \sqrt{-\frac{dX^\mu}{d\tau}\frac{dX^\nu}{d\tau}\eta_{\mu\nu}} \ \Rightarrow \ 
S=-\frac{1}{2\pi l_S^2}\int\ d^2\sigma \sqrt{-det\left(\frac{dX^\mu}{d\sigma^\alpha}\frac{dX^\nu}{d\sigma^\beta}\eta_{\mu\nu}\right)}.
\eeq 
The strings may be open and closed. The spectrum of string excitations 
\beq
l_S^2M=\sum_n\ N_n(+\bar N_n) \in Z, \ \ \ N_n=\alpha^\mu_{-n}\alpha^\mu_{n},
\eeq
 contains zero modes associated  with observed particles and heavy massive modes.  The lowest string states are:
$$ \begin{array}{lll}
 \mbox{open string} & \alpha^\mu_{-1}|0> \to A^\mu \to \int d^Dx \sqrt{-g}\ tr(F_{\mu\nu}F^{\mu\nu}) 
& \mbox{this state is associated with photon} \\
 \mbox{closed string }& \alpha^\mu_{-1} \bar \alpha^\nu_{-1}|0> \to  g^{\mu\nu}, ... \to \int d^Dx\sqrt{-g} \ R+ ...
 &  \mbox{this state is associated with graviton} 
 \end{array}$$
 The spectrum of open strings contains spin 0, 1/2 and 1 states associated with gauge and matter fields, the spectrum of closed strings contains spin 2 state associated with gravity. Besides the vibrational modes, strings contain also the modes connected with the winding  of the world line on a string. All together these modes define the full spectrum of a string. Thus, for a string on a circle with radius $R$ one has the  momentum states with $M^2=m^2/R^2$,  the winding states with $M^2=\omega^2 R^2/l_S^4$ and the full spectrum $M^2=m^2/R^2+\omega^2 R^2/l_S^4$. The string is characterized by a minimal size called the string length $l_S=\sqrt{\alpha'}$. It is assumed that this size is close to the Planck length.

Quantum theory of strings is formulated in critical dimension of space-time  where it is free from conformal anomalies. For the bosonic string this critical dimension is equal to 26 and for the fermion string to 10. Besides, the string spectrum may contain taxions, particles with negative mass squared. To get rid of these states, one considers a supersymmetric  fermion string which is free from taxions. Its spectrum starts from zero modes which are usually associated with point-like particles of local quantum field theory. 

To get from the string theory the effective 4-dimensional low energy theory containing massless modes, one needs to perform compactification of extra dimensions. The properties of the compact 6-dimensional manifold define the properties of the obtained low energy theory. Thus, the degeneracy of the compact manifold in size and shape manifested in the existence of the scalar fields called moduli, defines the values of the couplings, and different topologies define the symmetry group and the field content of the 4-dimensional theory. The gravity action defined in D dimensions
and the matter field action defined on a p-brane
\beq
S_D=\frac{1}{l_S^{D-2}}\int\ d^Dx\sqrt{-g}\ R+ \cdots+\frac{1}{l_S^{p-3}}\int\ d^{p+1}x\ \sqrt{-\gamma}\ tr(F_{\alpha\beta}F^{\alpha\beta})+\cdots
\eeq
being compactified to 4-dimensions take the form
\beq
S_4=\underbrace{\frac{V}{l_S^{D-2}}}\int\ d^4x\sqrt{-g_4}\ R_4+ \cdots+\underbrace{\frac{v}{l_S^{p-3}}}\int\ d^{4}x\ \sqrt{-g_4}\ tr(F_{\mu\nu}F^{\mu\nu})+\cdots 
\eeq\vspace{-0.6cm}

$ \hspace{2.0cm} \frac{1}{16\pi G_N} \hspace{3.7cm}\frac{1}{16\pi g_{YM}^2} $\vspace{0.5cm}

The existing multiple possibilities of multidimensional theories do not allow one at the moment to  choose the preferable scheme and to make definite predictions.

Phenomenologically, the most acceptable is the so-called heterotic string.  In this case, one has the unification of the gauge and the Higgs fields that allows in particular to predict the coupling constants and get the top-quark mass of the order of 170 GeV. In this theory one also gets the cancellation of anomalies  which is possible for a fixed gauge group of associated GUT: $SO(32)$ or $E_8\times E_8$. This theory possesses the right-handed neutrino and the Majorana mass term,
permits the proton decay. The effective low energy theory gives the desired unification with gravity and contains the mechanism of spontaneous supersymmetry breaking  via effects of supergravity
in the hidden sector.

The string theory contains not only strings but other extended objects of various dimensions.
The emerging picture of the world consists of branes, the open strings end up on the branes and the open strings propagate in the bulk.

\subsection{M-theory and the Theory of Everything }

\begin{figure}[htb]
\begin{center}\vspace{0.3cm}
\leavevmode
\includegraphics[width=0.8\textwidth]{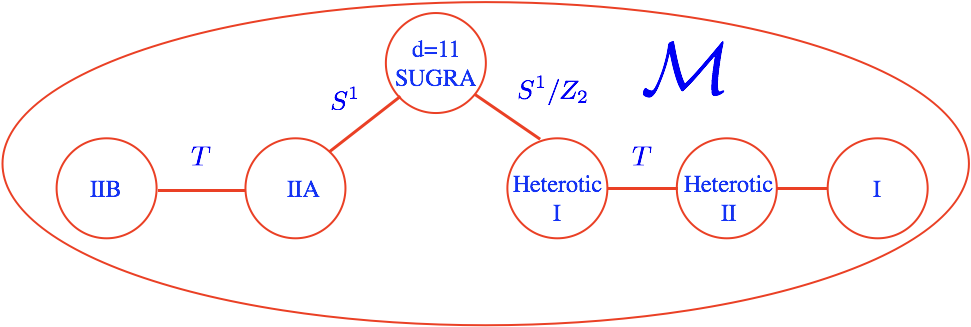}
\caption{The string landscape and  ${\cal M}$ theory}
\label{M}
\end{center}
\end{figure}

There are five types of consistent string theories  free from conformal and gauge anomalies and of taxions (type IIA, type IIB, type I, and two Heterotic)~\cite{heteroic_strings}.  All five string theories are only consistent in 10 space-time dimensions,  all five  have world-sheet supersymmetry and lead to space-time-supersymmetry in 10 dimensions. They are believed to be different vacua of a single unified "theory" called $M-theory$. However, there is no adequate formulation of this theory. The other vacuum of M-theory is 11-dimensional supergravity (see Fig.\ref{M}~\cite{Lukas_talk})
It is assumed that the ultimate unified theory will be the "theory of everything", i.e. will describe on a fundamental level all laws of Nature. The form of this theory, however, is still unknown. It is not clear which degrees of freedom are fundamental. Moreover, is is quite possible that there are different, dual to each other, descriptions of the same reality. The example of such a duality is the so-called AdS/CFT correspondence when some characteristics of a theory can be described as in the framework of the 4-dimensional conformal field theory and also in the  framework of classical gravity in the 5-dimensional de Sitter space~\cite{Maldacena}. Here we are still far from detailed predictions which allow experimental tests.

\section{Conclusion. The priority tasks of high energy physics}

The successes of the Standard Model and the enormous efforts for its tests and search for new physics at accelerators as well as in non-accelerator experiments define the future of high energy physics in the coming years. The experiments at the Large Hadron Collider are at the edge of modern knowledge. The success of these experiments is the success of all high energy physics. However, the peculiarity of the modern situation is that there is no field where we may expect the guaranteed discovery. We make the first steps into the unknown land and try to unveil the mystery. We have to be persistent and patient. There are many theoretical models which suggest new physics at different scales. Which of these models happens to be correct and adequate to Nature we have to find experimentally. Today we may talk about priority tasks. They are:
\begin{itemize}
 \item Investigation of the Higgs sector;
\item  Search for particles of Dark Matter;
\item  Study of  the neutrino properties in non-accelerator experiments;
\item Search for new physics (supersymmetry);
\item The areas that were left behind come to the front:  confinement, exotic hadrons, dense hadron matter
\end{itemize}
Further development of high energy physics crucially depends on the results of these searches.

\vspace{0.3cm} {\large \bf Acknowledgements} \vspace{0.1cm}

The authors would like to express their gratitude to the
organizers of the School for their effort in creating a pleasant atmosphere and support. This work was partly supported by RFBR grant \# 17-0232-00837. I am also grateful to M.Gavrilova for her valuable help in preparing the manuscript.

\end{document}